\documentclass{aa} 

\usepackage{graphicx}
\usepackage{txfonts}
\usepackage{xfrac}
\usepackage[colorlinks=true,urlcolor=blue,citecolor=blue,
            linkcolor=blue]{hyperref}
\usepackage{mhchem}

\def\la{\mathrel{\mathchoice {\vcenter{\offinterlineskip\halign{\hfil
$\displaystyle##$\hfil\cr<\cr\sim\cr}}}
{\vcenter{\offinterlineskip\halign{\hfil$\textstyle##$\hfil\cr
<\cr\sim\cr}}}
{\vcenter{\offinterlineskip\halign{\hfil$\scriptstyle##$\hfil\cr
<\cr\sim\cr}}}
{\vcenter{\offinterlineskip\halign{\hfil$\scriptscriptstyle##$\hfil\cr
<\cr\sim\cr}}}}}
\def\ga{\mathrel{\mathchoice {\vcenter{\offinterlineskip\halign{\hfil
$\displaystyle##$\hfil\cr>\cr\sim\cr}}}
{\vcenter{\offinterlineskip\halign{\hfil$\textstyle##$\hfil\cr
>\cr\sim\cr}}}
{\vcenter{\offinterlineskip\halign{\hfil$\scriptstyle##$\hfil\cr
>\cr\sim\cr}}}
{\vcenter{\offinterlineskip\halign{\hfil$\scriptscriptstyle##$\hfil\cr
>\cr\sim\cr}}}}}

\def\nH{n_{\langle\rm H\rangle}}
\def\NH{N_{\langle\rm H\rangle}}
\def\muH{\mu_{\rm H}}
\def\rhod{\rho_{\rm d}}
\def\pabl#1#2{\frac{\partial #1}{\partial #2}}
\def\vth{v_{\rm th}}
\def\cs{c_{\rm s}}
\def\vz{\langle v_z\rangle}
\def\tstop{\tau_{\rm stop}}
\def\tedd{\tau_{\rm eddy}}
\def\St{{S\hspace*{-0.25ex}t}}

\def\half{\hspace*{0.2pt}\sfrac{1\hspace*{-0.4pt}}{\hspace*{0.2pt}2}}

\begin{document} 

\title{Mixing and diffusion in protoplanetary disc chemistry}

\author{P.~Woitke\inst{1,2,3}
       \and
       A.~M.~Arabhavi\,\inst{2,3,4,5}
       \and
       I.~Kamp\inst{4}
       \and
       W.-F.~Thi\inst{6}
}

\institute{
           Space Research Institute, Austrian Academy of Sciences,
           Schmiedlstr.~6, A-8042, Graz, Austria
         \and
           Centre for Exoplanet Science, University of St Andrews, 
           St Andrews, UK
         \and
           SUPA, School of Physics \& Astronomy, University of St
           Andrews, St Andrews,  KY16 9SS, UK
         \and
           Kapteyn Astronomical Institute, University of Groningen,
           P.O.~Box 800, 9700 AV Groningen, The Netherlands
         \and
           Faculty of Aerospace Engineering, Delft University of
           Technology, Delft, The Netherlands
         \and  
           Max Planck Institute for Extraterrestrial Physics,
           Giessenbachstrasse, 85741 Garching, Germany
}

\date{Received 20/07/2022; accepted xx/xx/2022}

\abstract{We develop a simple iterative scheme to include vertical
  turbulent mixing and diffusion in {\sc ProDiMo} thermo-chemical
  models for protoplanetary discs. The models are carefully checked
  for convergence toward the time-independent solution of the
  reaction-diffusion equations, as e.g.\ used in exoplanet atmosphere
  models. A series of five T\,Tauri disc models is presented where we
  vary the mixing parameter $\alpha_{\rm mix}$ from 0 to $10^{\,-2}$ and
  take into account (a) the radiative transfer feedback of the
  opacities of icy grains that are mixed upward and (b) the feedback
  of the changing molecular abundances on the gas temperature
  structure caused by exothermic reactions and increased line
  heating/cooling. We see considerable changes of the molecular and ice
  concentrations in the disc. The most abundant species (\ce{H2},
  \ce{CH4}, \ce{CO}, the neutral atoms in higher layers, and the ices
  in the midplane) are transported both up and down, and at the
  locations where these abundant chemicals finally decompose, for
  example by photo processes, the release of reaction products has
  important consequences for all other molecules.  This generally
  creates a more active chemistry, with a richer mixture of ionised,
  atomic, molecular and ice species and new chemical pathways that are
  not relevant in the unmixed case. We discuss the impact on three
  spectral observations caused by mixing and find that (i) icy grains
  can reach the observable disc surface where they cause ice
  absorption and emission features at IR to far-IR wavelengths, (ii)
  mixing increases the concentrations of certain neutral molecules
  observable by mid-IR spectroscopy, in particular OH, HCN and
  \ce{C2H2}, and (iii) mixing can change the optical appearance of CO
  in ALMA line images and channel maps, where strong mixing would
  cause the CO molecules to populate the distant midplane.}


\maketitle


\section{Introduction}
\label{sec:intro}

Mixing and diffusion are important processes in chemical models for
planetary atmospheres, see e.g.\ \citet{Moses2011, Hu2012,
  Parmentier2013, Kreidberg2014, Charnay2015, Rimmer2016,
  Zhang2018}. In the deep layers of gas giant atmospheres, for
instance, where densities and temperatures are high and the chemical
reactions are fast, thermo-chemical equilibrium holds. With the
falling temperatures and decreasing densities in higher layers, the
chemical rates decrease and eventually become smaller than the
turbulent eddy-mixing rates. Consequently, the molecular
abundances become constant above a certain pressure level -- an effect
known as ``transport-induced quenching'' in planetary chemistry.  Only
at even higher altitudes, once the atmosphere becomes transparent to
the stellar UV irradiation, the molecular abundances start to change
again quickly due to photo-dissociation and photo-ionisation, beside
other high-energy processes induced by cosmic rays, X-rays, and
stellar energetic particles (SEPs), see e.g.\ \citet{Barth2021}.

Another well-known effect in planetary chemistry is that above a
certain layer called the {\sl homopause}, the gas-kinetic
micro-physical diffusion becomes more important than eddy-diffusion
due to turbulent mixing \citep{Hunten1973, Zahnle2016}. This
leads to molecule-dependent scale heights and changes of element
abundances above the homopause which is essential to understand the
transition into the exosphere.

In protoplanetary disc chemistry, despite the striking physical and
chemical similarities to planetary atmospheres, mixing and diffusion
have been rarely taken into account. The introduction of diffusive
transport terms to the chemical rate equations requires a transition
from first-order ordinary differential equations to second-order
partial differential equations. Since disc chemistry is at least 2D,
this means a considerable numerical and conceptual challenge.

\citet{Ilgner2004} pioneered the research on the impact of eddy mixing
on disc chemistry in the frame of (1+1)\,D viscous disc evolution
models for the innermost 10\,au. They considered the sulphur molecules
CS, SO, SO2, NS and \ce{HCS+} in particular, and concluded that
vertical mixing is the most important process among all transport
processes considered.

\citet{Semenov2006} used a similar approach but included gas-grain
processes. They focused on the CO molecule in the outer disc regions.
They concluded that CO can be considerably enhanced by vertical and
radial mixing in the cold disc regions, where the dust temperature is
lower than about 25\,K and CO is expected to freeze out. The mixing
rates were derived from an $\alpha$ disc model using
$\alpha\!=\!10^{-2}$. They discussed the effects of radial, vertical,
and ``full 2D'' mixing in a time-dependent disc model. Unfortunately,
since this paper only has a single equation, it remains somewhat
unclear how they actually solved the time-dependent partial
differential equations.

\citet{Semenov2011} studied the outer parts ($>$10\,au) of
DM\,Tau-like protoplanetary discs with given densities and
temperatures. They distinguished between ``responsive'' and
``unresponsive'' molecules and ice species. Surprisingly, with regard
to the results of this paper, CO, OH, \ce{H2O} as well as water an
ammonia ice were classified as ``unresponsive'', i.e.\ their column
densities were not altered much by the inclusion of mixing. In
contrast, some sulphur-bearing molecules and poly-atomic (organic)
molecules were found to be ``hypersensitive'', i.e.\ their column
densities were affected by more than 2 orders of magnitude. We note
that these criteria are solely based on changes of the vertical column
densities, which are not necessarily proportional to the observable
line fluxes.  In the T\,Tauri disc models presented in this paper,
many of the prominent (sub-)\,mm lines are optically thick in the
inner ($\la$\,100\,au) disc regions. In the mid-IR, the lines can also
be covered by dust opacity. Hence the column densities, which are
mostly set by the midplane concentrations, are no good predictors for
line fluxes in many cases.

\citet{Heinzeller2011} used a simple numerical scheme to include
advection caused by inward accretion and winds, and radial and
vertical eddy-mixing in the thermo-chemical disc models developed by
\citet{Nomura2005}.  In their appendix, it becomes clear that an
iterative approach was used, in which for each disc time integration
step the adjacent concentrations were simply taken from the
neighbouring cells at the previous time step and then assumed to be
constant, while the central concentrations are advanced in time, point
by point.  We will follow this basic idea in this paper.  We note,
however, that their figure 1 seems to show some artefacts in the form
of radial oscillations which could indicate that their numerical
scheme is not fully converging.  The results of \citet{Heinzeller2011}
are complex, as many effects (vertical and radial mixing, winds,
accretion) are discussed at the same time. They concluded that
vertical mixing has a considerable impact on the disc chemistry and
increases the abundances of many species in the warm molecular layer
of the disc observable with the Spitzer Space Telescope and the James
Webb Space Telescope (JWST).  The viscous parameter for the turbulent
mixing was set to $\alpha\!=\!10^{-2}$.

Common to the disc papers cited here is that the chemical mechanisms
and pathways, which cause the observed changes in the models due to
turbulent mixing, do not become fully clear.  In this paper, we will
concentrate on vertical mixing only, and will thoroughly explain and
test our numerical approach.  After an extended introduction to
mixing and diffusion in Sect.~\ref{sec:mixing}, we explain our
thermo-chemical disc model and test its convergence in
Sect.~\ref{sec:model}.  We systematically vary the $\alpha$-parameter
that sets the strength of the turbulent mixing, and take the feedback
of the changing concentrations on the gas temperature structure into
account.  We then show and discuss our results in
Sect.~\ref{sec:results}, and demonstrate the main chemical mechanism
that change the concentrations. We then continue in
Sect.~\ref{sec:impact} to study the impact caused by turbulent mixing
on a few observable quantities, such as ice emission and absorption
features in the spectral energy distribution (SED), mid-IR molecular
line emission features observable with JWST, and CO $J\!=\!2\to1$
channel maps observable with Atacama Large Millimeter Array (ALMA). We
conclude by summarising the main chemical mechanisms found and effects
caused by turbulent mixing in protoplanetary discs in
Sect.~\ref{sec:summary}.


\section{Chemistry with vertical mixing}
\label{sec:mixing}

\subsection{Basic equations}
The basic rate equation to solve for each chemical species is
\begin{equation}
  \pabl{n_i}{t} = P_i - L_i - \pabl{\Phi_i}{z} \ ,
  \label{eq:chemsys}
\end{equation}
where $n_i\ \rm[cm^{-3}]$ is the particle density of species $i$,
$P_i$ and $L_i$ $\rm[cm^{-3}s^{-1}]$ are the chemical production and
destruction rates, and $\Phi_i\ \rm[cm^{-2}s^{-1}]$ is the diffusive
particle flux according to \citet{Hu2012}, \citet{Zahnle2016} and
\citet{Rimmer2016}
\begin{equation}
  \Phi_i = -\,(K+D_i)\,n\,\pabl{}{z}\!\left(\frac{n_i}{n}\right) 
           +D_i\,n_i\,\frac{(\mu-m_i)\,g_z}{kT}
           \ ,
  \label{diff_flux}
\end{equation}
where $n\!=\!\sum_i n_i$ is the total particle density. $m_i$ is the
mass of gas species $i$ and $\mu\!=\!\sum n_i m_i/n$ the mean
molecular weight.\linebreak
$g_z\!=\!(GM_\star\,z)/(r^2+z^2)^{3/2}\!=\!z\,\Omega^2$ is the
gravitational acceleration in the vertical direction, $G$ the
gravitational constant, $M_\star$ the mass of the central star, $r$
the distance from the rotation axis, $z$ the height over the midplane,
and $\Omega$ is the Keplerian circular frequency. We have neglected
here the ``thermal diffusion factor'' $\alpha_T$, which describes a
second-order effect that the lightest molecules tend to diffuse
towards the warmest places.  The original derivation of
Eq.\,(\ref{diff_flux}) can be found in \citet{Hunten1973} with further
reading in \citet{Kuramoto2013}.

$K\ \rm[cm^2/s]$ is the eddy-diffusion coefficient caused by
turbulent mixing motions in the $z$-direction, sometimes
also denoted by $k_{zz}$.  $K$ is generally given by a characteristic
mixing velocity times a characteristic length
\begin{equation}
  K = \vz\,L \ ,
\end{equation}
where $L$ is the size of the largest eddy (``mixing length'') and
$\vz$ is the root-mean square of the fluctuating part of the vertical
gas velocities. According to the $\alpha$-disc model of
\citet{Shakura1973}, the typical vertical length scale is given by the
pressure scale height $L\approx H$ and the turbulent mixing velocity
is approximated as $\vz\approx\alpha_{\rm mix}\,\cs$,
\begin{equation}
  K = \alpha_{\rm mix}\,\cs\,H \ ,
  \label{eq:K}
\end{equation}
where $\cs\!=\!\sqrt{kT/\mu}$ is the local isothermal sound speed and
$T$ the temperature. Typical values used in the literature are
$\alpha\!\approx\!10^{-5}\,...\,10^{-2}$ using different constraints,
see reviews by \citet{Lesur2022} and \citet{Pinte2022}. While
theoretical accretion models often favour higher values to explain the
mass accretion rates, see e.g. \citet{Manara2016,Mulders2017},
observational contraints about turbulent line broadening
\citep{Flaherty2015} and dust settling \citep{Pinte2018} suggest lower
values. Eddy mixing transports all molecules in the same way. In
contrast, the gas-kinetic micro-physical diffusion coefficient
$D_i\ \rm[cm^2 s^{-1}]$, due to Brownian motions,
\begin{equation}
  D_i = \frac{1}{3} \frac{v_{{\rm th},i}}{\sigma_{\!ia}\,n} 
  \label{eq:Di}
\end{equation}
depends on species $i$ because the collisional cross section
$\sigma_{ia}$ depends on the ``size'' of the molecule $i$ and the relative
velocity depends on its reduced mass $1/m_{\rm red}\!=\!1/m_i+1/m_j$.  A
simple hard-sphere approximation would be \citep{Jeans1967}
\begin{eqnarray}
  D_i &\!\!\!=\!\!\!&\frac{1}{3}
         \left(\sum_j\frac{n_j\,\pi(r_i+r_j)^2}
                          {\sqrt{\frac{8\,kT}{\pi\,m_{\rm red}}}}\right)^{-1}
         \nonumber\\
     &\!\!\!=\!\!\!&\frac{1}{3} \sqrt{\frac{8\,kT}{\pi\,m_i}}
         \left(\sum_j\frac{n_j\,\pi(r_i+r_j)^2}
                          {\sqrt{1+\frac{m_i}{m_j}}}\right)^{-1}
      =~ \frac{b_{ia}}{n} \ ,
\end{eqnarray}
where the effective radii of the molecules $r_i$ can be measured from
viscosity experiments \citep{Jeans1967}, or calculated from
Lennard-Jones
potentials\footnote{\resizebox{86mm}{!}{\url{https://demonstrations.wolfram.com/BinaryDiffusionCoefficientsForGases/}}\label{foot1}}.
$b_{ia}$ is called the binary diffusion coefficient
$\rm[cm^{-1}s^{-1}]$ for collisions between species $i$ and a
``background atmosphere'' denoted by $a$. Some binary diffusion
coefficients for a few relevant combinations of H, H$_2$ and CO$_2$
are given in \citet{Hu2012}. Most atoms and molecules have effective
collisional radii $r_i$ between about 1\,\AA\ and 3\,\AA, see
Table~\ref{tab:molradii}. However, when the diffusion of charged
particles and/or diffusion in plasma is considered, the Coulomb cross
sections (Langevin cross-section) are larger, by orders of magnitude,
than the geometrical cross sections, see e.g.~section 2.5.2 in
\citet{Wiesemann2014}.

\begin{table}[!t]
  \caption{Effective molecular radii for diffusion coefficients
    calculated with the tools provided at Wolfram's
    webpage$^{\ref{foot1}}$ at $T\!=\!300\,$K}
  \label{tab:molradii}
  \centering
  \vspace*{-2mm}
  \begin{tabular}{c|c}
    \hline
    & \\[-2.2ex]
    particle $i$ & radius $r_i$\\[0.2ex]
    \hline
    & \\[-2.2ex]
    \ce{He}  & 1.084\,\AA\\
    \ce{H2}  & 1.366\,\AA\\
    \ce{O2}  & 1.811\,\AA\\
    \ce{CO}  & 1.886\,\AA\\
    \ce{N2}  & 1.901\,\AA\\
    \ce{CH4} & 2.075\,\AA\\
    \ce{H2O} & 2.178\,\AA\\
    \ce{CO2} & 2.304\,\AA\\
  \end{tabular}
  \vspace*{-2mm}
\end{table}

\subsection{Location of the homopause}
\label{sec:homopause}

\begin{figure}[!t]
  \includegraphics[width=88mm,trim=18 22 15 15,clip]{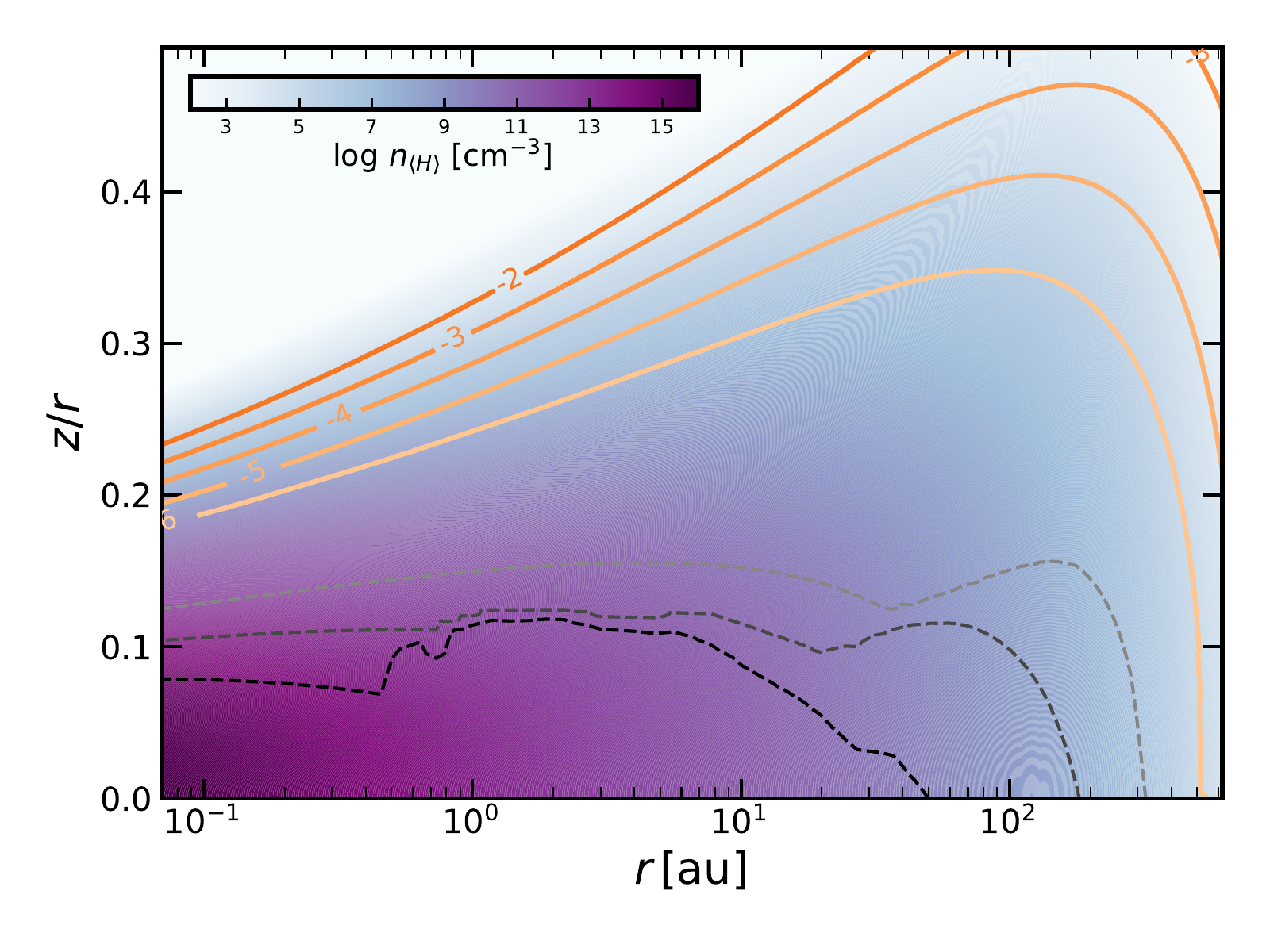}\\[-5mm]
  \caption{Location of the homopause, where the diffusion coefficients
    caused by eddy mixing and gas-kinetic motions are equal
    ($K\!=\!D_i$). The colours show the hydrogen nuclei density,
    $\nH$, as function of radius $r$ and relative height $z/r$.  The
    coloured contour lines show where the homopause is situated when
    $K$ is calculated with different values of $\log_{\rm
      10}\alpha_{\rm mix}$.  The dashed lines show where the vertical
    optical extinction $A_{\rm V,ver}$ equals 100 (black), 10 (dark
    grey) and 1 (light grey). The dust opacities correspond to our
    un-mixed model described in Sect.~\ref{sec:model}. The uneven $A_{\rm V}$
    contours are caused by the uneven spatial distribution of the
    ices, which are taken into account when calculating the opacities.}
  \label{fig:homopause}
  \vspace*{-1mm}
\end{figure}

Equations (\ref{eq:K}) and (\ref{eq:Di}) show that while the
eddy-diffusion coefficient $K$ is assumed to be constant as function
of height $z$, the molecular diffusion coefficient $D_i$ scales with
$1/n$, i.e.\ it increases with height.  In fact, in most parts of the
disc, we have $K\!\!\gg\!\!D_i$. However, in the region above the
$K\!\!=\!\!D_i$ layer, called the {\em homopause} (see
Fig.~\ref{fig:homopause}), the molecules, atoms and ions will tend to
arrange themselves with individual scale-heights, which could be
denoted by ``gravitational de-mixing''. Above the homopause, the
concentrations of the heavy molecules drops quickly, whereas the
lighter molecules keep on extending further up.  This
behaviour can be verified by considering the case of a non-reactive
($P_i\!=\!L_i\!=\!0$) mixture of ideal gases at constant temperature
without eddy mixing ($K\to 0$) in equilibrium, where both $n(z)$ and
$n_i(z)$ become time-independent functions of $z$.  The diffusive
fluxes vanish and we have
\begin{eqnarray*}
  && \Phi_i = 0 ~=~ -D_i\,n\,\frac{d}{dz}\!\left(\frac{n_i}{n}\right) 
           +D_i\,n_i\,\frac{(\mu-m_i)\,g_z}{kT}
           \\
  &\Rightarrow& \frac{d}{dz}\ln\frac{n_i}{n}
           ~=~ \frac{(\mu-m_i)\,g_z}{kT}
           ~=~ \frac{1}{H}-\frac{1}{H_i}
           \\
  &\Rightarrow& n_i(z) = n_i(0)\exp\Big(-\frac{z}{H_i}\Big)
           \quad\mbox{and}\quad n(z) = n(0)\exp\Big(-\frac{z}{H}\Big)
\end{eqnarray*}
with molecular and atmospheric scale heights $H_i=kT/(m_i g_z)$ and
$H=kT/(\mu g_z)$, respectively.

Figure~\ref{fig:homopause} shows the location of the homopause based
on the {\sc ProDiMo} model described in Sect.~\ref{sec:model} for a
couple of $\alpha_{\rm mix}$ values. A demonstration molecule with
mass $m_i\!=\!20$\,amu and radius $r_i\!=\!2\,$\AA\ is considered, in
a mixture of \ce{H+}, H, H$_2$, \ce{He+} and He, with radii 0.8\AA,
0.8\AA, 1.366\AA, 1.084\AA\ and 1.084\AA, respectively. The surprising
conclusion from Fig.~\ref{fig:homopause} is that gravitational
de-mixing seems likely to play a significant role in the upper disc
layers from which e.g.\ disc winds are launched, which can therefore
be expected to contain only little amounts of heavy elements.
We have neglected plasma effects here, i.e. increased Coulomb
collisional cross sections, where the homopause would be located
higher in the disc.


\subsection{Can mixing change the element abundances?}

An important question is whether the element abundances 
can be affected by diffusive transport and hence become spatially
dependent quantities. The element abundances are given by
\begin{equation}
  \nH\,\epsilon_k = \sum_i s_{i,k} n_i
  \label{eq:elem}
\end{equation}
where $s_{i,k}$ is the stoichiometric factor of element $k$ in
molecule $i$, $\epsilon_k$ are the element abundances with respect to
hydrogen, and $\nH$ is the total hydrogen nuclei particle
density. $\epsilon_{\rm H}\!=\!1$ by definition. The changes of the
element abundances follow from Eq.~(\ref{eq:chemsys}) after
multiplication with $s_{i,k}$ for a selected element $k$ and summation
over all chemical species $i$
\begin{equation}
  \pabl{}{t}\big(\nH\,\epsilon_k\big) = \sum_i s_{i,k}\Big(P_i - L_i\Big)
                       - \sum_i \pabl{}{z}\Big(s_{i,k}\Phi_i\Big) \ .
  \label{eq:elem2}
\end{equation}
Chemical reactions conserve the element abundances, hence the first
term in Eq.\,(\ref{eq:elem2}) vanishes. Concerning the diffusive flux
$\Phi_i\rm\ [cm^{-2}s^{-1}]$ let us first consider the case
$K\!\gg\!D_i$ where turbulent mixing dominates. From
Eq.~(\ref{diff_flux}) we then have
\begin{equation}
  \Phi_i = -K\,n\,\pabl{}{z}\!\left(\frac{n_i}{n}\right) \ .
\end{equation}
The diffusive element fluxes are hence given by
\begin{eqnarray}
  \Phi_k &\!=\!& \sum_i s_{i,k}\,\Phi_i
  ~=~ -\sum_i s_{i,k}\,K\,n\,\pabl{}{z}\!\left(\frac{n_i}{n}\right) 
  \label{eq:flux2} \\
  &\!=\!& -K\,n\,\pabl{}{z}\!\left(\frac{\sum_i s_{i,k}\,n_i}{n}\right)
  ~=~ -K\,n\,\pabl{}{z}\!\left(\frac{\nH\,\epsilon_k}{n}\right) \ ,
  \label{eq:flux3}
\end{eqnarray}
where we have used that $K$ does not depend on molecule.  If we can
further assume\footnote{We define $\muH$ as the proportionality
  constant between gas mass density $\rho$ and total hydrogen nuclei
  density $\nH$,
  i.e.\ $\rho\!=\!\muH\,\nH\!=\!\nH\sum_k\epsilon_k\,m_k$, where $m_k$
  is the mass of element $k$, hence
  $\muH\!=\!\sum_k\epsilon_k\,m_k$. Since the mass density equals
  $\rho\!=\!\mu\,n$, where $\mu$ is the mean molecular weight, we find
  in general
  $$\pabl{}{z}\Big(\frac{\nH\epsilon_k}{n}\Big) =
  \pabl{}{z}\Big(\frac{\nH}{n}\Big)\,\epsilon_k +
  \frac{\nH}{n}\pabl{\epsilon_k}{z} =
  \pabl{}{z}\Big(\frac{\mu}{\muH}\Big)\,\epsilon_k +
  \frac{\nH}{n}\pabl{\epsilon_k}{z}\ .$$ If we can assume
  $\nH\!\propto\!n$, then \ $\muH\!\propto\!\mu$, the first term
  vanishes and we find $\Phi_k=-K\,\nH\pabl{\epsilon_k}{z}$ as used in
  the main text. However, the mean molecular weight $\mu$ changes with
  chemical composition whereas $\muH$ remains the same, and thus,
  strictly speaking, Eq.\,(\ref{eq:chem3}) and the conclusions below
  that equation seem incorrect. See further discussion of this point
  in Sect.~\ref{sec:smallSt}.} that $\nH\!\propto\!n$ we find
\begin{equation}
  \pabl{}{t}\big(\nH\,\epsilon_k\big)
  ~=~ -\pabl{}{z}\Phi_k
  ~=~
  \pabl{}{z}\bigg(K\,\nH\,\pabl{\epsilon_k}{z}\bigg)
  \ .
  \label{eq:chem3}
\end{equation}
Equation (\ref{eq:chem3}) states that eddy diffusion works against
spatial gradients of element abundances. No matter how complicated the
chemical processes are, eddy diffusion will always diminish such
gradients $(\Phi_k\!\to\!0~\Rightarrow~\pabl{\epsilon_k}{z}\!\to\!0$),
making sure that the element abundances will eventually become
constant throughout the disc.

This result seems a bit counter-intuitive at first, for example when we
consider a particular molecule only, which is, for example, abundant in
deep layers and mixed up where it dissociates, one would
expect this process to enrich the higher layers with the elements the
molecule was composed of.  However, this is not how eddy mixing
works. Mixing rather exchanges molecules for other molecules in the local
neighbourhood, and if we properly sum up the total effect on the
element abundances, via $\Phi_k=\sum_i s_{i,k}\,\Phi_i$, the net
effect is exactly zero, unless there is a spatial gradient in element
abundances in the first place, in which case eddy diffusion will
reduce that gradient.

However, this conclusions is only valid as long as $K$ is independent
of molecules/particle to be mixed by diffusion.  Once we add $D_i$
with $D_i\!\ga\!K$ (above the {\sl homopause}), we can no longer pull
out the net diffusion coefficient $(K+D_i)$ from the sum in
Eq.~(\ref{eq:flux2}). In that case, diffusion will change the
element abundances. An example for such long-term effects by
gravitational de-mixing was already described in
Sect.~\ref{sec:homopause}.

Another relevant case occurs when we consider the difference between
gaseous and ice species. The ice species are transported with the dust
grains, and the grains are only moved by the larger turbulent
eddies. In order to arrive at an effective dust particle diffusion
coefficient, the advective effect of all individual turbulent eddies
has to be averaged. This procedure has been carried out, with
different methods and approximations, by \citet{Dubrulle1995},
\citet{Schraepler2004} and \citet{Youdin2007}. The result of
\citet[][see their Eq.\,27]{Schraepler2004}, is
\begin{equation}
  K_{\rm ice} = \frac{K}{1+\St}
  \label{eq:Kice}
\end{equation}
where $\St$ is the Stokes number given by
\begin{equation}
  \St = \frac{\tstop}{\tedd}
  = \frac{\frac{\rhod\,a}{\rho\,\vth}}
  {\frac{H}{\alpha_{\rm mix}\,\cs}}
  = \sqrt{\frac{\pi}{8}}\,\frac{\rhod\,a\,\alpha_{\rm mix}}{\rho\,\,H}
\end{equation}
where $\tstop$ is the stopping distance, $\tedd$ the turnover
timescale of the largest turbulent eddy, $\rhod$ the dust material
density, $\vth$ the thermal velocity, $\cs$ the sound velocity, $H$
the pressure scale height and $a$ is the grain radius. The factor
$\sqrt{\pi/8}$ is because $\cs^2=kT/\mu$ but $\vth^2=8kT/(\pi\mu)$.

It is questionable, however, in how far Eq.\,(\ref{eq:Kice}) describes
the actual dynamical motions of dust grains with large Stokes numbers
in disks. The main reason is that such particles will start to migrate
inward, i.e.\ they develop a systematic velocity with respect to the
gas, which goes beyond the mixing approach. Such motions have been
shown to be able to change the elemental abundances accross ice lines
of key volatiles such as CO \citep{Krijt2018}.  Furthermore, it is not
obvious how to choose $a$ as we actually need the cumulative effect of
all grains of different sizes on the diffusive flux, where it also
must be debated how the ice is distributed onto the bare grains as
function of size, and how the size dust distribution is altered by ice
formation, see e.g.\ \citet{Arabhavi2022}.

Assuming that we can determine $K_{\rm ice}$ anyway, and
assuming that all vertical element fluxes $\Phi_k$ must vanish
everywhere in the disc in the time-independent limiting
case, we can derive a local condition for each element $k$ from
Eq.\,(\ref{eq:flux2}):
\begin{equation}
  \Phi_k = -K\,n\,\pabl{}{z}\left(\frac{\sum_{\rm gas} s_{ik} n_i}{n}\right)
  -K_{\rm ice}\,n\,\pabl{}{z}\left(\frac{\sum_{\rm ice} s_{ik} n_i}{n}\right)
        ~=~ 0 \ .
\end{equation}
Defining the element abundances present in the gas and ice,
respectively, as
\begin{equation}
  \epsilon_k=\epsilon_k^{\rm gas}\!+\!\epsilon_k^{\rm ice}
  \;\;,\;\;
  \nH\,\epsilon_k^{\rm gas}\!=\!\sum\limits_{i,\rm gas} s_{ik} n_i
  \;\;,\;\;
  \nH\,\epsilon_k^{\rm ice}\!=\!\sum\limits_{i,\rm ice} s_{ik} n_i
\end{equation}
we find the following result that is independent of $\alpha_{\rm mix}$
\begin{equation}
    \pabl{}{z}\left(\frac{\nH\,\epsilon_k^{\rm gas}}{n}\right)
    +\frac{K_{\rm ice}}{K}\,
    \pabl{}{z}\left(\frac{\nH\,\epsilon_k^{\rm ice}}{n}\right)
    ~=~ 0 \ .
  \label{eq:egrad}
\end{equation}

\subsubsection{Small Stokes numbers}
\label{sec:smallSt}

In the limiting case of small dust radii $a$ and/or large gas
densities $\rho$, we have $\St\ll 1$, i.e.\ well-coupled grains, and
hence $K_{\rm ice}\approx K$. Thus, from Eq.\,(\ref{eq:egrad}), we obtain
\begin{equation}
  \frac{\nH\,\epsilon_k}{n}=\ce{const} \ .
  \label{eq:econst}
\end{equation}
previously referred to as the case of ``constant element abundances''.
Clearly, when all species are transported in equal ways, and hence
mixed with the same efficiency, the element abundances cannot change.
However, Eq.\,(\ref{eq:econst}) actually states a somewhat different
result.  Diffusion will make sure that, on the long run, the
concentrations of elements as measured with respect to the total gas
particle density $n$ become constant.  At the locations of the major
phase transitions $\ce{H+}\!\leftrightarrow\!\ce{H}$ and
$\ce{H}\!\leftrightarrow\!\ce{H2}$, where the mean molecular weight
changes by roughly a factor of 2 each, the element abundances must
change as well, in the opposite direction, approximately
$\epsilon_k\!\propto\!1/\mu$. Otherwise, when $\mu$ is constant, we
have $n\propto\nH$ and hence indeed
$\epsilon_k=\ce{const}$. Considering a complete vertical column, from
ionised down to molecular conditions, Eq.\,(\ref{eq:econst}) states
that the helium abundance, for instance, should be about $4\times$
larger in the ionised regions as compared to the molecular regions.

\subsubsection{Large Stokes numbers}
\label{sec:largeSt}

In the limiting case of large Stokes numbers ($a\to\infty$ and/or
$\rho\to 0$) we have $K_{\rm ice}\ll K$, and thus from
Eq.\,(\ref{eq:egrad})
\begin{equation}
  \frac{\nH\,\epsilon_k^{\rm gas}}{n}=\ce{const} \ .
  \label{eq:echange}
\end{equation}
Equation~(\ref{eq:echange}) states a remarkable result. If an element
can freeze out anywhere in the column, a rather low level of
$\epsilon_k^{\rm gas}$ will be enforced there, and this low level of
$\epsilon_k^{\rm gas}$ must be identical everywhere in the column to
make the diffusive fluxes vanish, irrespective of the local
$\epsilon_k^{\rm ice}$.  This effect is commonly know as ``cold
trap'', see e.g.~\citet{Krijt2016} and \citet{Oberg2021}.  Since a cold surface
cannot move, but air does, the gaseous water molecules in a room will
continue to diffuse toward and condense onto that cold surface, until
eventually the supersaturation of the air over the cold surface
approaches unity. If the surface temperature is kept lower than air
temperature, this will eventually cause an under-saturation of water
vapour in the warm gas in the room.

In a protoplanetary disc, we have the same principle situation.
Certain molecules will freeze out quickly in the cold midplane,
establishing very low local $\epsilon_k^{\rm gas}$.  After just a few
vertical mixing timescales (see Eq.\,\ref{eq:taumix}), we would hence
expect that most molecules in the upper layers should simply be mixed
down into the midplane and freeze out there, leaving behind very low
concentrations also in the upper layers.

But this conclusion would be in obvious conflict with ALMA observations, for
example \citep{Pinte2018,Rab2020}, where we clearly see CO molecules
only above the icy midplane at 100\,au distances. We conclude that
the cold trap effect cannot work in discs at full effect, possibly because:
\begin{itemize}
  \item The actual $\alpha_{\rm mix}$ in discs is much lower than
    $10^{-2}$, such that the mixing timescale according to
    Eq.\,(\ref{eq:taumix}) approaches the disc lifetime. For an
    assumed disc lifetime of $5\times 10^6\,$yrs, this would
    suggest $\alpha_{\rm mix}<3\times 10^{-5}$ at 100\,au, and even
    smaller $\alpha_{\rm mix}$ at smaller radii.
  \item The effective Stokes numbers (after averaging over dust size
    distribution) result to be small in Eq.\,(\ref{eq:Kice}), such that
    the limiting case described by Eq.\,(\ref{eq:echange}) is not
    reached. 
  \item A combination of both.
\end{itemize}
Even if these factors result in large deviations from
Eq.\,(\ref{eq:echange}), the cold trap effect could still cause
important effects with respect to the case of constant element
abundances (Eq.\,\ref{eq:econst}) over long timescales, which however
will be difficult to prove observationally. In principle, one can
solve Eq.\,(\ref{eq:egrad}) numerically, in an iterative way, to make
it consistent with the ice chemistry, to obtain $\epsilon_k^{\rm
  gas}(z)$ and $\epsilon_k^{\rm ice}(z)$.

Further investigations are clearly required here, both about the
influence of molecular diffusion on the upper disc layers and about
the cold trap effect, where Eq.\,(\ref{eq:egrad}) could serve as a
useful starting point.  However, in the remaining parts of this paper,
we will not discuss these complications any further, but concentrate
on the most basic case, where we have vanishing molecular mixing
$D_i\!\to\!0$ and no difference between the diffusive transport of gas
and dust $K_{\rm ice}\!\to\!K$, in which case the element abundances
can be assumed to be constant throughout in the disc. We also neglect
here vertical gradients of the mean molecular weight $\mu$.

\subsection{Implementation of mixing rates in {\sc ProDiMo}}

A direct numerical solution of the reaction-diffusion equations
(Eq.\,\ref{eq:chemsys}), which is a system of coupled stiff partial
differential equations of second order, is difficult and can be
computationally very demanding, see \citet{Semenov2006, Semenov2011}
and \citet{Hu2012}.  Here, we follow an iterative approach similar to
\citet{Heinzeller2011} and \citet{Rimmer2016}.  We consider height
grid points $\{z_k\,|\,k=1,...,K\}$.  As shown in
Appendix~\ref{app:numerical}, the divergence of the diffusive particle
flux $i$ at point $k$ can be numerically discretised as
\begin{equation}
  -\pabl{\Phi_{i,k}}{z}
  = {\cal A}\,n_{i,k+1} ~-~ {\cal B}\,n_{i,k} ~+~ {\cal C}\,n_{i,k-1} 
\end{equation}
where $n_{i,k}$ is the particle density of species $i$ at height grid
point $k$, and where the coefficients ${\cal A}$, ${\cal B}$, ${\cal
  C}$ are given by the geometry of the grid points, the eddy and
molecular diffusion coefficients, $K$ and $D_i$, and the total
particle densities $n=\sum n_i$ at the neighbouring height grid points
$k\!-\!1$, $k$, $k\!+\!1$.  Our approach is then very simple.  We
consider the neighbouring concentrations $n_{i,k+1}$ and $n_{i,k-1}$ as
fixed during the solution of the chemistry at height point $k$.
Hence, we can solve
\begin{equation}
  \pabl{n_i}{t} = \tilde{P_i} - \tilde{L_i} \ ,
\end{equation}
with the existing time-independent chemistry solver in the {\sc
  ProDiMo} model, where
\begin{eqnarray}
  \tilde{P_i} &=& P_i + {\cal A}\,n_{i,k+1} + {\cal C}\,n_{i,k-1} \label{eq:Pi}\\
  \tilde{L_i} &=& L_i + {\cal B}\,n_{i,k}  \label{eq:Li}\ .
\end{eqnarray}
Since {\sc ProDiMo} solves the chemistry and heating/cooling balance
on all grid points in a particular order (first downward -- then
outward, following the two main irradiation directions), the constant
$n_{i,k+1}$ are taken from the same chemistry iteration (from the last
point above just solved), whereas the constant $n_{i,k-1}$ are taken
from the previous chemistry iteration. In
Appendix~\ref{app:convergence} we show that we can actually solve
simple diffusion problems with this basic iterative scheme. It is not
very efficient nor elegant, but it allows us to quickly implement 
vertical mixing in {\sc ProDiMo} without changing the code structure
at all.

The contribution of the diffusive mixing to the production and
destruction of species $i$ on grid point $k$ is formulated in terms of
two additional pseudo reactions. The MI reaction (``mixing in'') is a
spontaneous process with constant rate, with no entries in the
Jacobian. The MO reaction (``mixing out'') has a rate that is
proportional to $n_{i,k}$
\begin{eqnarray}
  \mbox{MI:}\hspace{15mm}
  &\longrightarrow& \ce{mol}_i
  \hspace*{5mm}\mbox{($k_{\rm MI}={\cal A}\,n_{i,k+1} + {\cal C}\,n_{i,k-1}$)}\\
  \mbox{MO:}\hspace{7.5mm}
  \ce{mol}_i &\longrightarrow& 
  \hspace{11.5mm}\mbox{($k_{\rm MO}=\cal{B}$)} \ ,
\end{eqnarray}
where $k_{\rm MI}$ and $k_{\rm MO}$ are the rate constants applied
in the chemical network.
The iteration of the chemistry is then continued until the solution
relaxes towards a steady state where all particle densities at all
grid-points $n_{i,k}$ do not change any further.
In that case, the neighbouring concentrations $n_{i,k+1}$ and
$n_{i,k-1}$ correctly describe the spatial gradients in the system and
we have reached the time-independent solution of
\begin{equation}
  0 = P_i - L_i - \pabl{\Phi_i}{z} 
\end{equation}
for each chemical species $i$ on all grid points $k$.

\section{The disc model}
\label{sec:model}

The physical and chemical setup of the disc model considered in
this paper was described in detail in \citet{Woitke2016}, see their
table~3. The central star is a 2\,Myr old T\,Tauri star with a mass of
0.7\,$M_\odot$ and a luminosity of 1\,$L_\odot$, the disc has a mass
of 0.01\,$M_\odot$, with a gas/dust ratio of 100, and the dust grains
have sizes between 0.05\,$\mu$m and 3\,mm, with a powerlaw
size-distribution of index -3.5. The dust component, approximated by
100 size bins, is settled according to the prescription of
\citet{Dubrulle1995} with $\alpha_{\rm settle}\!=\!10^{-2}$. In a
consistent $\alpha$-model, $\alpha_{\rm mix}\!=\!\alpha_{\rm settle}$
should be used, but here we chose to keep $\alpha_{\rm settle}$
constant to study the changes that are solely caused by the
introduction of mixing to the chemistry, using different choices of
$\alpha_{\rm mix}$.  All other disc shape, mass, radial extension,
dust material and opacity parameters, as well as the cosmic-ray, UV
and X-ray irradiation parameters are listed in table~3 in
\citet{Woitke2016}.

We use the {\sl large DIANA standard} chemical network
\citep{Kamp2017} in this work, which has 235 chemical species, with
reaction rates mostly based on the UMIST\,2012 database
\citep{McElroy2013}, but including a simple freeze-out and desorption
ice chemistry, X-ray processes including doubly ionised species
\citep{Aresu2011}, excited molecular hydrogen, and PAHs in 5 different
charging states, altogether 2843 reactions. Surface chemistry
\citep{Thi2020a,Thi2020b} is not included.

Since the up-mixing of ices formed in the disc midplane is a strong
effect in the models presented in this paper, we used the consistent
calculation of ice opacities as implemented by \citet{Arabhavi2022},
including \ce{H2O}-ice, \ce{CO2}-ice, \ce{NH3}-ice, \ce{CH4}-ice,
CO-ice and \ce{CH3OH}-ice.  We assume that these ices form a layer of
equal thickness on top of each refractory grain ($q\!=\!0$ in Eq.\,(3)
of \citet{Arabhavi2022}. The thickness and composition of this ice
layer depends on the position in the disc and is adjusted according to
the local results of the chemistry. Thus, the dust opacity used in the
continuum radiative transfer depends on the results of the chemistry,
which requires global iterations to solve.  However, since we do up to
1000 global iterations anyway, to solve the reaction-diffusion
equations, and each iteration includes continuum radiative transfer
and chemistry \& heating/cooling balance, these additional feedbacks
are automatically resolved.

We choose to use 80 radial grid points and 120 vertical grid points
for our models, a compromise between having acceptable computation times
and the need to resolve the vertical structures as good as possible,
to study the vertical mixing. The computation time for one global
iteration of one model is about 40\,min user time on 32 CPUs. For 1000
global iterations, we needed about 1 month of user time.  All
computations have been performed with {\sc ProDiMo} git version
{\tt 2f79b977} (11$^{\rm th}$ February 2022).

Although it would be relatively straightforward to include radial
mixing as well, we have only included vertical mixing in this paper.
We have also chosen to concentrate entirely on turbulent eddy mixing and
drop all terms for gas-kinetic molecular diffusion ($D_i\!=\!0$),
both of which might be investigated in forthcoming papers.

\section{Results}
\label{sec:results}

\subsection{Convergence of the Numerical Method}

Before we discuss the physical and chemical results of our disc models
with vertical mixing, we first need to make sure that our simple
iteration scheme actually converges.

We measure the convergence by means of the following two criteria,
which evaluate the differences of the various chemical concentrations
between the current and the previous global iteration in different
ways
\begin{eqnarray}
  {\rm crit\,1} &\!\!=\!\!& \max_i\!\Big\{\,\big|\log_{10} m_{i,j}/m_{i,j-1}\big|\,\Big\}\label{eq:crit1}\\
  {\rm crit\,2} &\!\!=\!\!& \left(\frac{1}{N_{\rm p} N_{\rm sp}}
                  \sum\limits_{k=1}^{N_{\rm p}} \sum\limits_{i=1}^{N_{\rm sp}}
                  \Big(\log_{10}
                  n_{k,i,j}/n_{k,i,j-1}\Big)^2\right)^{1/2}\label{eq:crit2}\ ,
\end{eqnarray}
where $m_{i,j}$ is the disc volume integrated mass of species $i$ at
iteration $j$, and $n_{k,i,j}$ is the particle density of species $i$
on 2D grid point $k$ at iteration $j$. $N_{\rm p}$ is the number of 2D
grid points, and $N_{\rm sp}$ is the number of chemical species in the
network.

\begin{figure}[!t]
  \centering
  \vspace*{-1mm}  
  \includegraphics[width=70mm,trim=15 20 10 10,clip]
                  {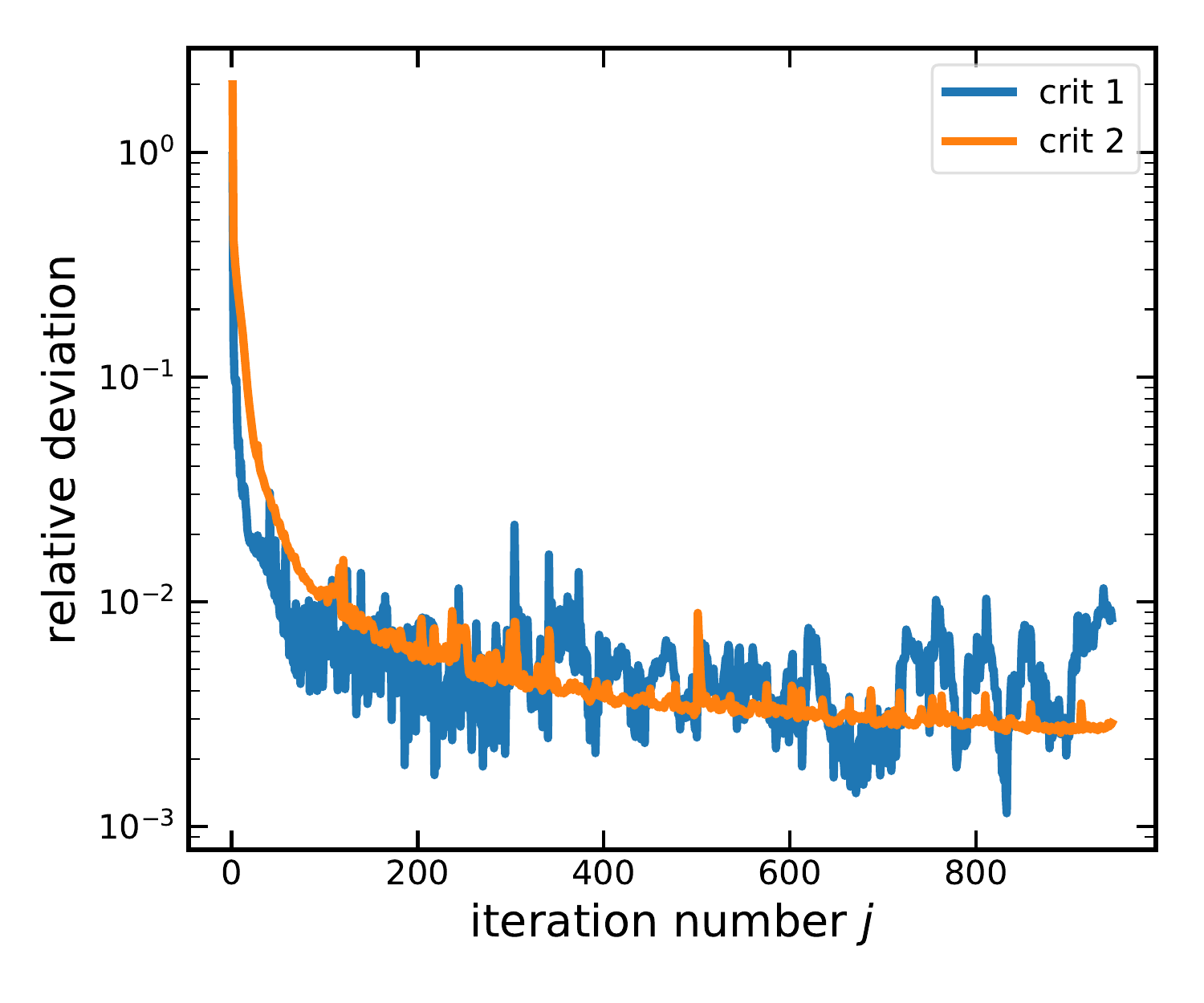}\\[-1mm]
  \caption{Measuring the convergence of the iterative solution
    technique to solve the chemistry with vertical mixing, here for
    the model with medium strong mixing, $\alpha_{\rm
      mix}\!=\!10^{-3}$. See text for explanations.}
  \label{fig:iteration1}
  \vspace*{-1mm}
\end{figure}

Figure~\ref{fig:iteration1} plots these two convergence criteria as
function of iteration number for the model with medium strong mixing
$\alpha_{\rm mix}\!=\!10^{-3}$. The figure shows that, due to our
very simple iteration method, the convergence is rather
slow.  We need a couple of 100 iterations to bring the relative
changes down to about 0.003\,dex, but even at 1000 iterations, there
are still fluctuations of a similar order of magnitude.

\begin{figure*}
  \vspace*{-2mm}
  \centering
  \begin{tabular}{cc}
  \hspace*{5mm} {\small\sf start of iterations} &
  \hspace*{5mm} {\small\sf end of iterations}\\
  \hspace*{-3mm}  
  \includegraphics[width=78mm,trim=13 61 15 32,clip]
                  {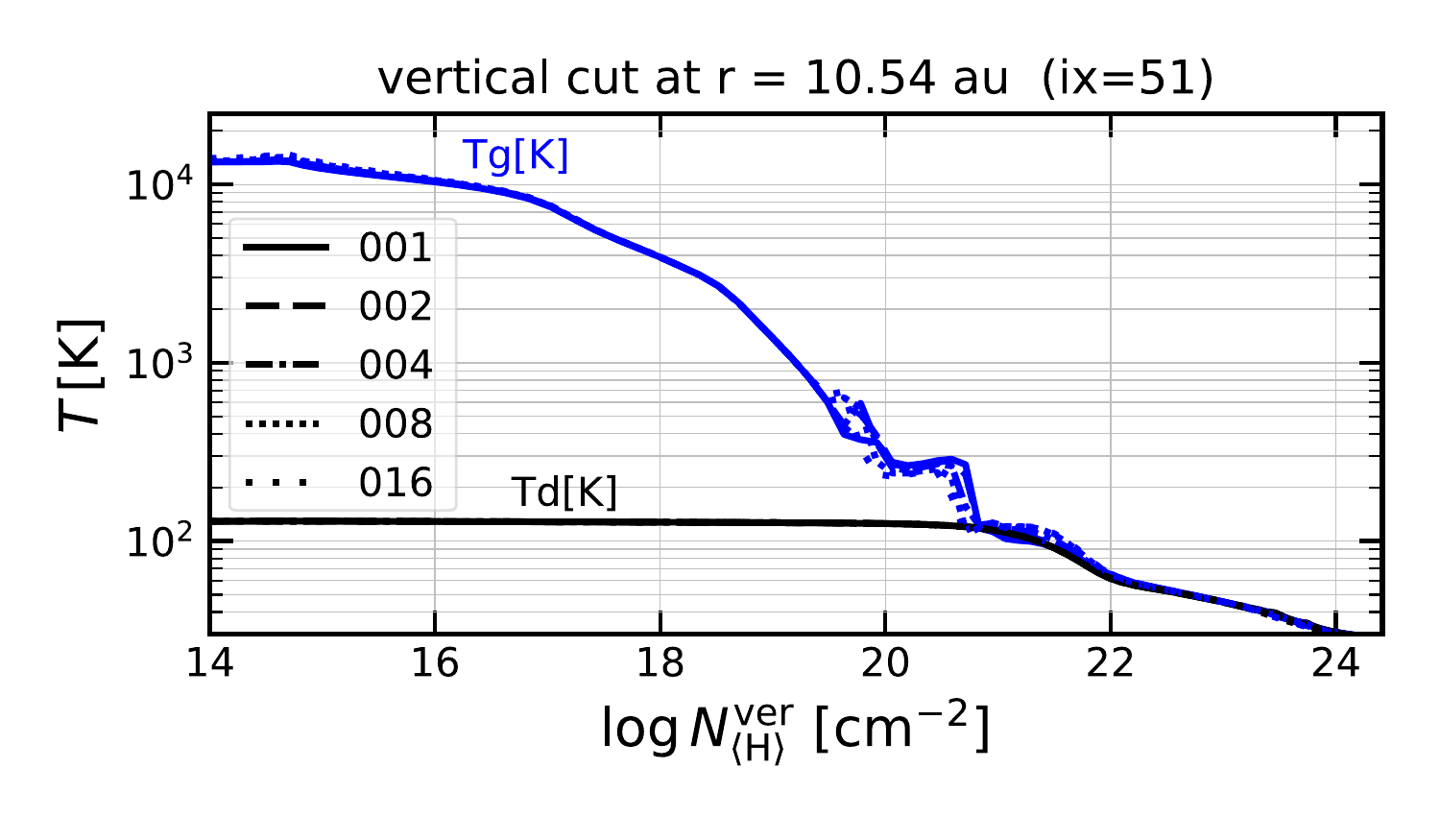}
  &
  \hspace*{-3mm}  
  \includegraphics[width=78mm,trim=13 61 15 32,clip]
                  {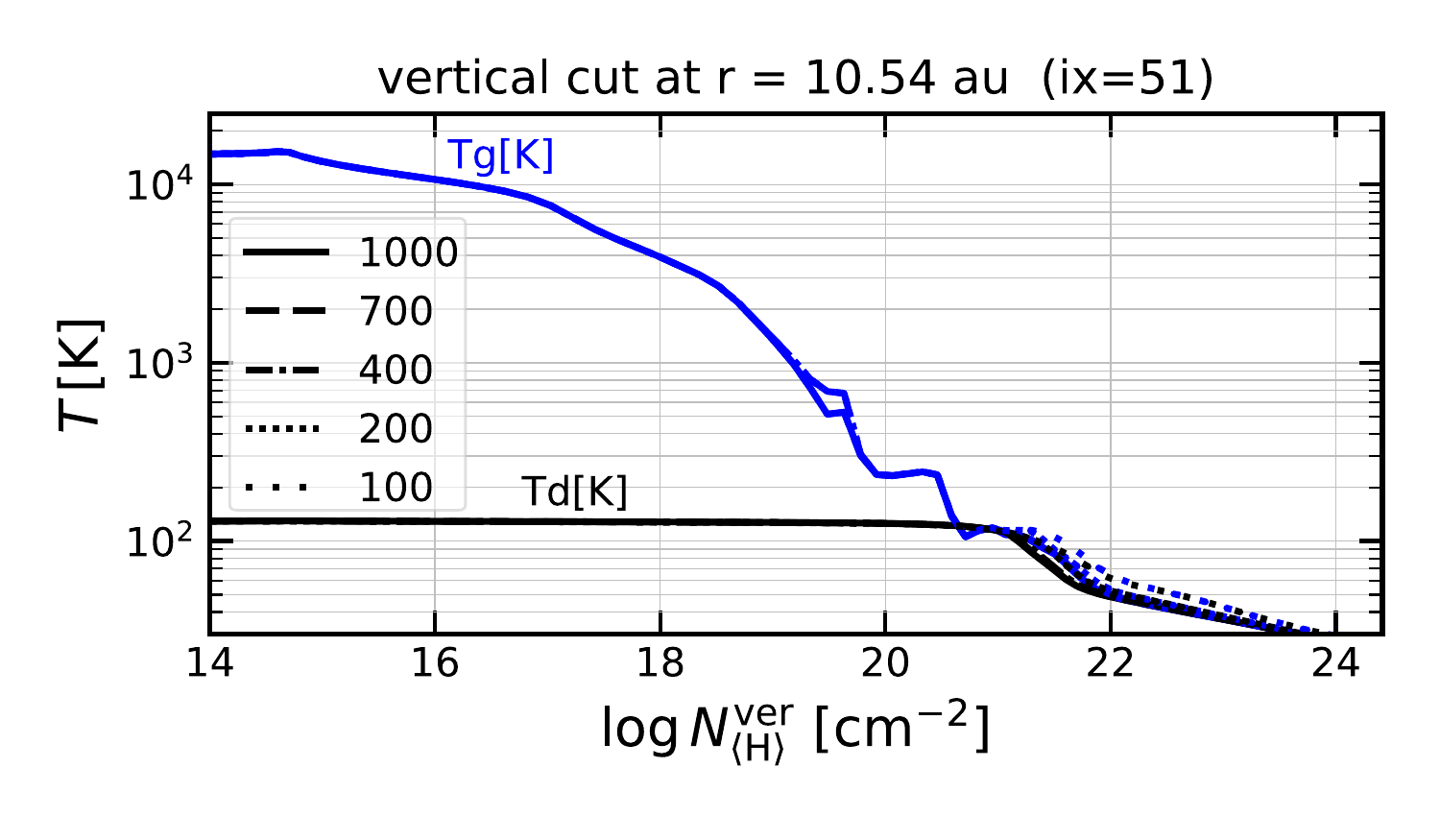}\\[-1mm]
  \hspace*{-3mm}  
  \includegraphics[width=78mm,trim=20 20 15 32,clip]
                  {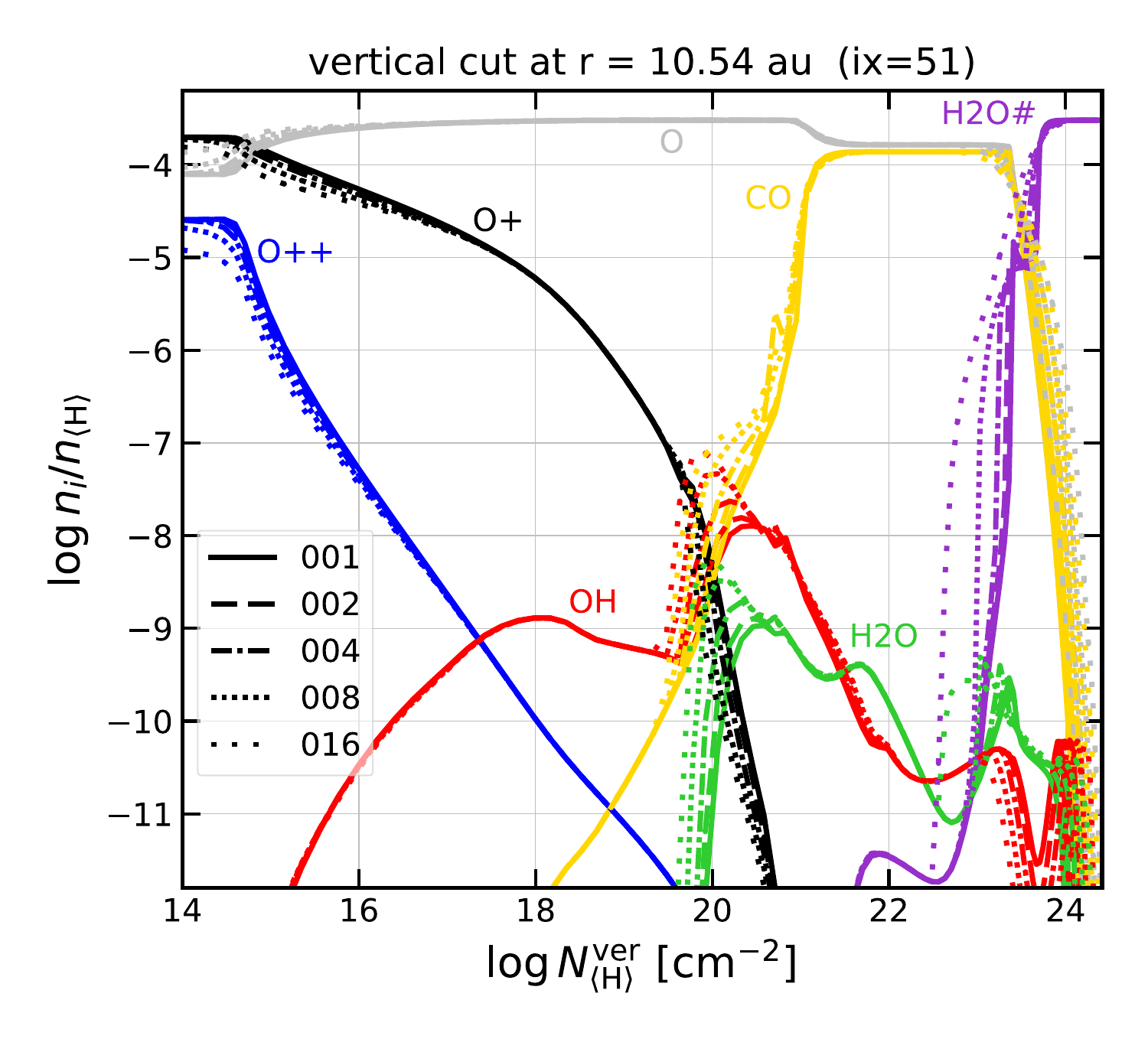}
  &
  \hspace*{-3mm}  
  \includegraphics[width=78mm,trim=20 20 15 32,clip]
                  {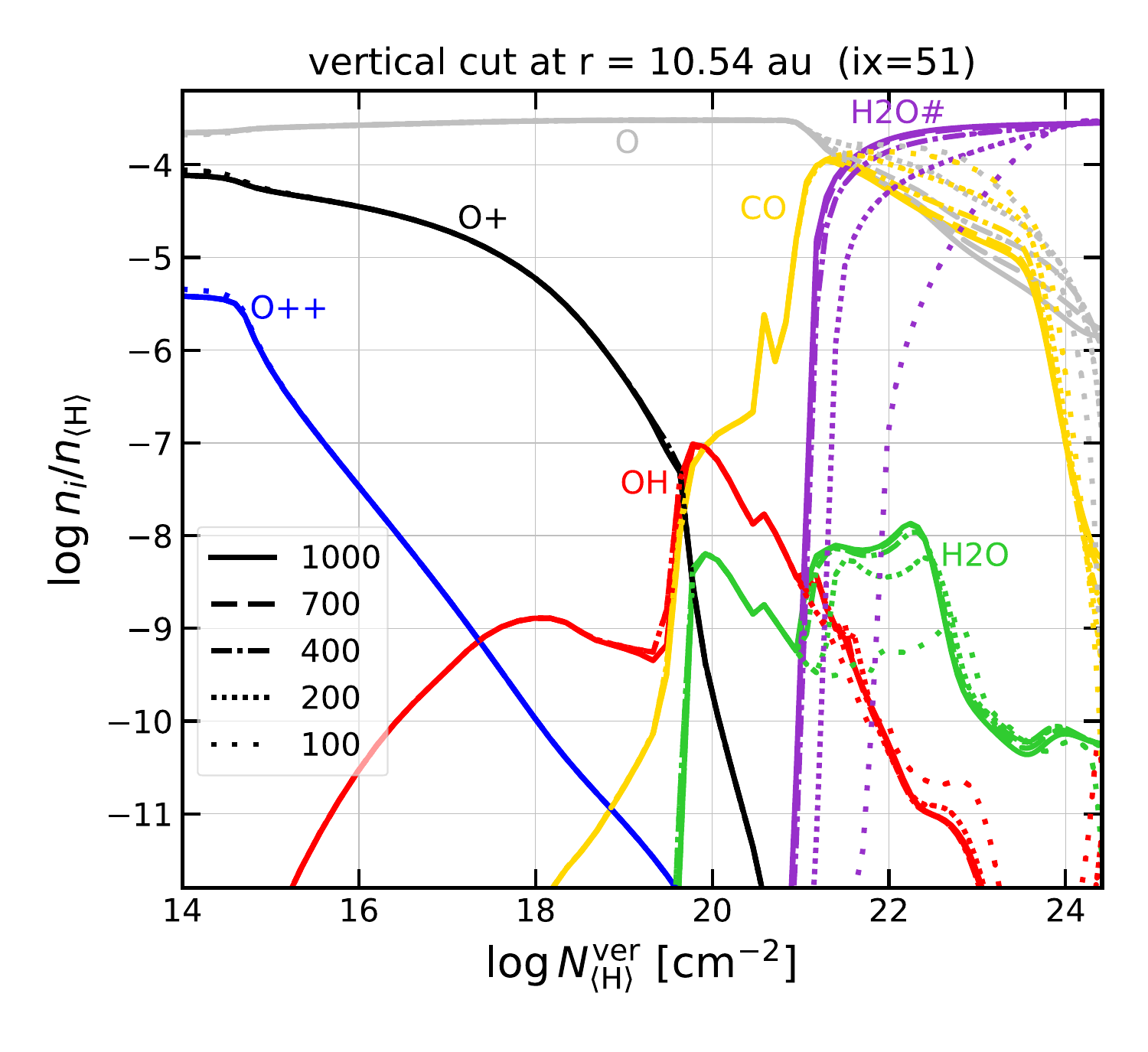}
  \end{tabular}\\[-1mm] 
  \caption{Monitoring the convergence of the iterative solution method
    along a vertical cut at $r\!=\!10\,$au as function of the vertical
    hydrogen nuclei column density $N_{\rm\langle
      H\rangle}\!=\!\int_z^\infty \nH(z)\,dz$ for the model with
    medium strong mixing $\alpha_{\rm mix}\!=\!10^{-3}$. The upper
    plots show the changing gas and dust temperature as function of
    iteration number. The lower plots shows the concentrations of a
    few selected oxygen-bearing gas and ice species. The left row
    shows the first few iterations, and the right row shows the
    convergence towards iteration 1000, which is considered as the
    final solution. The different full, dashed and dotted lines
    correspond to the different numbers of iterations carried out as
    indicated.}
  \label{fig:iteration2}
\end{figure*} 

Figure~\ref{fig:iteration2} shows some details how the temperature and
a few selected molecular concentrations change during the course of
the iterations, considering a vertical cut through the disc at
10\,au. We observe that the chemical structure changes considerably
within the first few iterations, but then the changes per iteration
settle down, and eventually the differences between iteration 400
(dash-dotted lines) and iteration 1000 (full lines) are barely
recognisable on the right side of Fig.~\ref{fig:iteration2}. The
temperature structure also changes. The gas and dust temperatures cool
down because (i) more \ce{OH} and \ce{H2O} form which lead to
increased line cooling and (ii) the optical depths increase due to the
presence of icy grains in upper regions. We conclude that our method
does converge, albeit very slowly, which agrees with the expected
convergence behaviour as discussed in Appendix~\ref{app:convergence}.

\subsection{Impact on the chemical and temperature structure}

\begin{figure*}
  \hspace*{0mm}
  \resizebox{185mm}{!}{
  \begin{tabular}{ccc}
    $r\!=\!1\,$au &
    $r\!=\!10\,$au &
    $r\!=\!100\,$au \\
    \hspace*{-4mm}
    \includegraphics[height=32.1mm,trim=9  15 15 31,clip]
                    {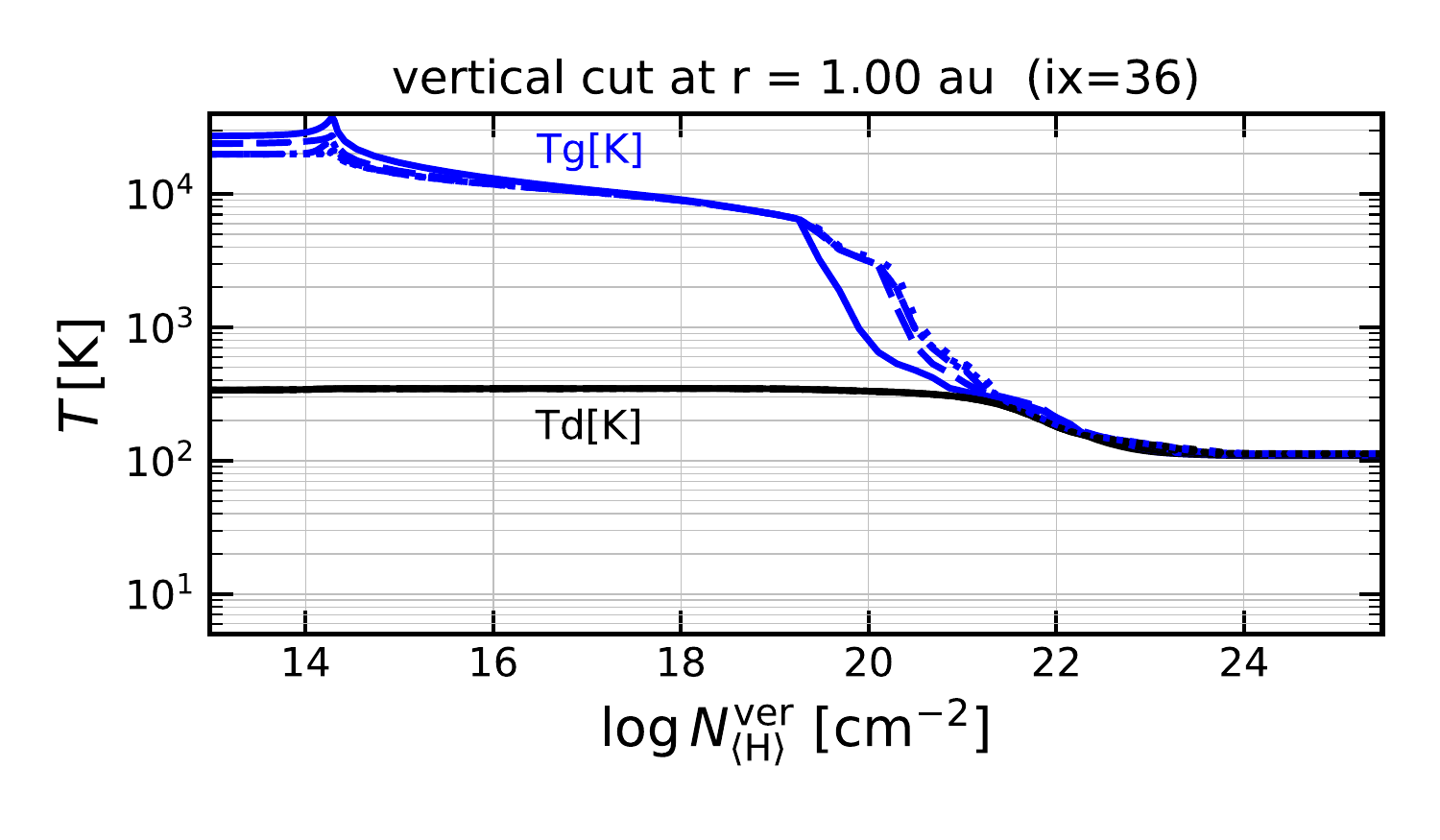} &
    \hspace*{-5mm}
    \includegraphics[height=32.1mm,trim=35 15 15 31,clip]
                    {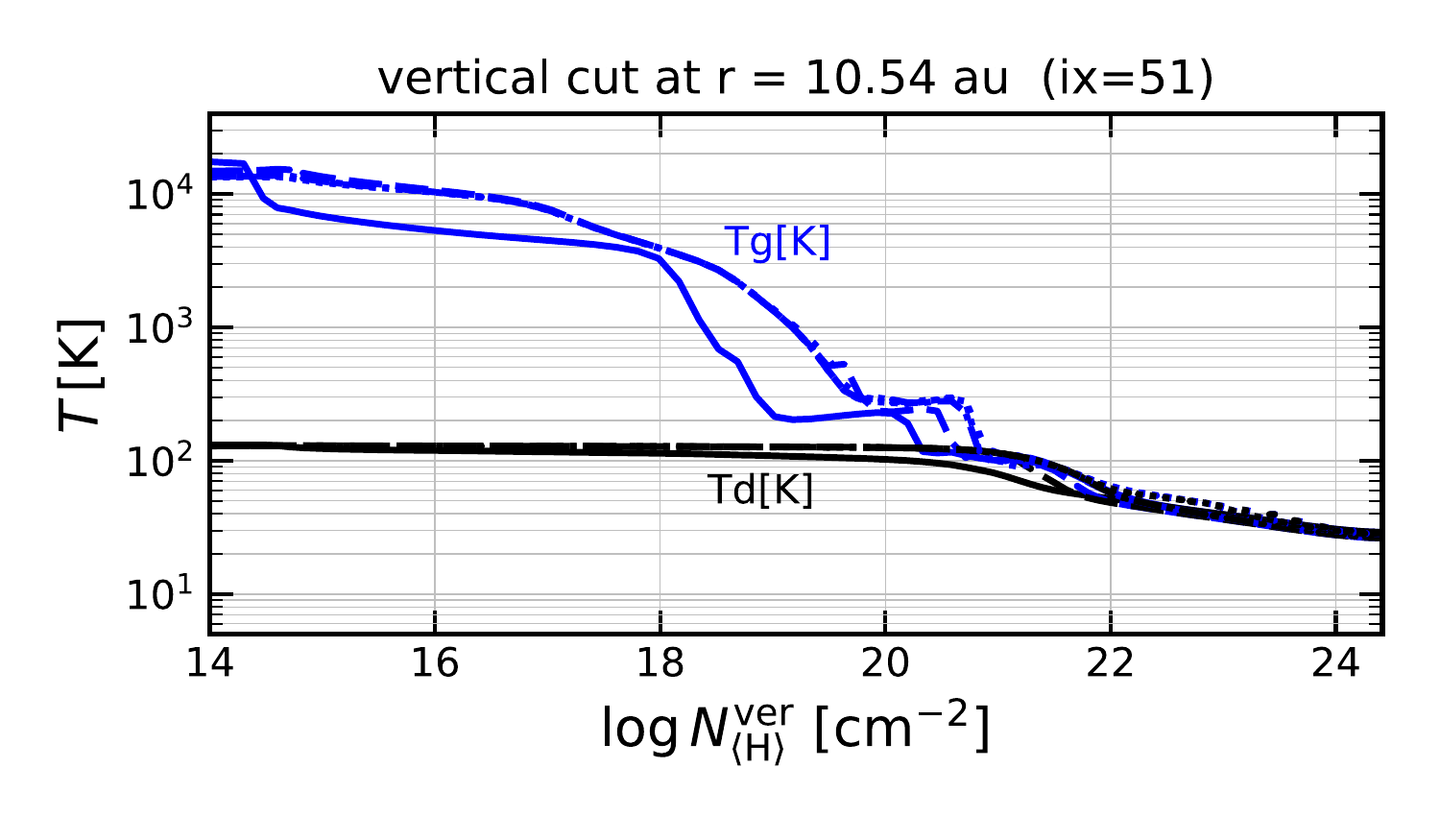} &
    \hspace*{-5mm}
    \includegraphics[height=32.1mm,trim=35 15 15 31,clip]
                    {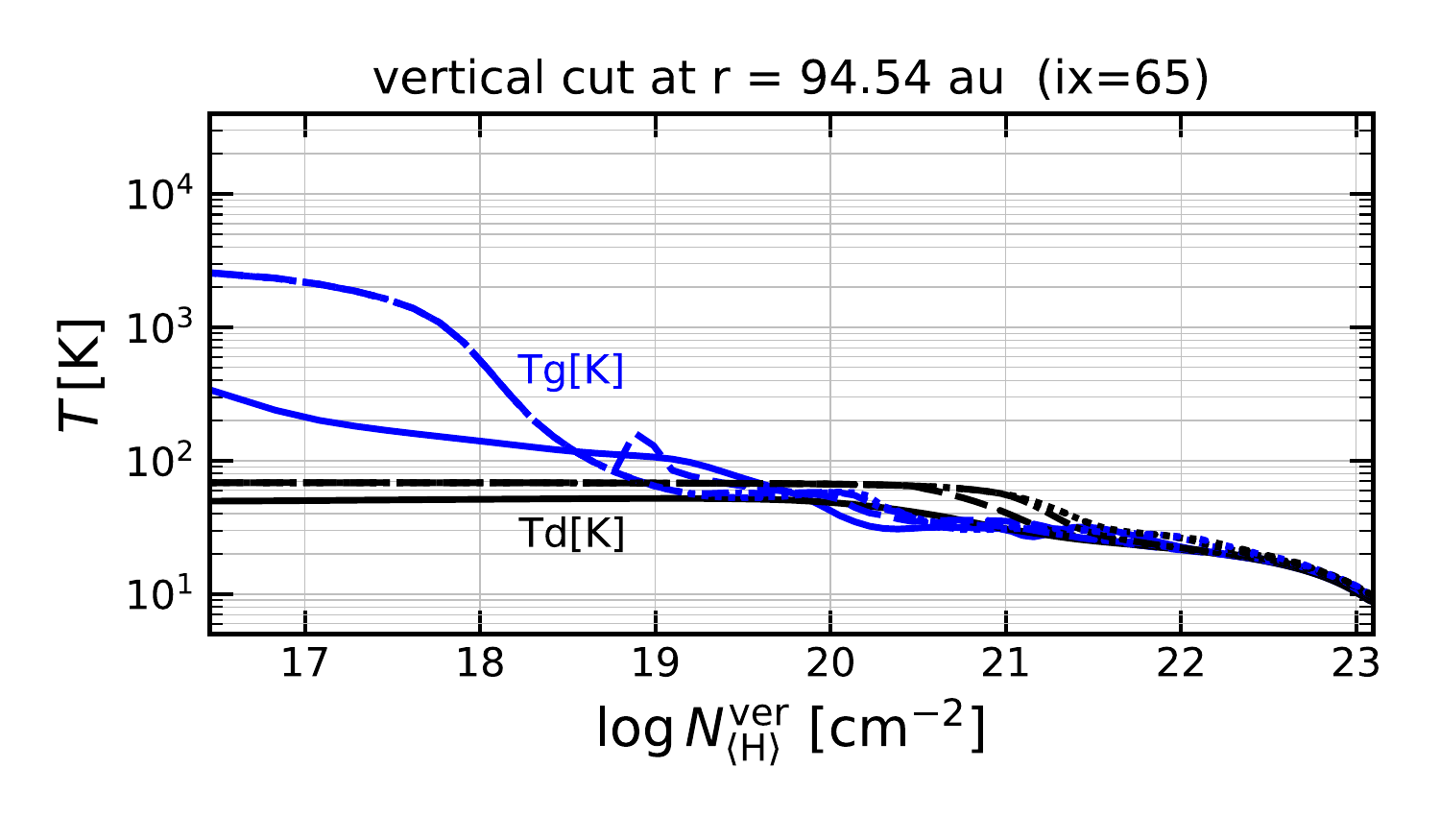} \\[-2mm]
    \hspace*{-4mm}
    \includegraphics[height=47.25mm,trim=7  61 15 31,clip]
                    {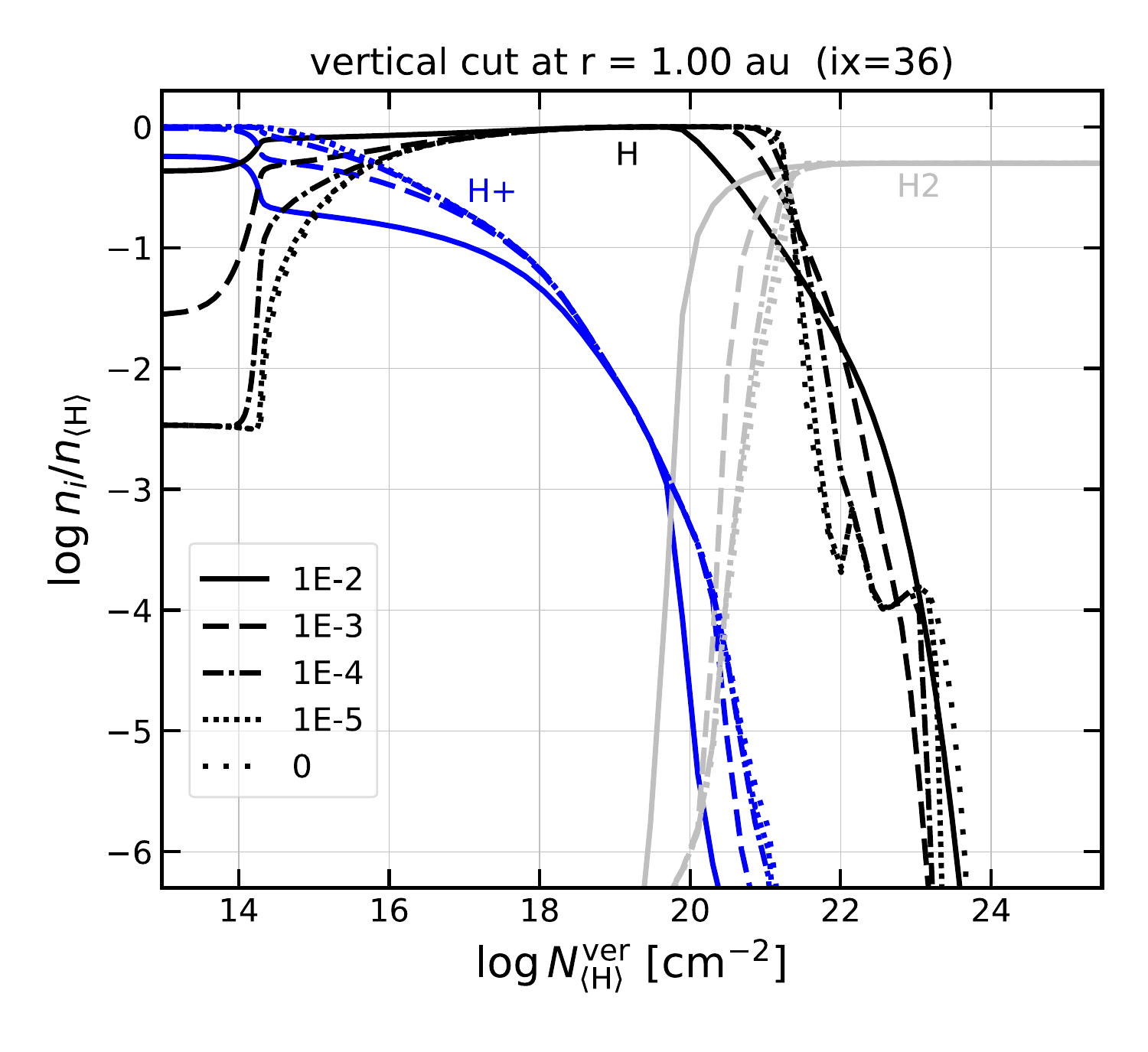} &
    \hspace*{-5mm}
    \includegraphics[height=47.25mm,trim=34 61 15 31,clip]
                    {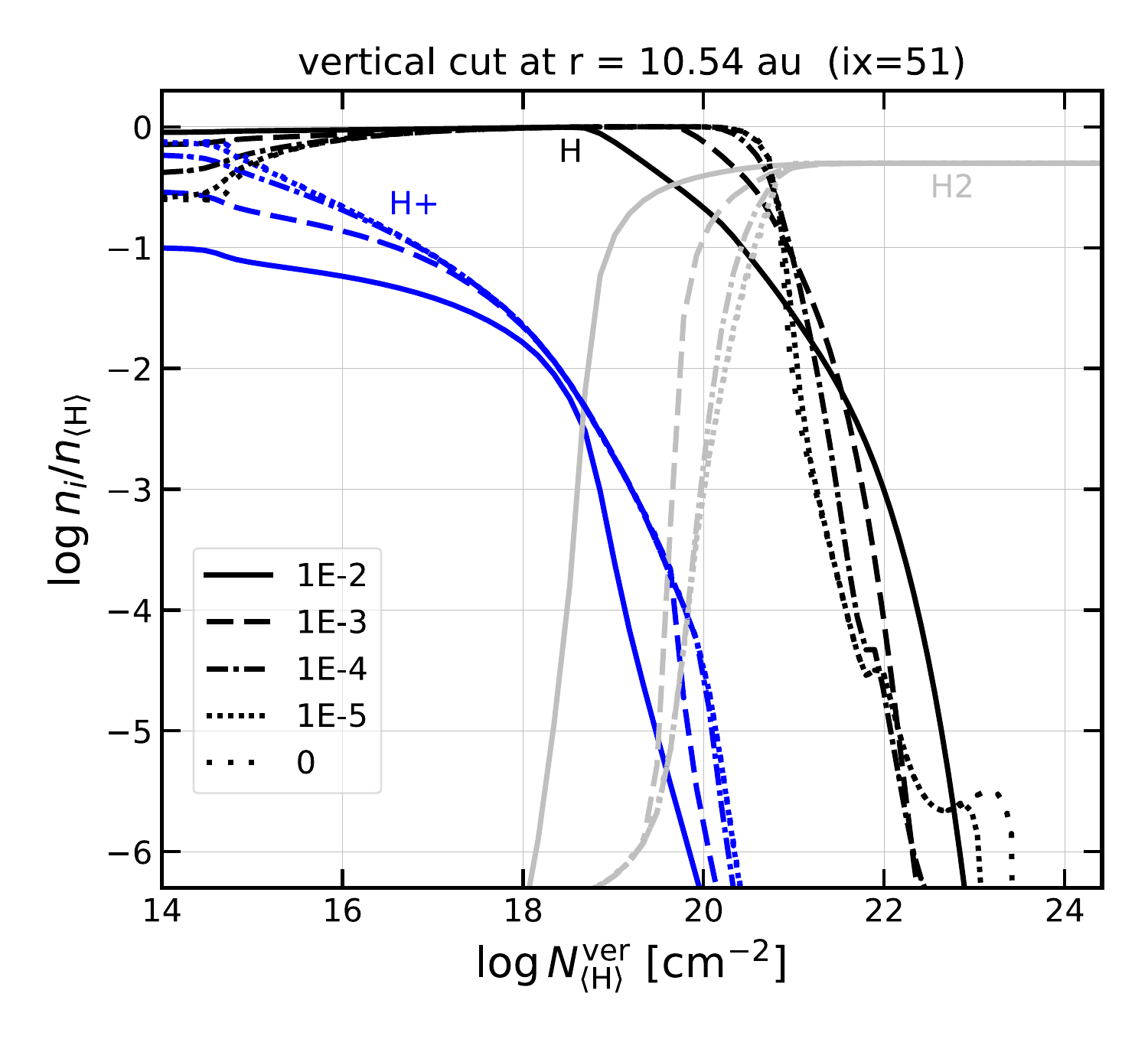} &
    \hspace*{-5mm}
    \includegraphics[height=47.25mm,trim=34 61 15 31,clip]
                    {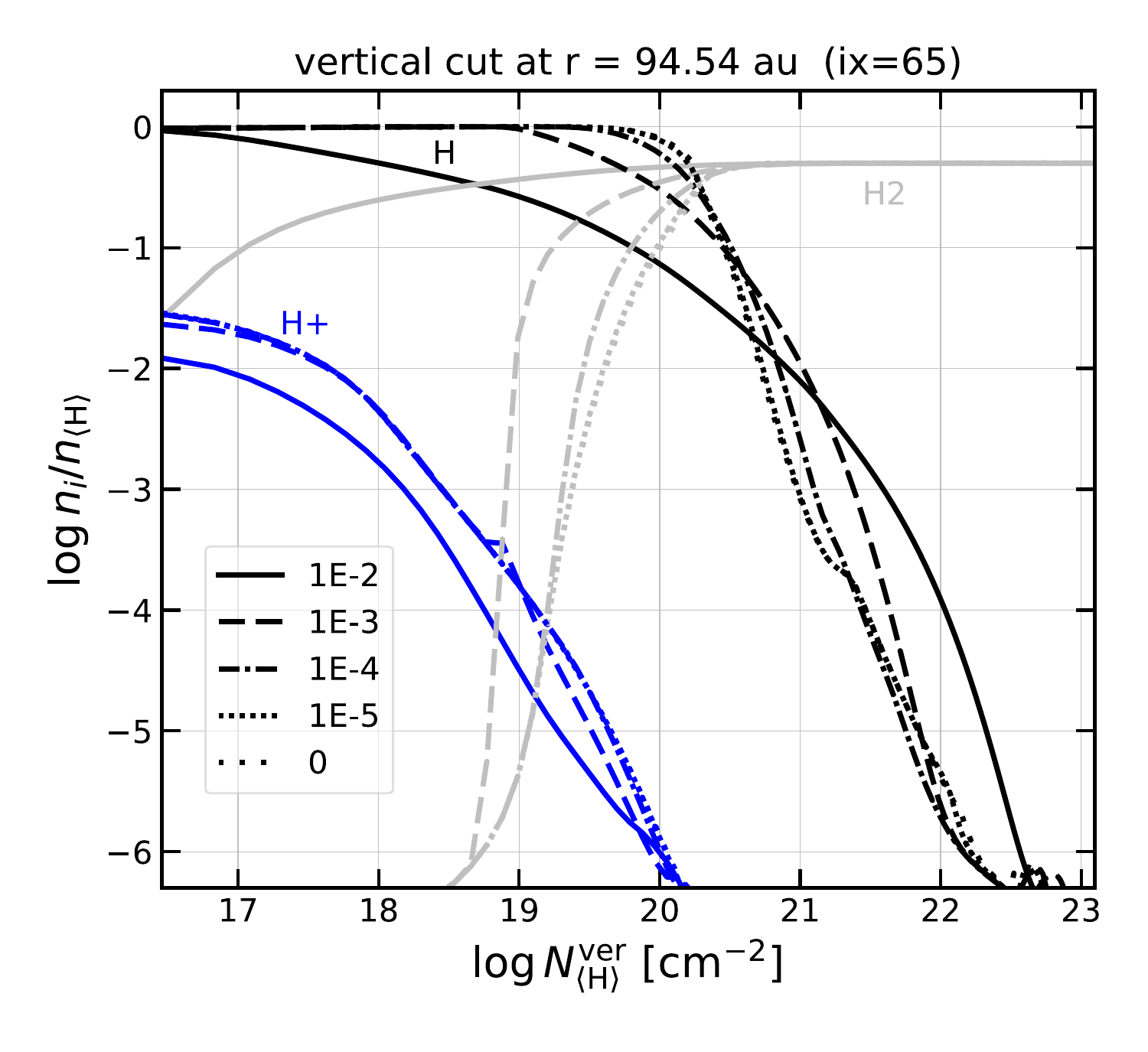} \\[-1mm]
    \hspace*{-4mm}
    \includegraphics[height=48.5mm,trim=16 61 15 31,clip]
                    {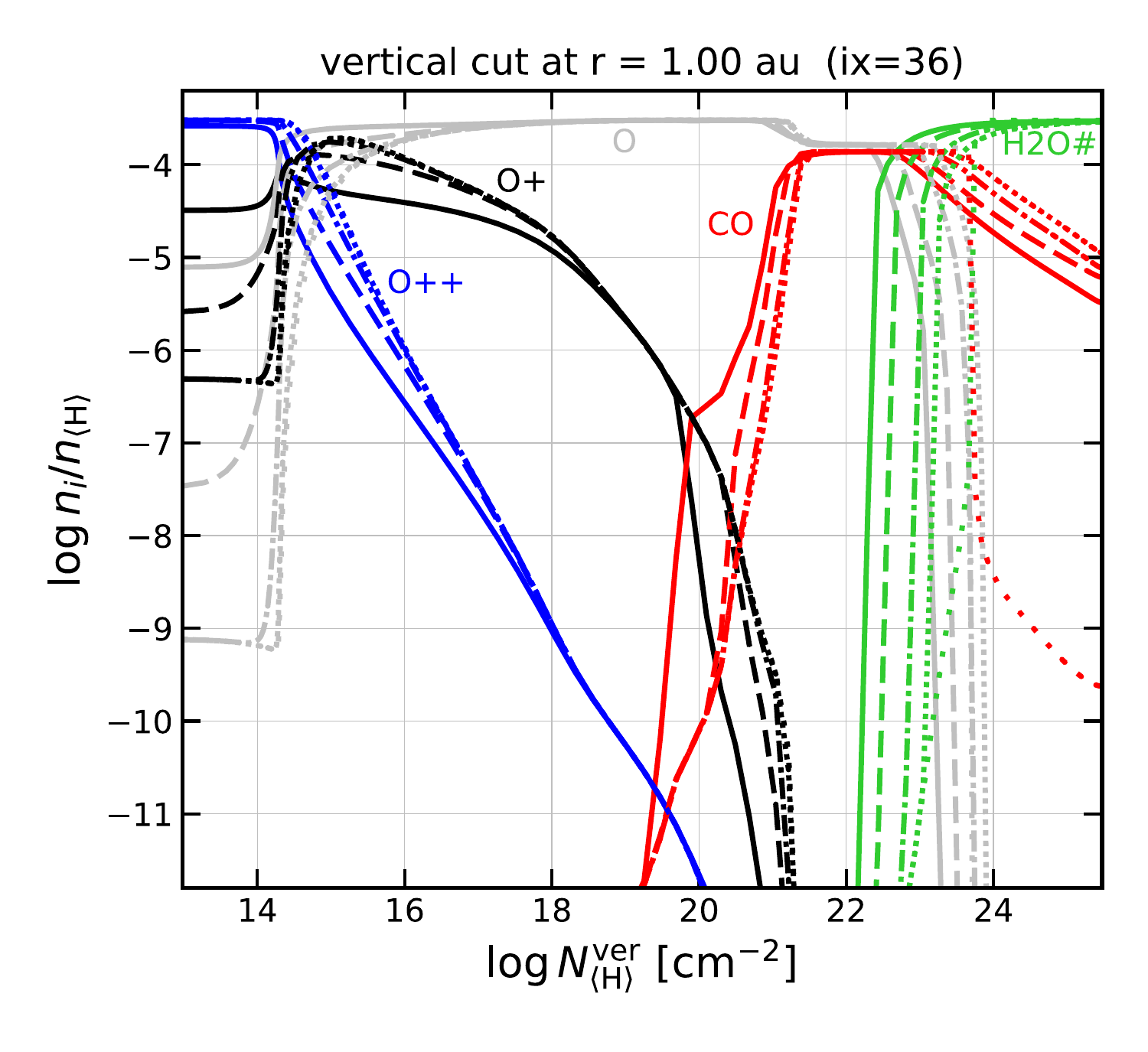} &
    \hspace*{-5mm}
    \includegraphics[height=48.5mm,trim=42 61 15 31,clip]
                    {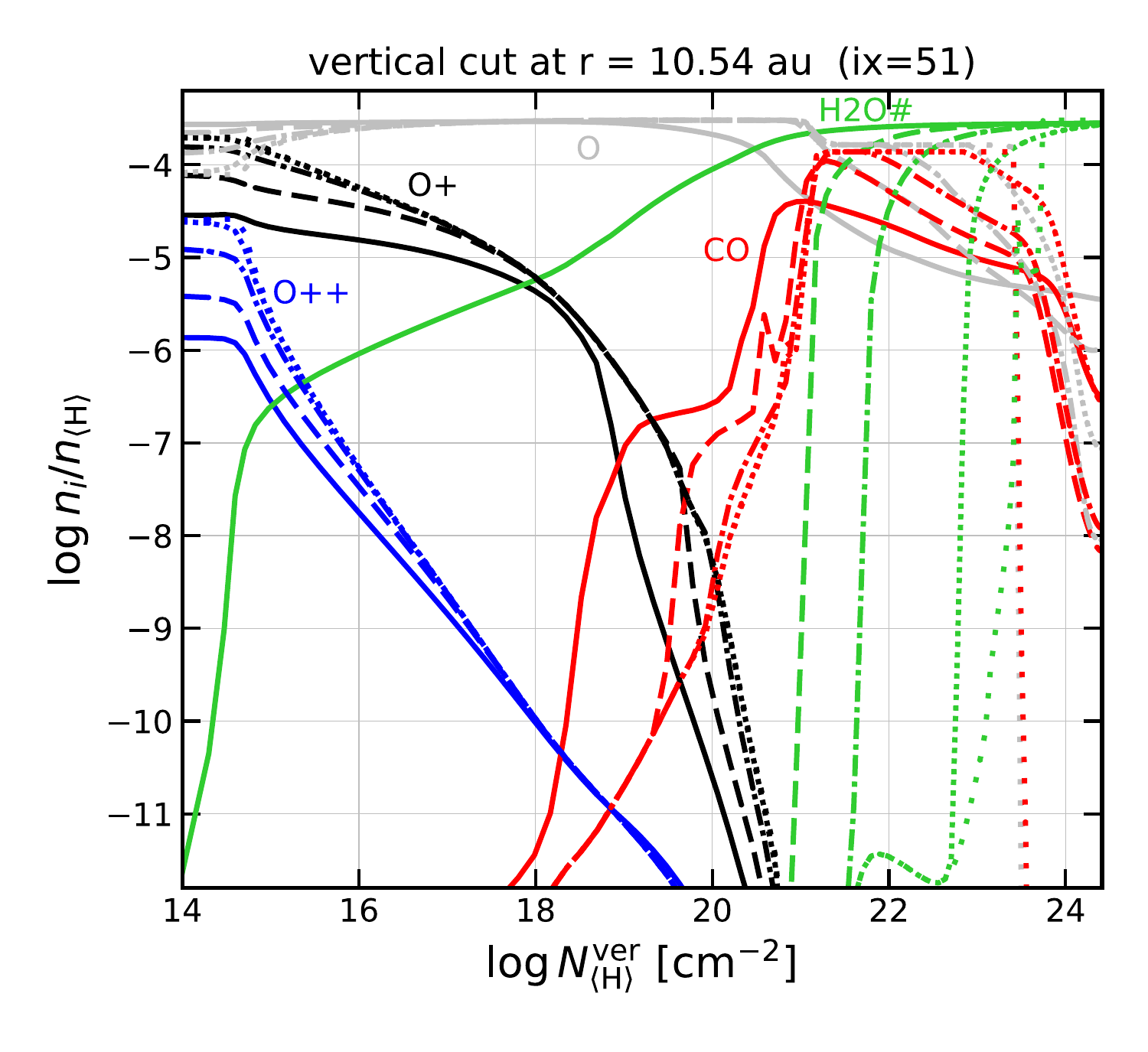} &
    \hspace*{-5mm}
    \includegraphics[height=48.5mm,trim=42 61 15 31,clip]
                    {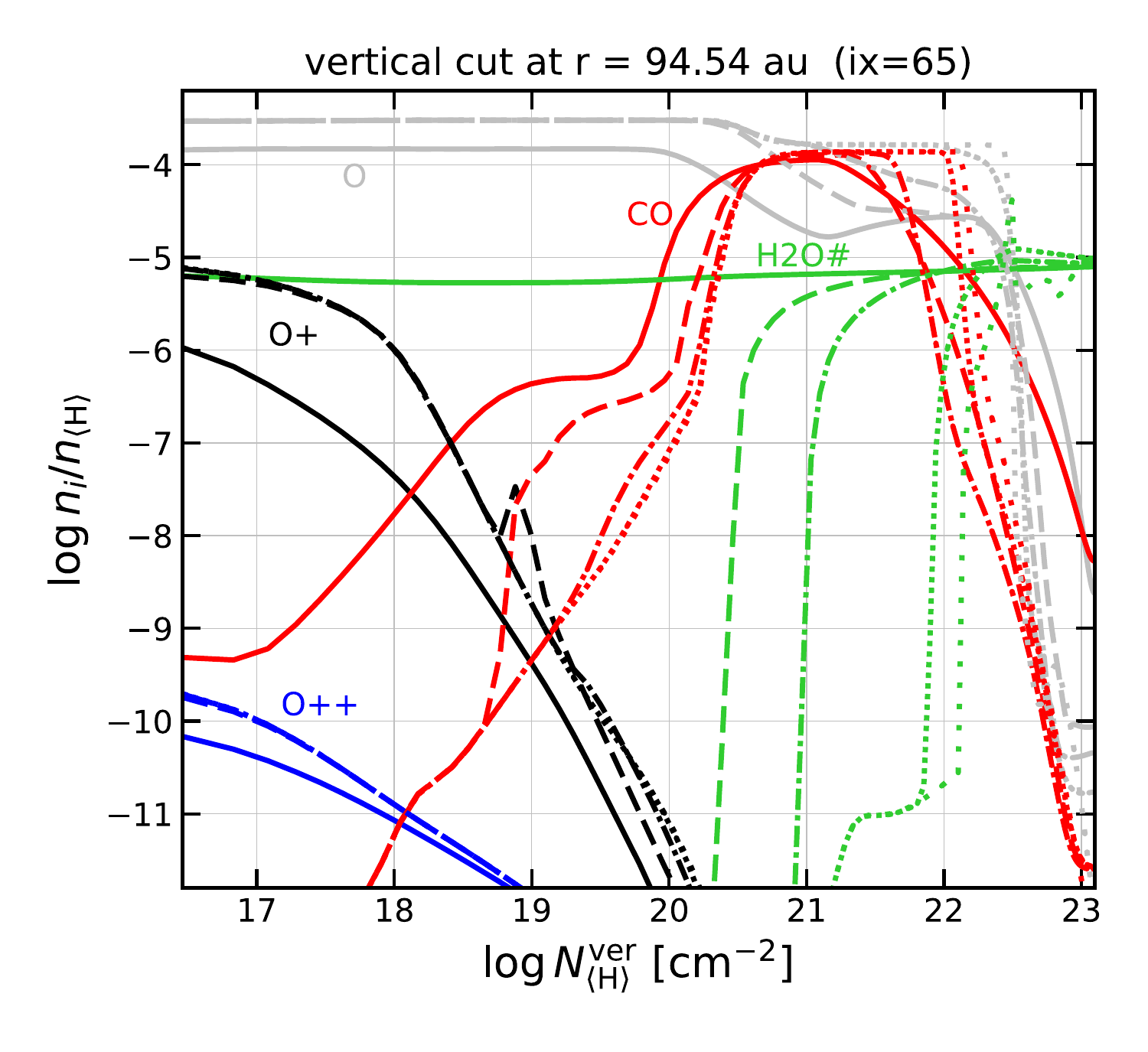} \\[-1mm]
    \hspace*{-4mm}
    \includegraphics[height=48.5mm,trim=16 61 15 31,clip]
                    {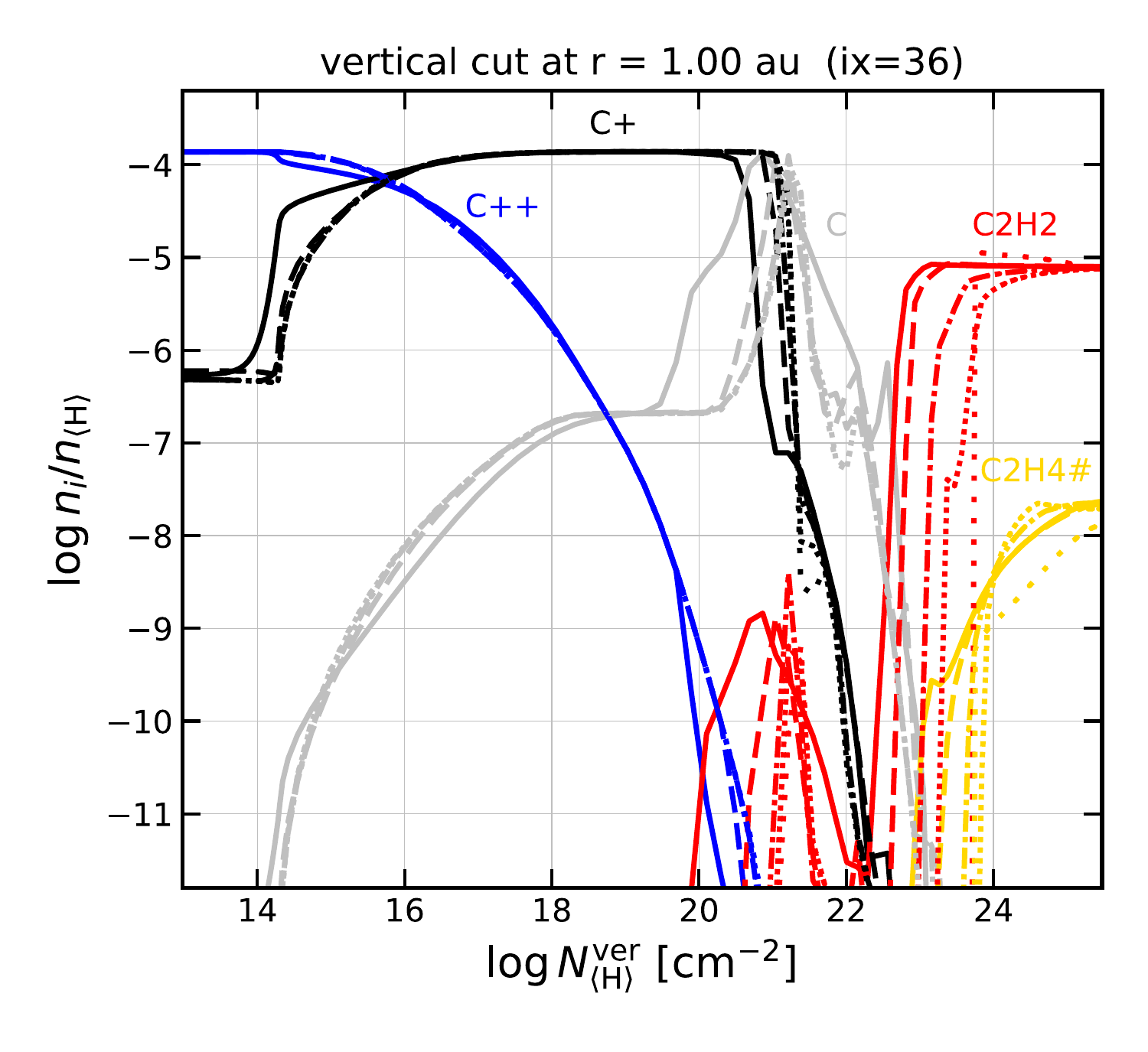} &
    \hspace*{-5mm}
    \includegraphics[height=48.5mm,trim=42 61 15 31,clip]
                    {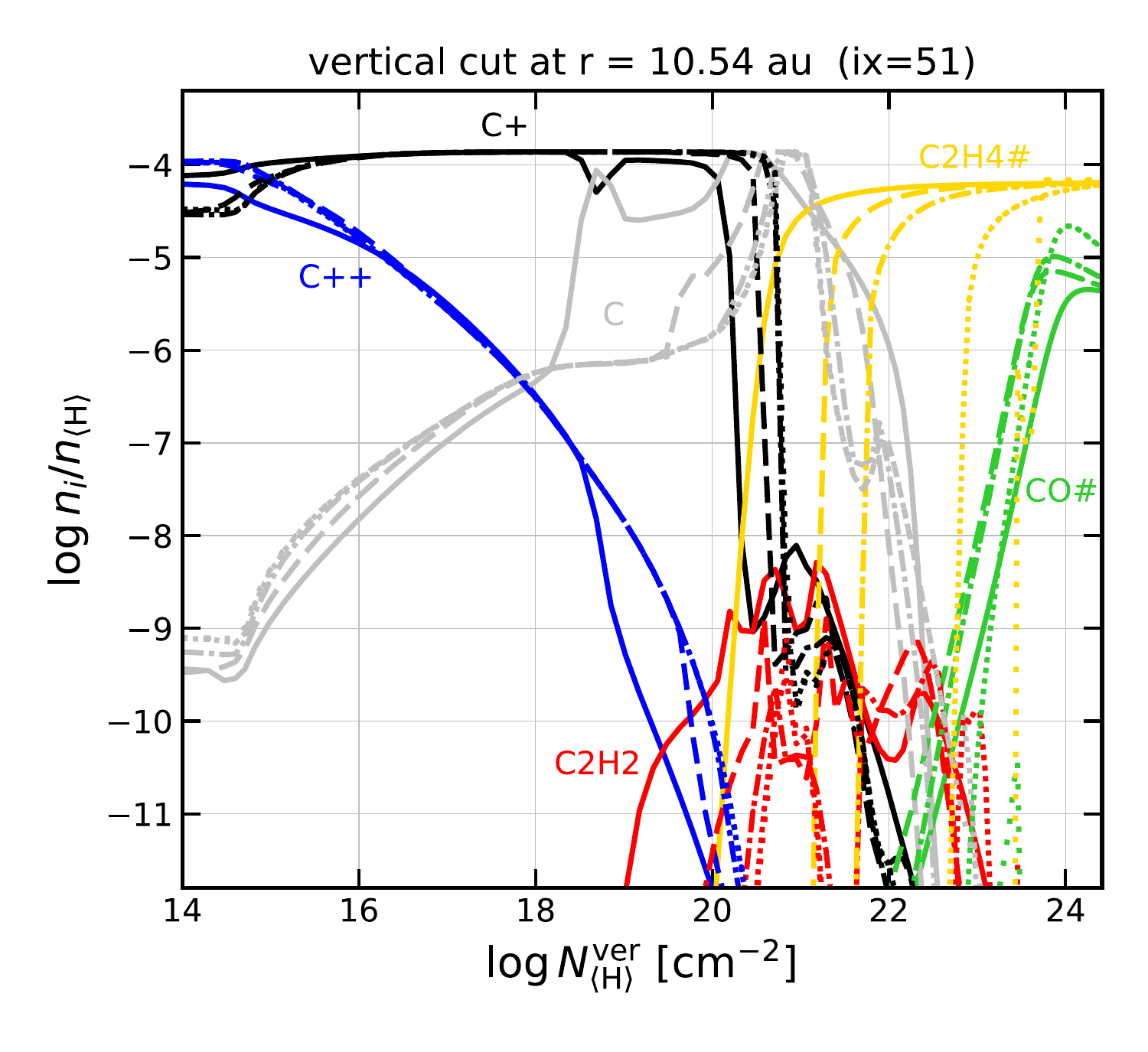} &
    \hspace*{-5mm}
    \includegraphics[height=48.5mm,trim=42 61 15 31,clip]
                    {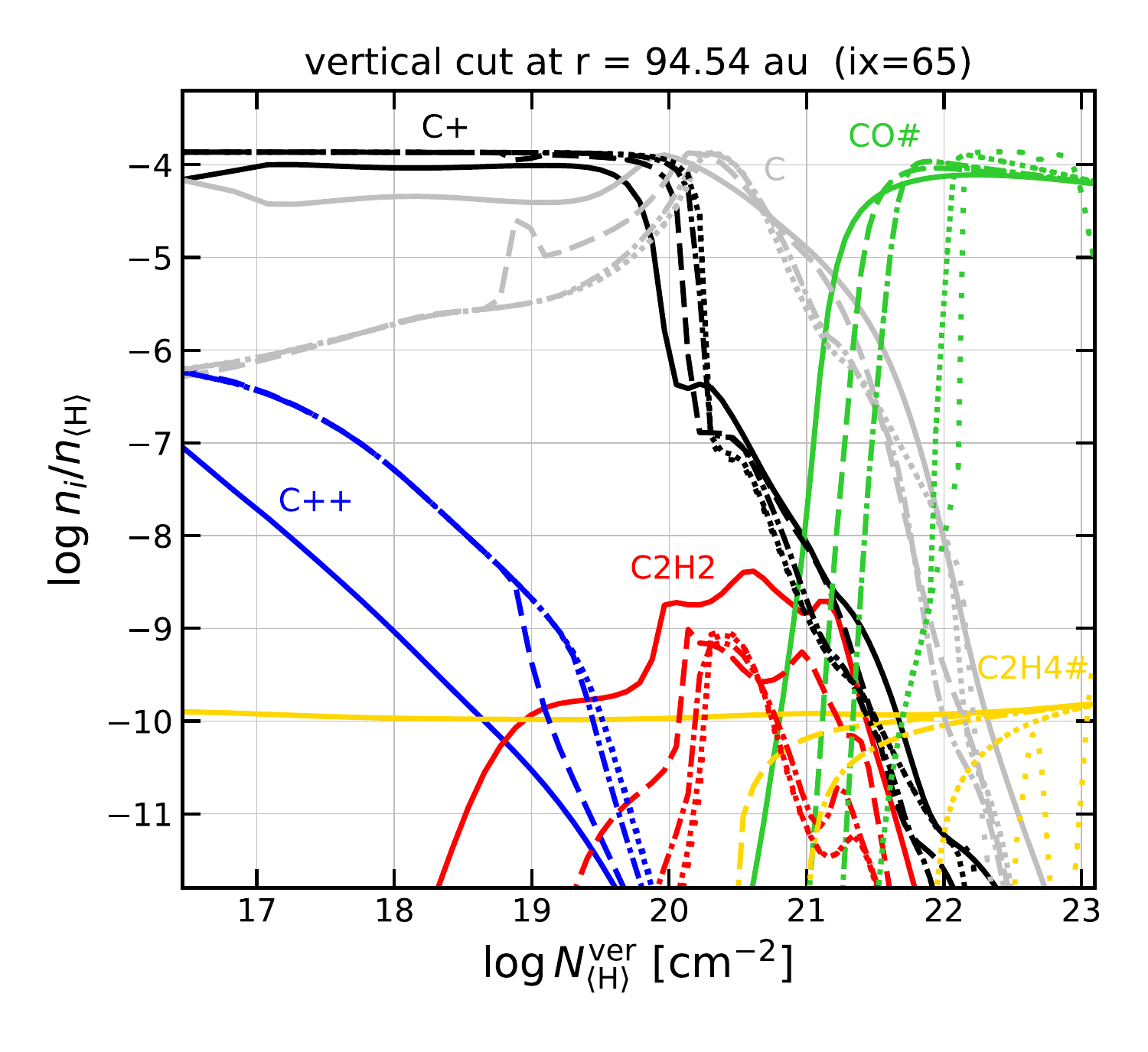} \\[-1mm]
    \hspace*{-4mm}
    \includegraphics[height=55mm,trim=16 21 15 31,clip]
                    {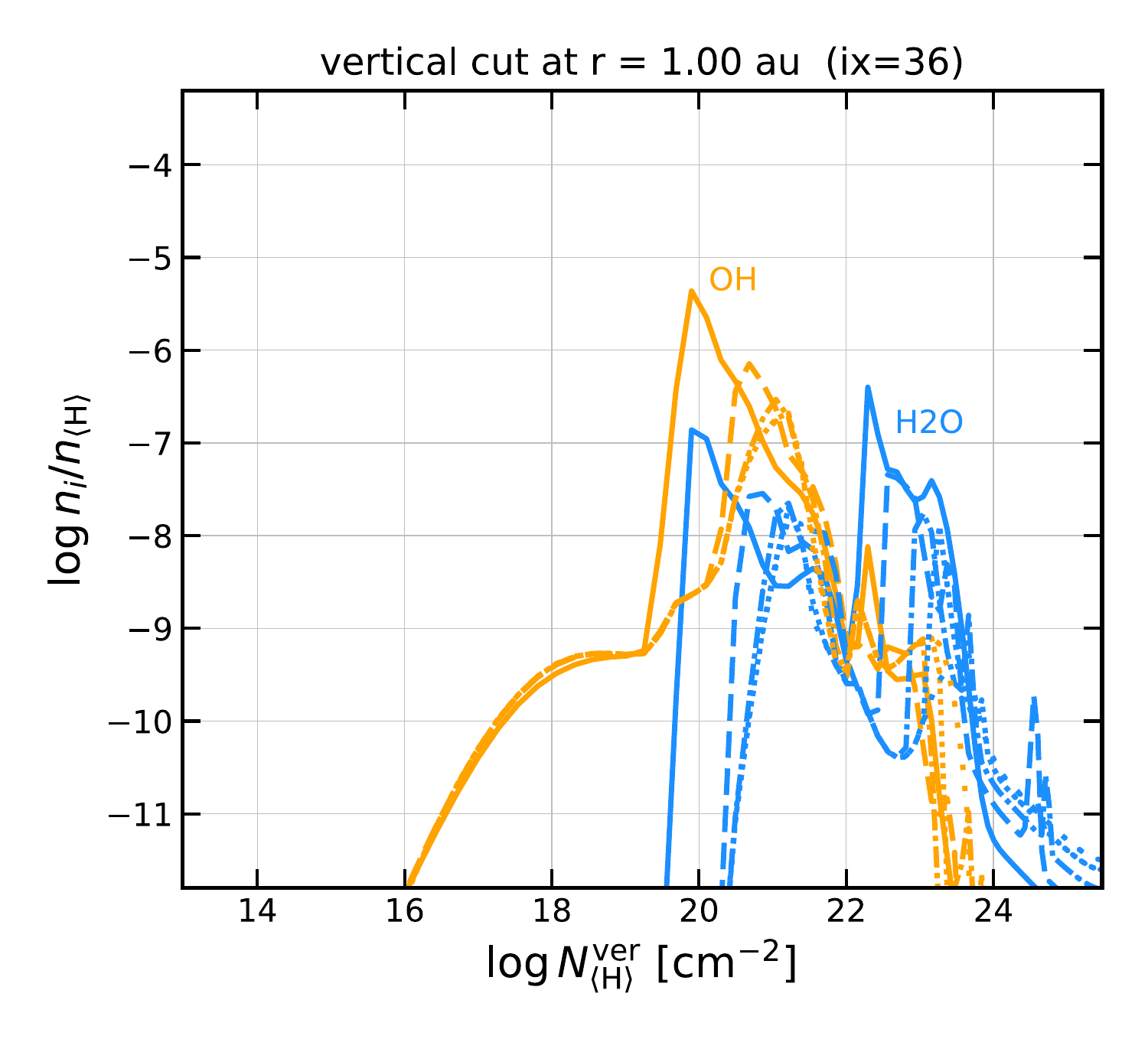} &
    \hspace*{-5mm}
    \includegraphics[height=55mm,trim=42 21 15 31,clip]
                    {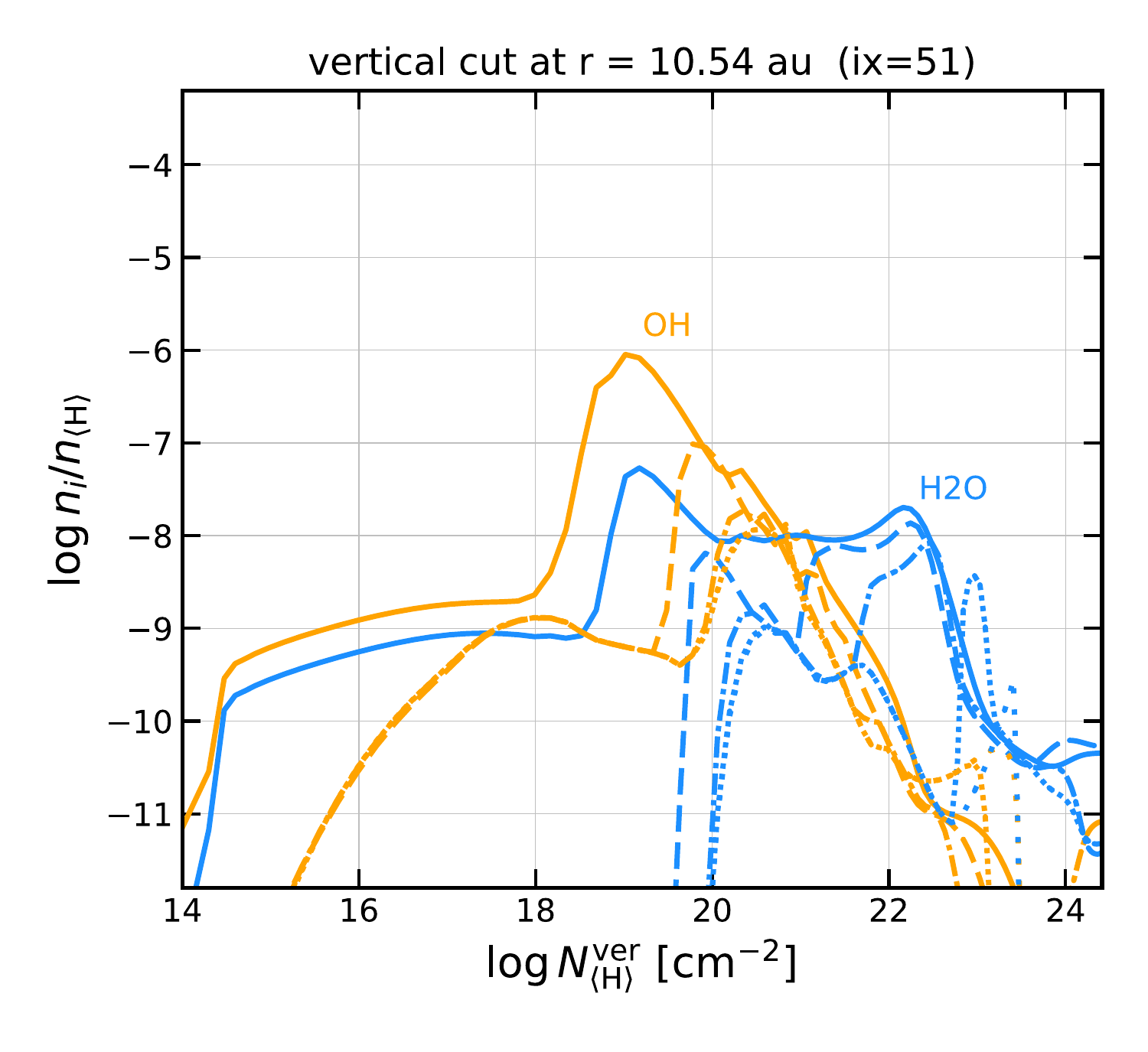} &
    \hspace*{-5mm}
    \includegraphics[height=55mm,trim=42 21 15 31,clip]
                    {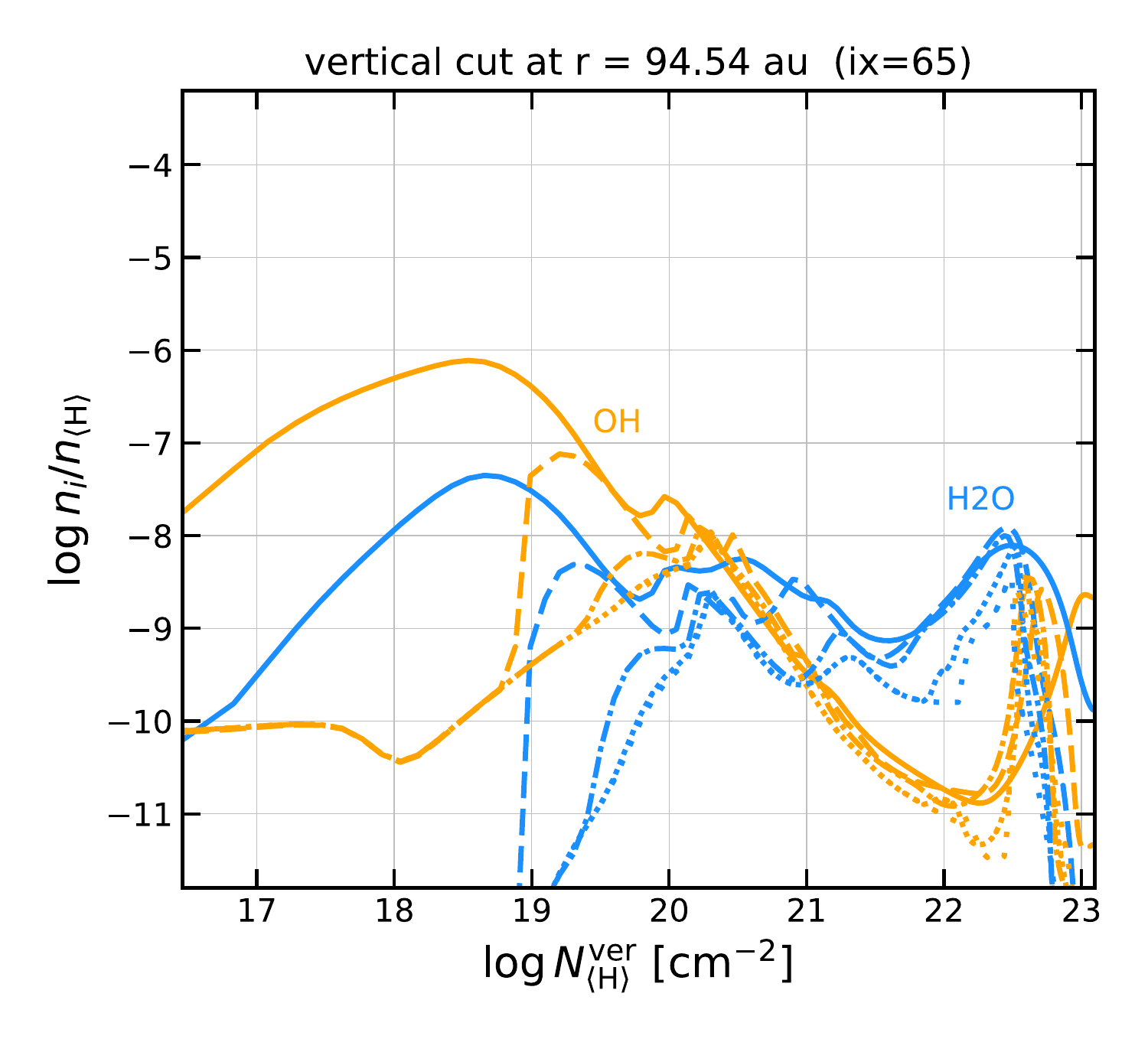} \\[-1mm]
  \end{tabular}}
  \caption{Resulting temperatures and chemical concentrations as
    function of vertical hydrogen column density $N_{\rm\langle
      H\rangle}$, for radial distances 1\,au, 10\,au, and 100\,au,
    from left to right. The top row shows the gas and dust
    temperatures, the second row the main hydrogen species, the third
    row the oxygen chemistry, and the last row the carbon chemistry.
    Only a few selected species are plotted. In each part figure, the
    various dashed and dotted lines correspond to the mixing parameter
    $\alpha_{\rm mix}$ as explained in the legends.}
  \label{fig:chem_results}
\end{figure*}

\subsubsection{Overview of principle effects}

Figure \ref{fig:chem_results} shows our resulting temperature and
chemical structures. Each plot shows a sequence of well-iterated
results for $\alpha_{\rm mix}\in\{0,10^{-5},10^{-4},10^{-3},10^{-2}\}$
using different line styles.  The results are shown by three vertical
cuts at constant distance to the star, at 1\,au, 10\,au and 100\,au,
each as function of hydrogen nuclei column density $N_{\rm\langle
  H\rangle}\rm\,[cm^{-2}]$.

The figure shows a number of remarkable changes of
the chemical structure that are caused by the inclusion of vertical
mixing, which will be detailed in the forthcoming sections.
\begin{itemize}
\item[a)] The ices that predominantly form in the midplane are mixed
  upward by a distance that corresponds to several orders of magnitude
  in column density.
\item[b)] Molecules, which are mainly formed in the warm intermittent
  disc layers, where the temperatures are a few 100\,K, are mixed
  both downward and upward.
\item[c)] The increased molecular abundances of OH and H$_2$O cause
  reinforced line cooling at column densities
  $\NH$\,$\sim$\,$10^{19}...\,10^{22}\rm\,cm^{-2}$, which lowers the gas
  temperature and improves the conditions for the
  formation and survival of more molecules. These layers are the
  regions directly observable with infrared line spectroscopy.  We see
  that the maximums of the OH and H$_2$O concentrations shift leftward
  and upward by about 1-2 orders of magnitude in these plots.  These
  changes are larger for more distant regions and/or larger mixing
  parameter $\alpha_{\rm mix}$.
\item[d)] Neutral atoms are mixed up into the highest layers, where
  they absorb X-rays and FUV radiation, which leads to additional
  absorption of UV and X-rays, causing an increase of temperature, and
  an increase of the shielding for the upper disc layers at larger
  radii.
\item[e)] Ions such as O$^+$, which are predominantly formed in high
  layers, are increasingly able to meet molecules and even icy
  grains which are mixed upward.  This leads to a very rich
  ion-molecule chemistry, with new chemical pathways, that are
  normally unimportant in models without mixing.
\end{itemize}

\subsubsection{Beating UV photo-desorption by mixing}
\label{sec:icemix}

\begin{figure*}
  \hspace*{0mm}
  \resizebox{185mm}{!}{
  \begin{tabular}{ccc}
    $\alpha_{\rm mix}\!=\!0$ &
    $\alpha_{\rm mix}\!=\!10^{-3}$ &
    $\alpha_{\rm mix}\!=\!10^{-2}$ \\
    \hspace*{-8mm}
    \includegraphics[height=25.5mm,trim=16 48 10 15,clip]
                    {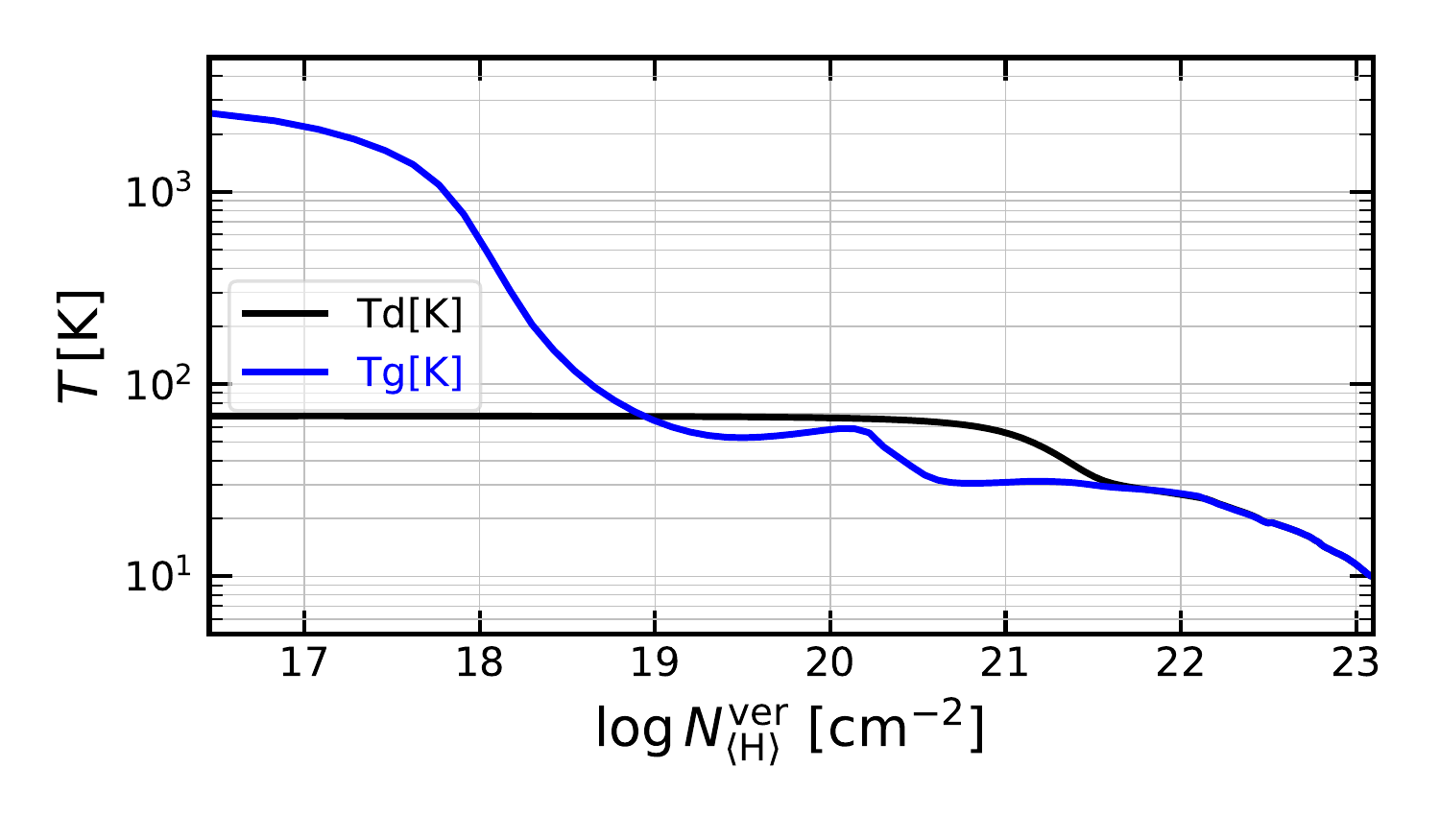} &
    \hspace*{-8mm}
    \includegraphics[height=25.5mm,trim=35 48 10 15,clip]
                    {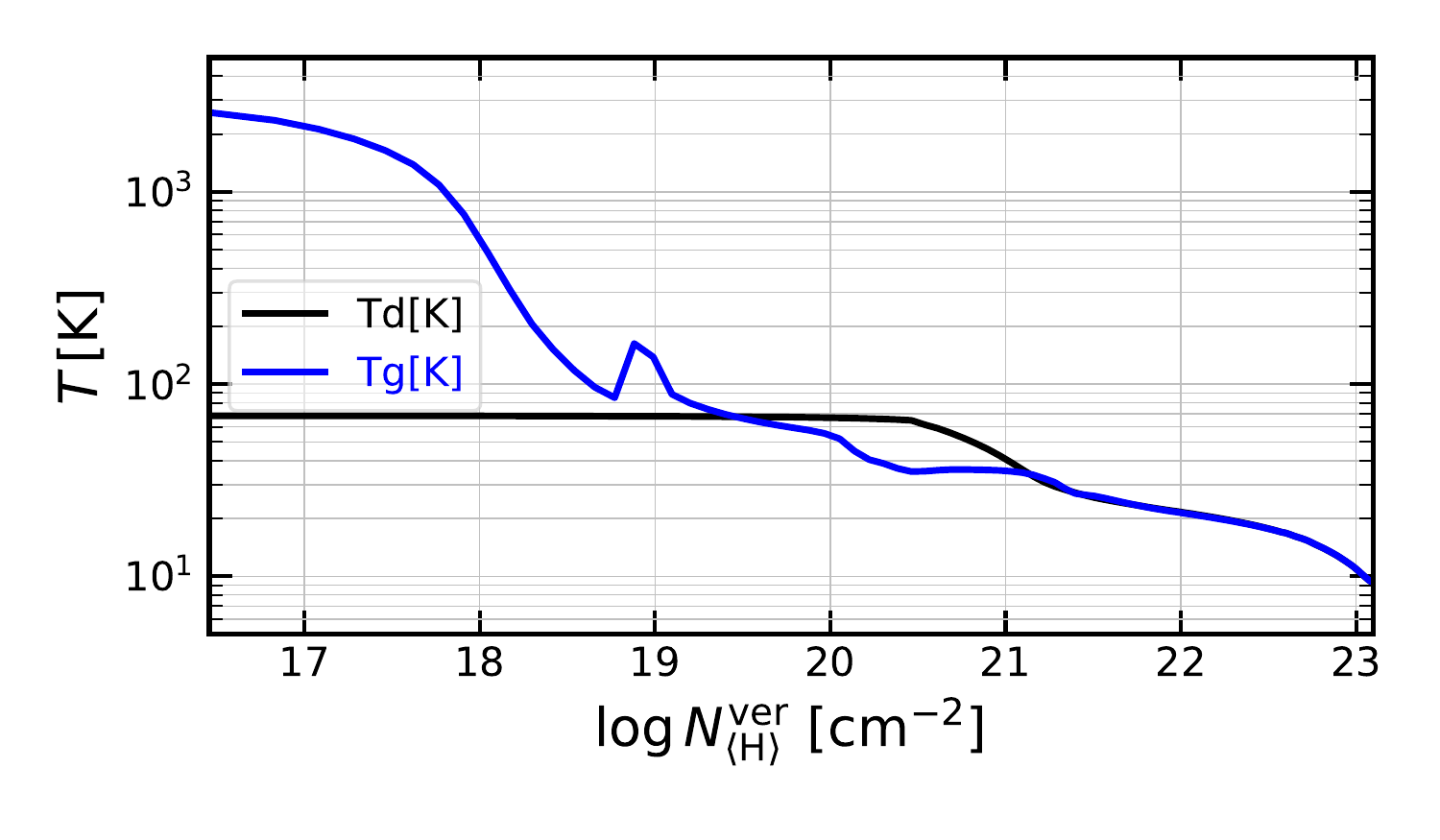} &
    \hspace*{-12.2mm}
    \includegraphics[height=25.5mm,trim=35 48 10 15,clip]
                    {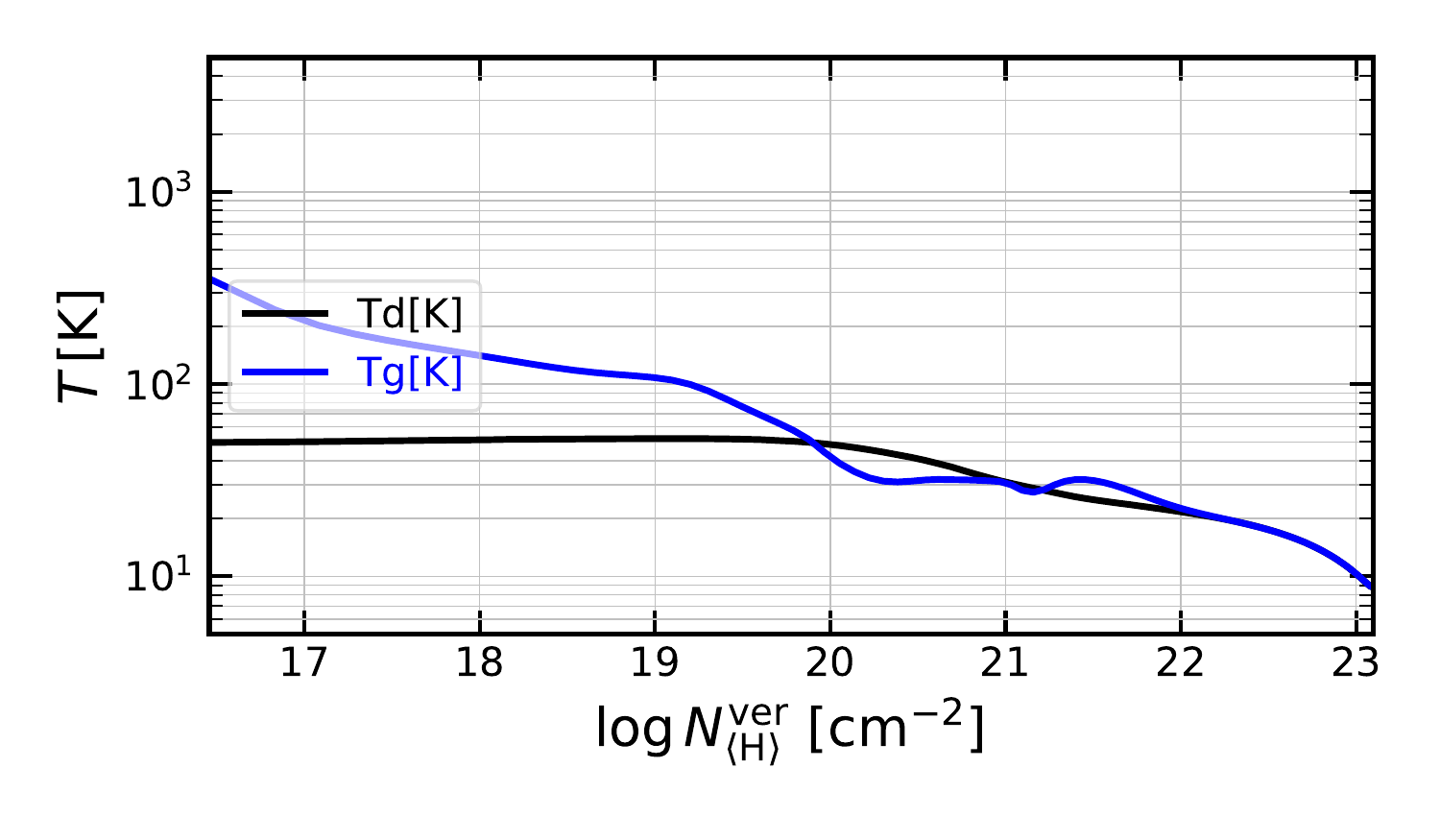} \\[-1mm]
    \hspace*{-4mm}
    \includegraphics[height=29mm,trim=18 48 42 17,clip]
                    {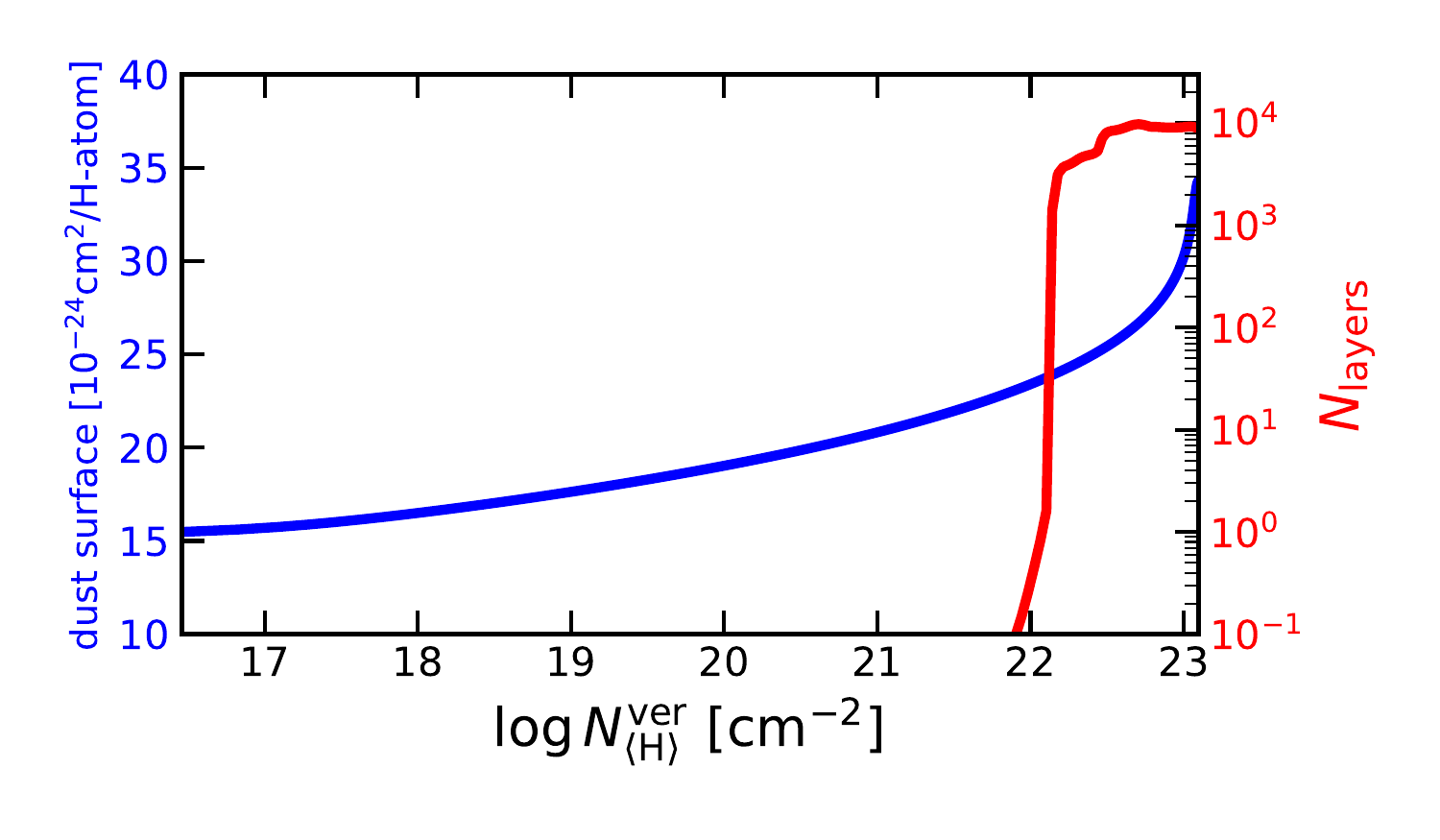} &
    \hspace*{-4mm}
    \includegraphics[height=29mm,trim=35 48 42 17,clip]
                    {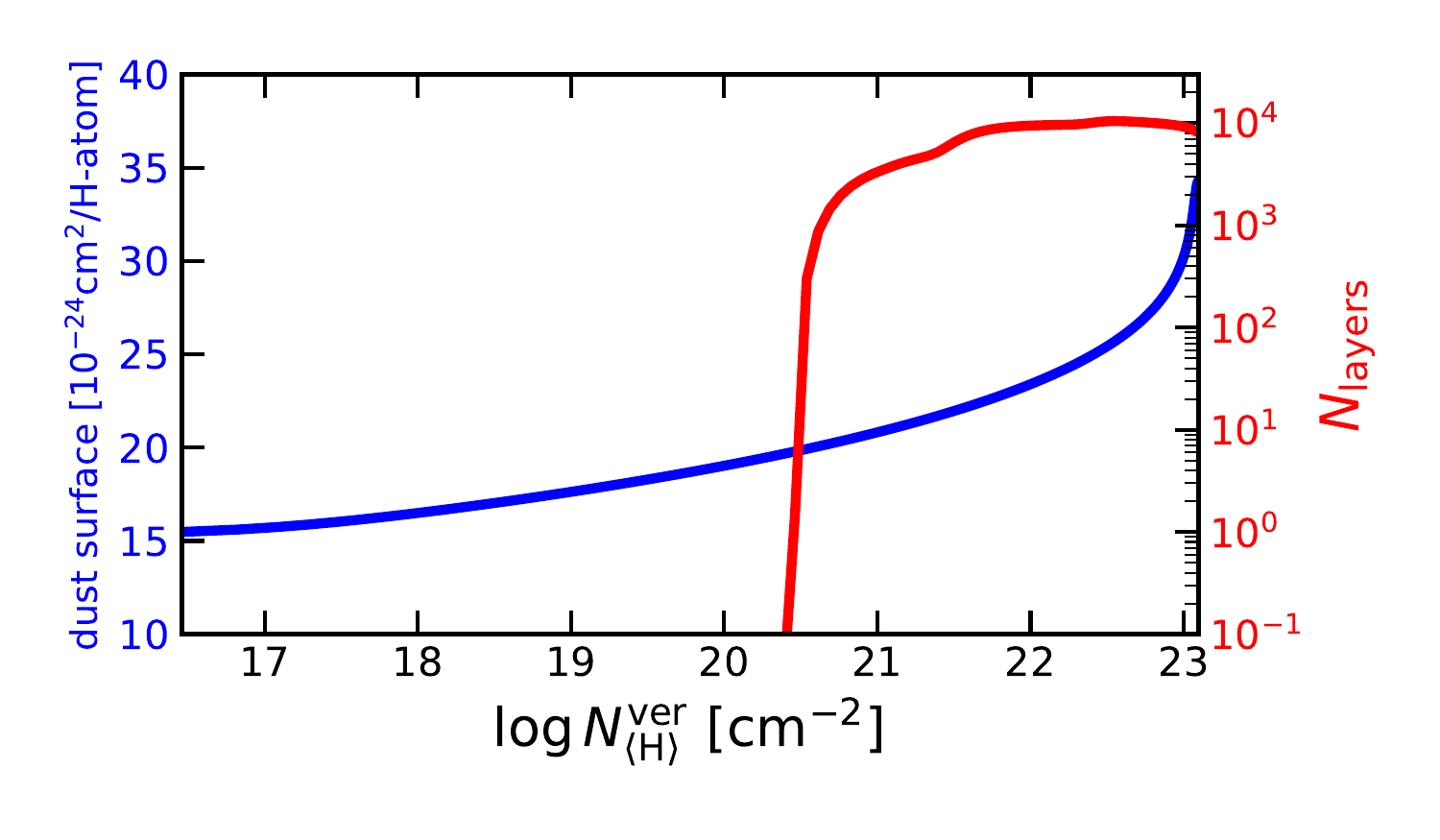} &
    \hspace*{-4mm}
    \includegraphics[height=29mm,trim=35 48 15 17,clip]
                    {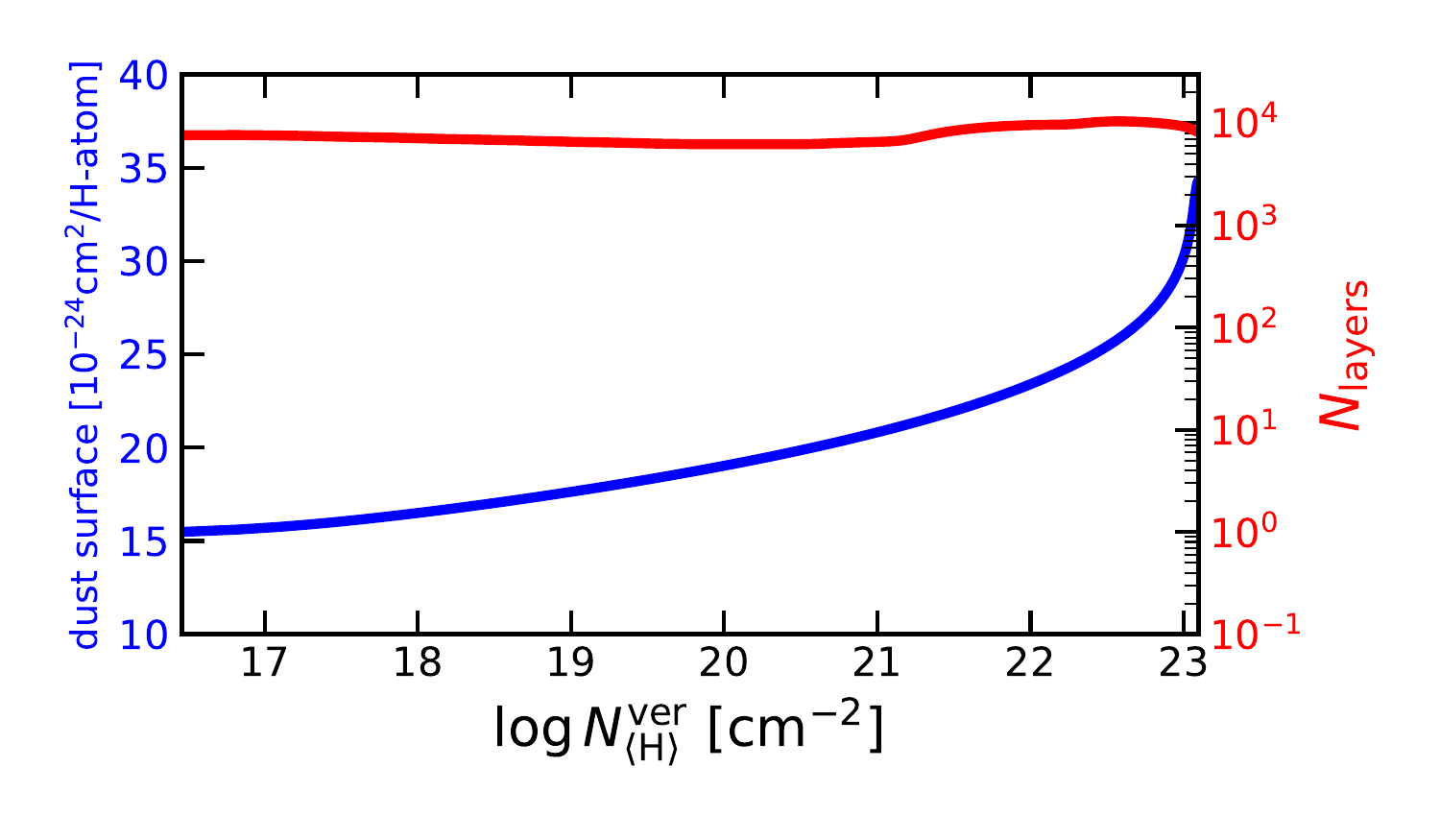} \\[-1mm]
    \hspace*{-9.5mm}
    \includegraphics[height=49.3mm,trim=20 22 15 17,clip]
                    {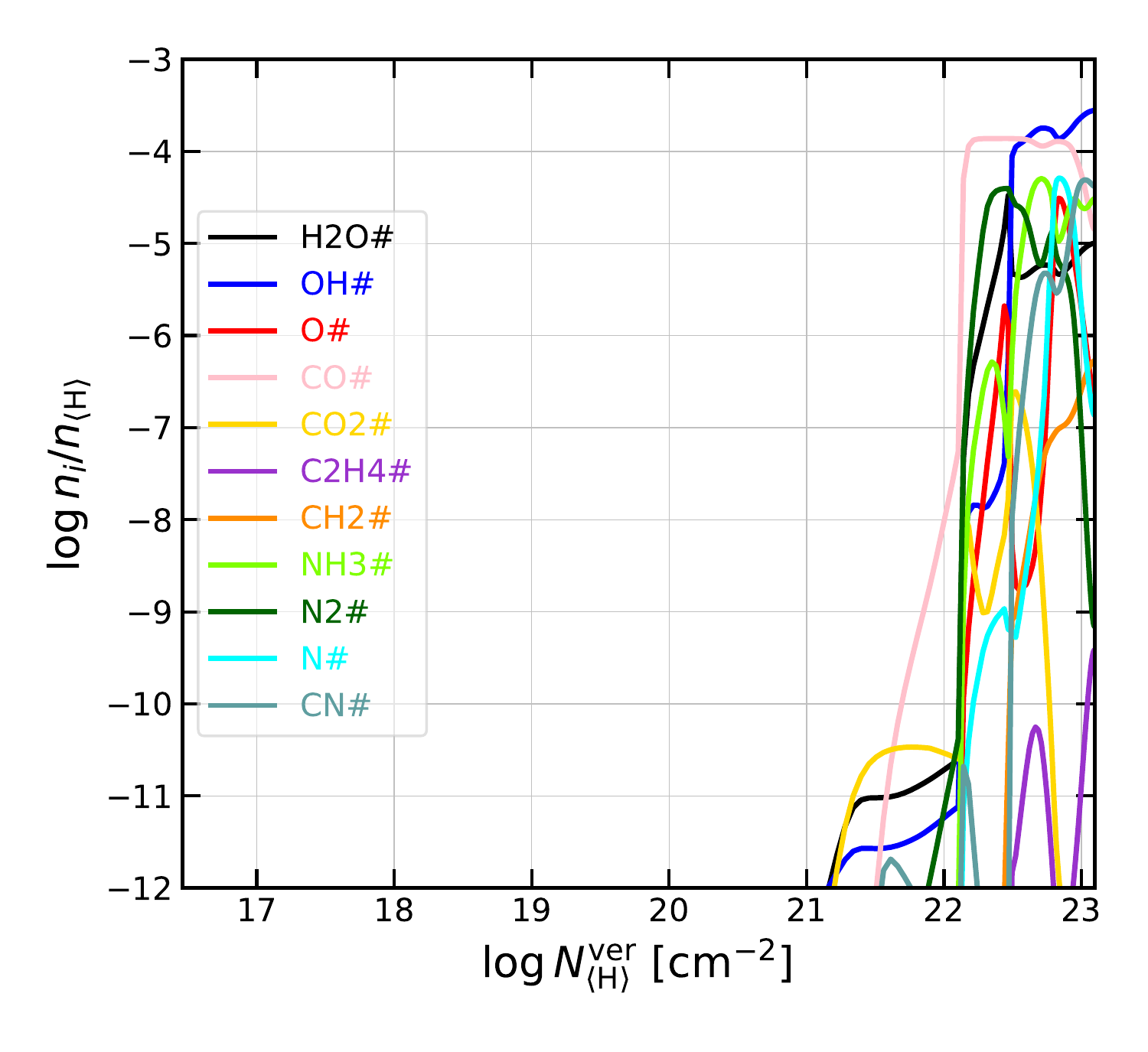} &
    \hspace*{-8.5mm}
    \includegraphics[height=49.3mm,trim=43 22 15 17,clip]
                    {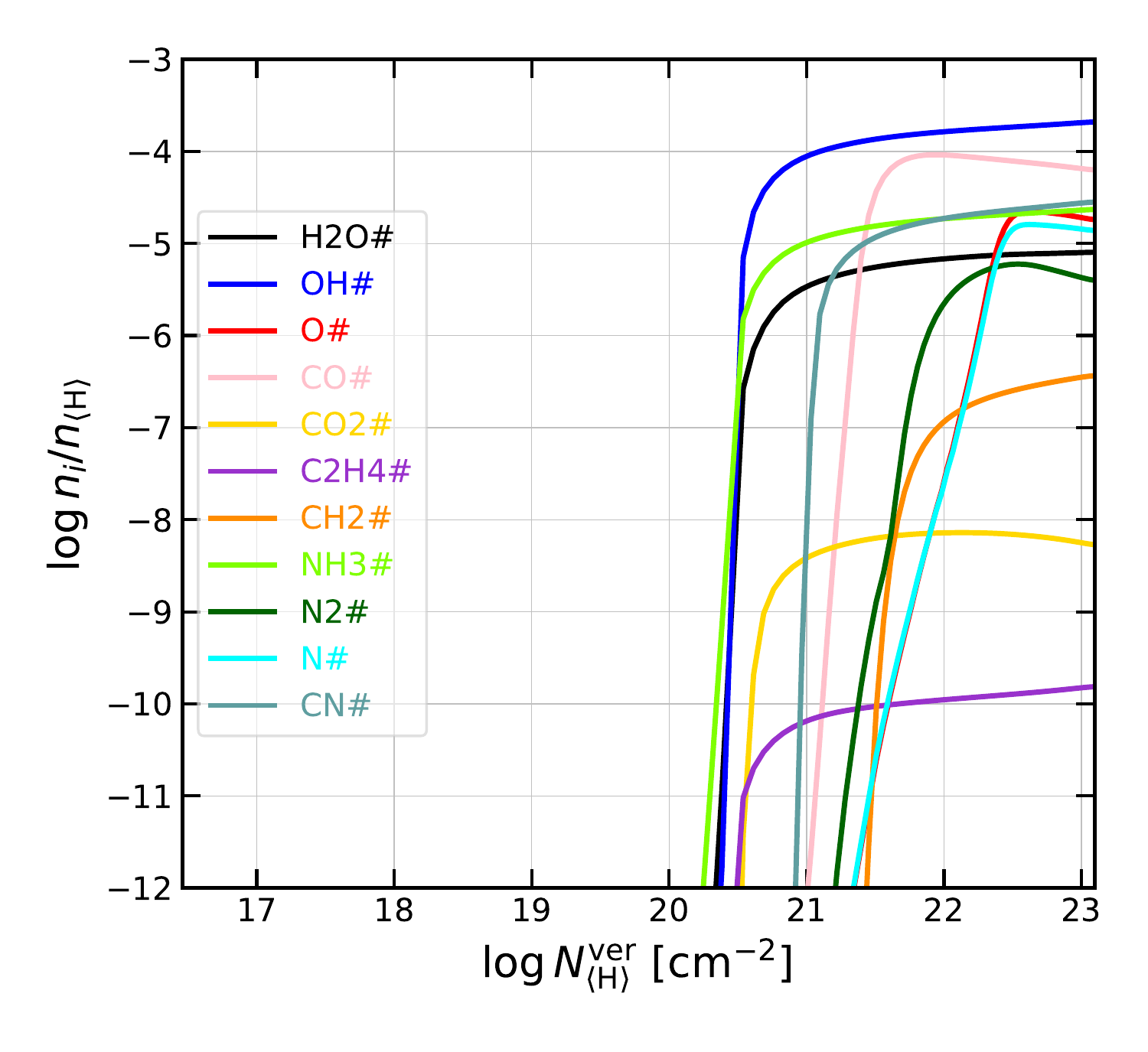} &
    \hspace*{-12.5mm}
    \includegraphics[height=49.3mm,trim=43 22 15 17,clip]
                    {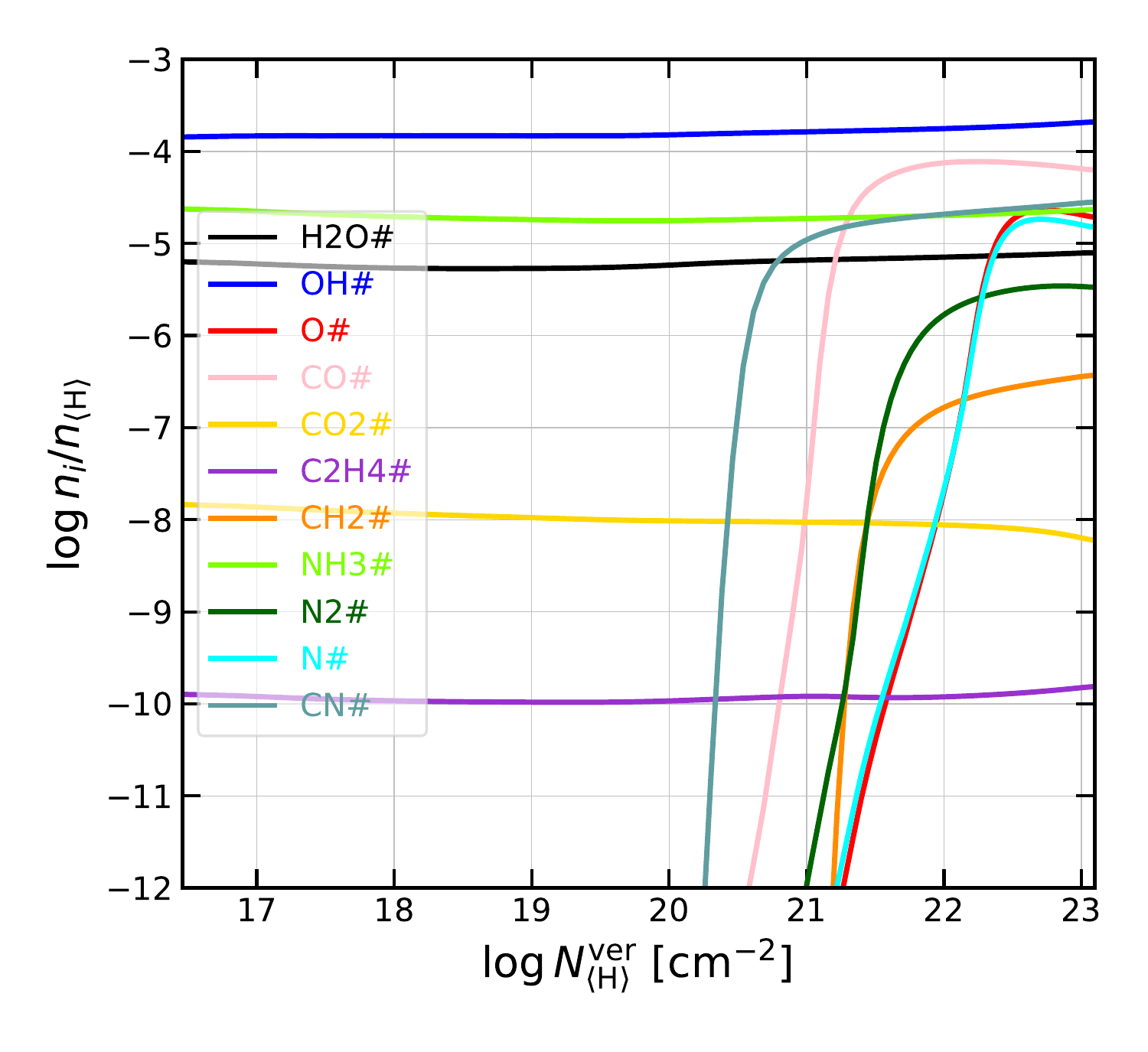} \\
  \end{tabular}}
  \caption{Vertical spreading of icy grains at $r\!=\!100\,$au for
    different mixing strengths $\alpha_{\rm mix}$. Results are shown
    as function of the vertical hydrogen nuclei column density
    $N_{\rm\langle H\rangle}$. The top row of plots shows the
    computed gas and dust temperatures, $T_{\rm g}$ and $T_{\rm
      d}$. The middle row of plots shows the bare grain surface area
    per H nucleus$\,=\!4\pi\langle a^2\rangle n_d/n_{\rm\langle
      H\rangle}$ after dust settling in blue, and the
    number of ice layers on each grain $N_{\rm layers}$ in red. The
    lower plots show the concentrations of selected ice species.}
  \label{fig:ice_ext}
\end{figure*}

At large radial distances $\sim$\,100\,au, the model with the highest
mixing strength $\alpha_{\rm mix}\!=\!10^{-2}$ exhibits some rather
peculiar results, see Fig.~\ref{fig:ice_ext}. The concentrations of
the thermally most stable ice species such as \ce{H2O}\#, \ce{NH3}\#,
\ce{CO2}\#, \ce{OH}\# and \ce{C2H4}\# (and some others not shown)
become about constant and remain high even at the very top of the
model, where they face the unattenuated stellar radiation field.  This
is in contrast to the less thermally stable ices such as \ce{CO}\# and
\ce{N2}\# (and others).

In models without mixing, the ices are generally not stable in the
optically thin disc regions because of UV photo-desorption. In the
model with strong mixing $\alpha_{\rm mix}\!=\!10^{-2}$, however, the
upward mixing rates seem to be large enough to replenish the ice
species faster than the UV desorption can destroy them. This is in
contrast to thermal desorption. Once the ices become thermally
unstable, due to the increasing dust temperature in higher layers,
they will sublimate very quickly and mixing cannot change that.  We
see this behaviour very clearly in the upper part of
Fig.~\ref{iceSED}, where the upper boundary of \ce{CO}\# does not
surpass the dust temperature $T_{\rm d}\!=\!25\,$K line, and the
\ce{H2O}\# upper boundary does not surpass the $T_{\rm d}\!=\!150\,$K
line, which corresponds to the desorption energies of these ices.

With an assumed stellar luminosity $L_\star\!=\!1\,L_\odot$ and
$f_{\rm UV}\!=\!0.01$, we obtain a UV flux of about $6\times10^{10}$ 
photons/cm$^2$/s between 912\AA\ and 2050\AA\ at $r\!=\!100\,$au 
in the optically thin part high above the disc, which is 
about a factor of $\chi\!=\!300$ larger than the standard Draine flux
in the ambient interstellar medium $F_{\rm Draine}\!=\!1.921\times
10^8\rm\ photons/cm^2/s$ \citep{Draine1996}. With the symbols
introduced in section~5 of \citet{Woitke2009a}, the UV
photo-desorption rate of an ice species is
\begin{equation}
  \frac{dn_{i\#}}{dt} = -\pi\langle a^2\rangle\,n_d\,Y_i\,\chi F_{\rm
    Draine}\times\left\{\begin{array}{cc}
  \displaystyle
  \frac{n_{i\#}}{n_{\rm act}}         & ,\ n_{\rm tot}^{\rm ice} < n_{\rm act}\\[3mm]
  \displaystyle
  \frac{n_{i\#}}{n_{\rm tot}^{\rm ice}} & ,\ n_{\rm tot}^{\rm ice} \geq n_{\rm act}
  \end{array}\right. \ ,
  \label{photo-desorption}
\end{equation}
where $n_d$ is the number density of dust particles, $\pi\langle
a^2\rangle n_d$ the total dust cross section per $\rm cm^3$, $Y_i$ the
UV photo-desorption yield, and $n_{\rm act}=4\pi\langle
a^2\rangle\,n_d\,2\, n_{\rm surf}$ is the total number of active
adsorption sites on all grains per $\rm cm^3$. Following
\citet{Aikawa1996}, we assume that the uppermost two mono ice layers
are active, meaning that frozen molecules can be hit directly by UV
and can escape freely from these layers, with a surface density of
$n_{\rm surf}\!=\!1.5\times10^{15}$ sites per $\rm cm^2$ of solid
surface.

The first case in Eq.\,(\ref{photo-desorption}) corresponds to a
situation where the grain surface is almost unoccupied and the ratio
$n_{i\#}/n_{\rm act}$ expresses the probability that a UV photon
absorbed by a grain hits a site occupied by ice species $i$. In the
second case, the surface is more than fully occupied and the ratio
$n_{i\#}/n_{\rm tot}^{\rm ice}$ expresses the probability to find ice
species $i$ in the ice mixture present, where
the total ice density is $n_{\rm tot}^{\rm ice}=\sum_i n_{i\#}$.
Hence, thick ice mantles can only be reduced layer by layer via
photo-desorption, i.e.\ at a constant rate (``zero-order
photo-desorption'').  Figure~\ref{fig:ice_ext} shows that, according
to the assumed dust size distribution and dust/gas ratio, the grains
in the midplane are covered by about $N_{\rm layers}=n_{\rm tot}^{\rm
  ice}/(4\pi\langle a^2\rangle n_d\,n_{\rm surf}) \approx 10^4$ ice
mono-layers.

The timescale to completely UV photo-desorb such grains in the
optically thin upper disc layers can be obtained by summing up
Eq.\,(\ref{photo-desorption}) over all ice
species $i$, applying the second case 
\begin{equation}
  \tau_{\rm UVdesorb} = \frac{n_{\rm tot}^{\rm ice}/\nH}
      {\frac{\pi\langle a^2\rangle n_d}{\nH}\,\bar{Y}\,\chi F_{\rm
          Draine}}
      \approx \big(35000\,{\rm yrs}\big)
      \left(\frac{300}{\chi}\right)
  \label{eq:tau_deorb}    
\end{equation}
where we have assumed a total initial ice concentration of $n_{\rm
  tot}^{\rm ice}/\nH\!\approx\!10^{-3.5}$, a mean photo-desorption
yield of $\bar{Y}\!\approx\!10^{-3}$ and a total dust surface per
H-atom of $4\pi\langle a^2\rangle
n_d/\nH\!\approx\!2\times10^{-23}\rm\,cm^2$ (compare
Fig.~\ref{fig:ice_ext}). We note that $\chi\!\approx\!300$ is what we
find in our model at $r\!=\!100\,$au.  $\chi\propto 1/r^2$
(i.e.\ $\tau_{\rm UVdesorb}\propto r^2$) is valid until the
interstellar UV photons become more relevant than the diluted stellar
UV, which happens at about 1700\,au in this case.

In comparison, the turbulent eddy-diffusion timescale is
\begin{equation}
  \tau_{\rm mix} = \frac{H_p}{\alpha_{\rm mix} c_s}
  \approx \big(18000\,{\rm yrs}\big)
  \left(\frac{r}{100\,{\rm au}}\right)^{1.4}
  \left(\frac{10^{-2}}{\alpha_{\rm mix}}\right)
  \label{eq:taumix}
\end{equation}
where we assumed a scale height of $H_p\!=\!({\rm 10\,au}) (r/{\rm
  100\,au})^{1.15}$, a temperature structure as $T=({\rm
  20\,K})(r/{\rm 100\,au})^{-0.5}$, a constant mean molecular
weight\footnote{To clarify, these assumptions about $\mu$ and $T(r)$
  are only used for this timescale discussion. In our proper disc
  models, we use the resulting mean molecular weight and gas
  temperature from the previous global iteration.}  of
$\mu\!=\!2.3\,$amu, and $c_s\!=\!\sqrt{kT/\mu}$.

\begin{figure*}
  \begin{tabular}{c|c}
    \hline
    &\\[-2.1ex]
    $\alpha_{\rm mix}\!=\!0$ &
    $\alpha_{\rm mix}\!=\!10^{-3}$ \\
    \hline
  \hspace*{-2mm}\begin{minipage}{91mm}
  \includegraphics[width=91mm,trim=160 230 70 35,clip]
                  {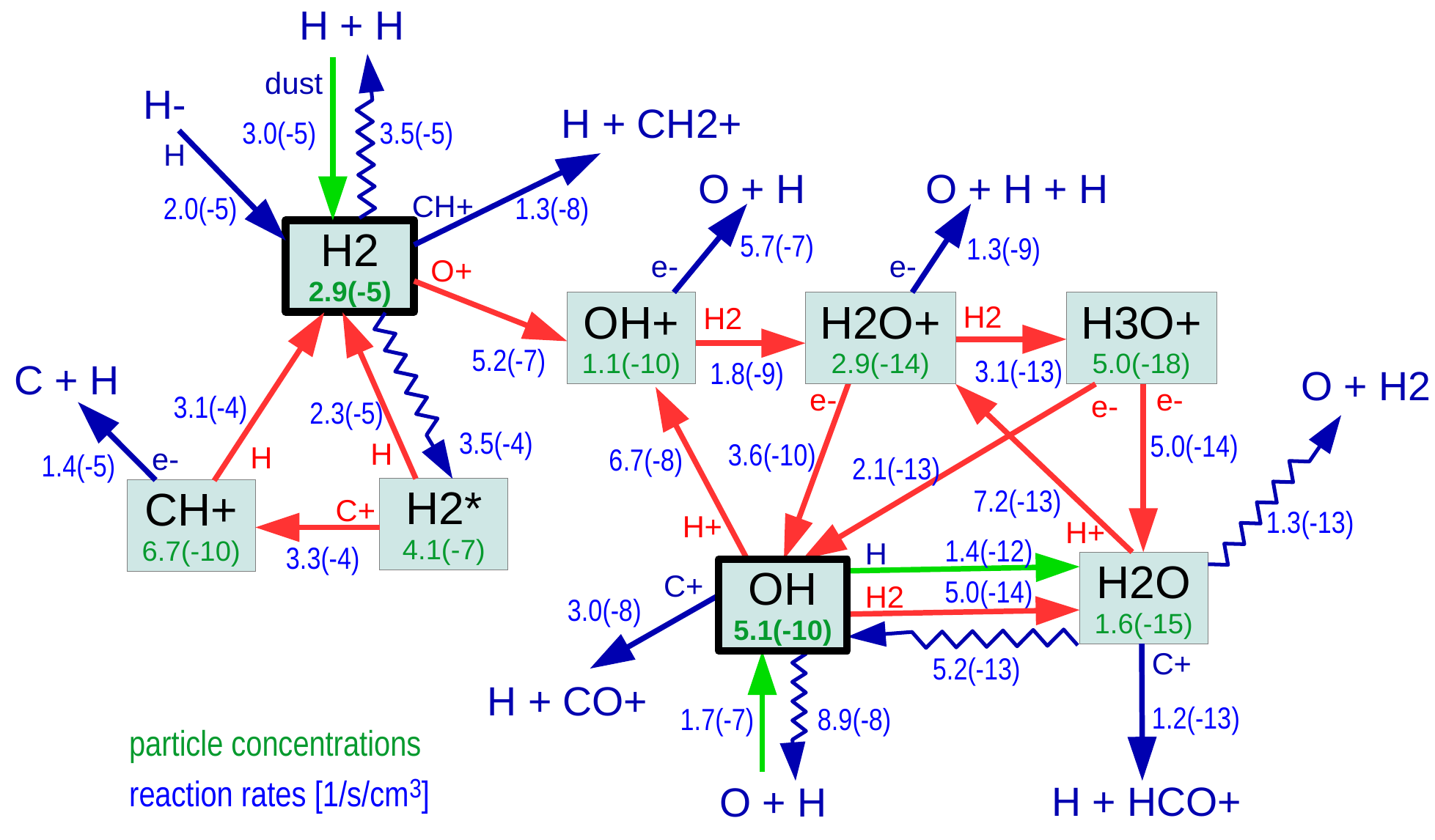}\end{minipage}\hspace*{-1mm} &
  \hspace*{-2mm}\begin{minipage}{91mm}
  \includegraphics[width=91mm,trim=160 230 70 35,clip]
                  {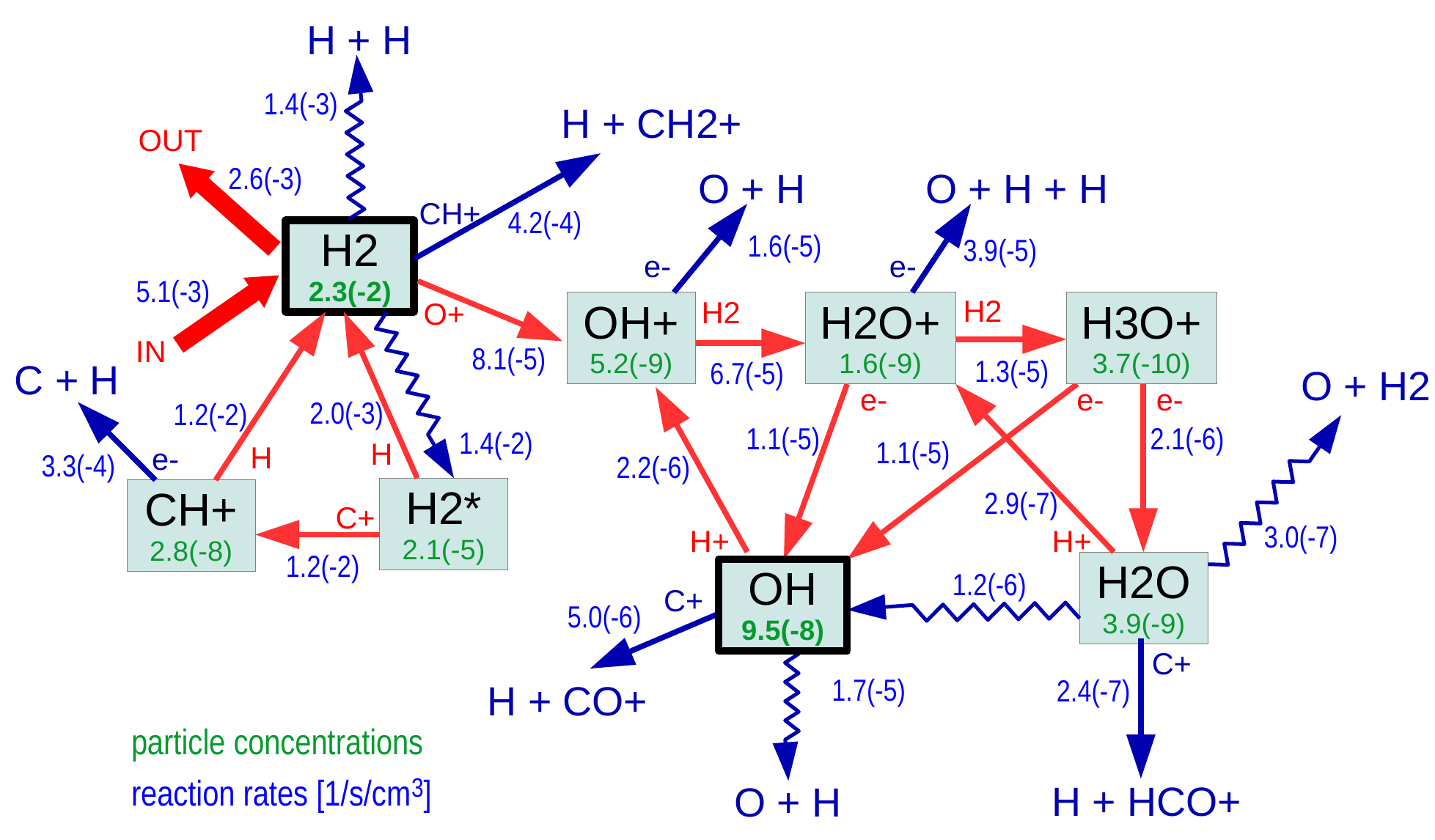}\end{minipage}\hspace*{-2mm}\\
  \hline
  \end{tabular}                
  \caption{Chemical pathways to create \ce{OH} and \ce{H2O} molecules
    without mixing (left) and with $\alpha_{\rm mix}\!=\!10^{-3}$
    (right) at one selected point in the disc at $r\!=\!10\,$au
    located in the upper disc layers, see text. The green numbers
    represent the particle concentrations and the blue numbers the
    rates $\rm[cm^{-3}s^{-1}]$ in the time-independent case.  The big
    red ``IN'' and ``OUT'' arrows on the right denote the in-mixing
    and out-mixing rates for \ce{H2}.  Wiggled arrows indicate
    photo-dissociation.  Light green arrows on the left mark the
    \ce{H2}-formation on grains and the radiative association
    reactions\ \ ${\rm O+H\to OH}+h\nu$\ \ and\ \ ${\rm OH+H\to
    H_2O}+h\nu$.\ \ $\rm H_2^\star$ denotes electronically excited molecular
    hydrogen. The figure does not follow up the production of
    \ce{CH2+}, \ce{CO+} and \ce{HCO+} marked in blue. The notation
    $a(-b)$ means $a\times10^{\,-b}$.}
  \label{fig:OHpath}
\end{figure*} 

The conclusion from this timescale comparison is that only for very
large $\alpha_{\rm mix}$ and large radii, the up-mixing of icy grains
can indeed be faster than UV desorption. The two timescales meet at
about 30\,au in our model with $\alpha_{\rm mix}\!=\!10^{-2}$, which is
in agreement with where we start to see icy grains in the uppermost
disc layers in our models. In all other cases, the icy grains stay
confined to the midplane due to UV photo-desorption.  However, the
vertical spread of the icy grains is distinctively wider as compared
to un-mixed models, and strongly depends on parameter $\alpha_{\rm
  mix}$. The entirely flat concentrations of the most stable ices
in the $\alpha_{\rm mix}\!=\!10^{-2}$ model are still somewhat
surprising. Possibly, the strong increase of the gas temperature, from
a few 10\,K to a few 100\,K, causing larger $c_s$ and hence shorter
mixing timescales, is responsible for this behaviour.

\subsubsection{Chemical pathways}
\label{sec:chempath}

In order to better understand the chemical processes induced by
mixing, let us examine one point in our disc model with $\alpha_{\rm
  mix}\!=\!10^{-3}$ at 10\,au, located at the maximum of the
concentration of \ce{OH}, that is the maximum of the red curve in
Fig.~\ref{fig:iteration2} (right part) at $N_{\rm\langle
  H\rangle}\!\approx\!1(+20)\rm\,cm^{-2}$. The notation
$a(+b)$ means $a\times10^{+b}$. Here we have a hydrogen nuclei density
of $\nH\!=\!2.4(+7)\rm\,cm^{-3}$. The gas and dust temperatures are
$T_g\!=\!347\,$K and $T_d\!=\!126\,$K, and the point is optically thin
with radial and vertical visual extinctions $A_V^{\rm rad}\!=\!0.024$
and $A_V^{\rm ver}\!=\!6.3(-4)$, respectively.  The most abundant
(``available'') species are \ce{H} $[9.5(-1)]$, \ce{H2} $[2.3(-2)]$,
\ce{O} $[3.0(-4)]$, \ce{e-} $[1.5(-4)]$, \ce{C+} $[1.3(-4)]$, \ce{N}
$[7.9(-5)]$, and \ce{H+} $[2.1(-5)]$, where the numbers in squared
brackets are the concentrations $n_i/\nH$.

Figure~\ref{fig:OHpath} shows how the up-mixing of \ce{H2} fuels the
ion-neutral chemistry in the upper disc regions, resulting in an
enhanced production of \ce{OH} and \ce{H2O}. The left side of the
figure shows the rates and concentrations in the case without mixing,
where OH is mainly produced via the radiative association reaction
${\rm O + H \to OH} + h\nu$, water is mainly produced by the
follow-up radiative association reaction ${\rm OH + H \to H_2O} + h\nu$,
and \ce{H2} is produced partly by the formation on grains and partly
via $\rm H + H^- \to H_2 + e^-$.

On the right side of Fig.~\ref{fig:OHpath}, which shows the results 
at the same point for the case with mixing $\alpha_{\rm
  mix}\!=\!10^{-3}$, the \ce{H2}-concentration is enhanced by a
factor of about 1000. Here, the starting point for the production of
\ce{OH} and \ce{H2O} is the reaction
\begin{equation}
  \rm H_2 ~+~ O^+ ~\longrightarrow~ OH^+ ~+~ H   \ ,
  \label{H2_O+}
\end{equation}
where the \ce{O+} is produced by charge exchange with \ce{H+}, which
results from secondary X-ray reactions with super-thermal electrons
$\rm H + e^- \to H^+ + e^- + e^-$. Although reaction~(\ref{H2_O+}) is
only the forth most important destruction reaction of \ce{H2} (after
out-mixing, photo-dissociation and reaction with \ce{CH+}), it
triggers the production of \ce{OH} and \ce{H2O} via the formation of
\ce{OH+}.  Quick H-addition reactions with \ce{H2} subsequently
produce \ce{H2O+} and \ce{H3O+}, before recombinations with free
electrons create the neutral molecules \ce{OH} and \ce{H2O}. This
\ce{H2O} formation path is nomally initiated by reactions of O with
\ce{H3+}, for example $\rm O + H_3^+ \to OH^+ + H_2$, but here we find
reaction~(\ref{H2_O+}) to be more efficient in producing \ce{OH+},
because the mixing leads to a coexistence of \ce{H2} and \ce{O+}.
Since \ce{H2} is required several times along these formation
pathways, even a small increase of the \ce{H2} concentration, caused
by up-mixing, can lead to a much higher production of \ce{OH} and
\ce{H2O} molecules. The resulting \ce{OH} and \ce{H2O} concentrations
are higher by about 2.5 and 6 orders of magnitude, respectively, at
this point, when mixing is taken into account with $\alpha_{\rm
  mix}\!=\!10^{-3}$. The reaction cycles are completed by
photo-dissociation and dissociative recombination reactions, which
again destroy the molecules and replenish the atoms.

Noteworthy, \ce{H2} is the only molecule that is strongly affected by
mixing directly in Fig.~\ref{fig:OHpath}. All other molecules shown in
Fig.~\ref{fig:OHpath} have the same reaction coefficients for the MI
and MO quasi reactions, yet their low concentrations render these
rates insignificant. The only other species directly affected by
mixing are those atoms and ions listed as ``abundant'' in the
beginning of this paragraph.  This finding can be explained by
considering the chemical formation/destruction timescales (particle
densities over rates). For low-abundance molecules, particle densities
are low but rates are similar, i.e.\ the chemical timescales are
short.  Only for the most abundant molecules the chemical timescale
can exceed the mixing timescale. The other species with low abundances
and short chemical timescales tend to rather adjust their
concentrations quickly to these changed circumstances.

\begin{figure*}
  \begin{tabular}{c|c}
    \hline
    &\\[-2.1ex]
    $\alpha_{\rm mix}\!=\!0$ &
    $\alpha_{\rm mix}\!=\!10^{-3}$ \\
    \hline
  \hspace*{-2mm}\begin{minipage}{94mm}
  \includegraphics[width=94mm,trim=40 112 35 0,clip]
                  {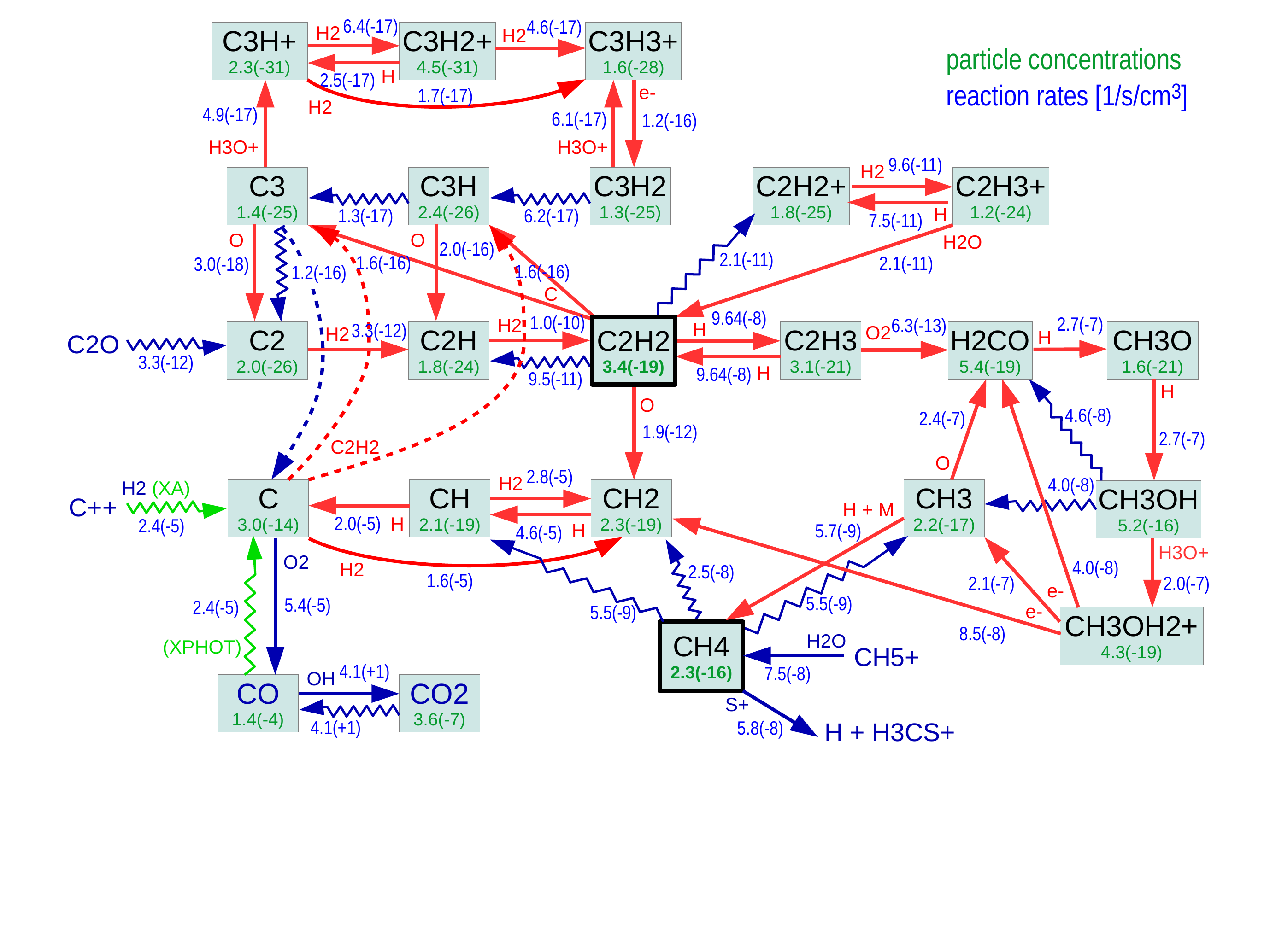}\end{minipage}\hspace*{-1mm} &
  \hspace*{-5mm}\begin{minipage}{92mm}
  \includegraphics[width=92mm,trim=50 112 35 5,clip]
                  {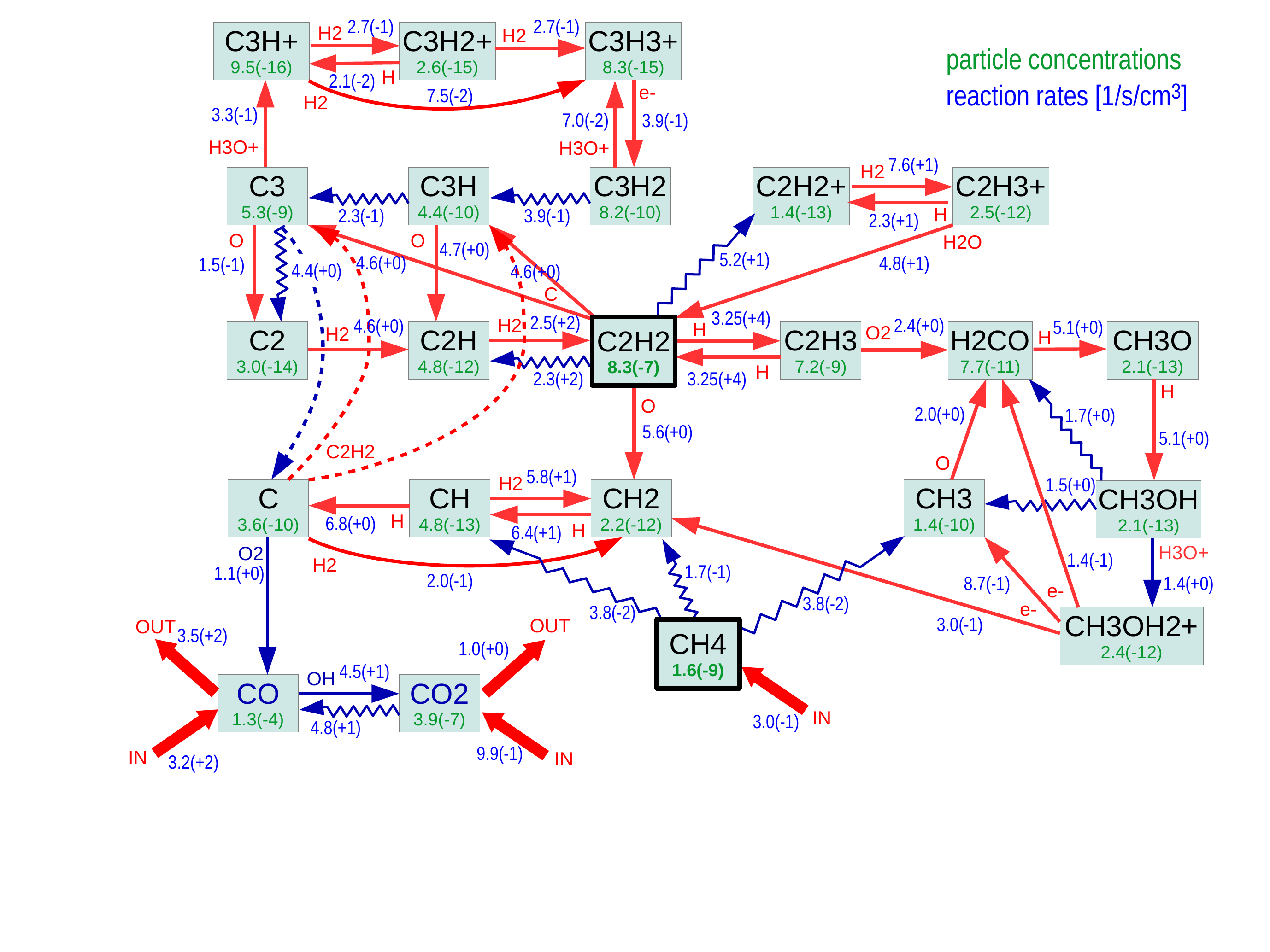}\end{minipage}\hspace*{-2mm}\\
  \hline
  \end{tabular}                
  \caption{Chemical pathways to create \ce{C2H2} and other
    hydro-carbon molecules without mixing (left) and with $\alpha_{\rm
      mix}\!=\!10^{-3}$ (right) at one selected point in the disc at
    $r\!=\!0.15\,$au, see text. Note the vastly different
    concentrations of all hydro-carbon molecules. As in
    Fig.~\ref{fig:OHpath}, the green numbers represent the particle
    concentrations and the blue numbers the rates $\rm[cm^{-3}s^{-1}]$
    in the time-independent case.  The big red ``IN'' and ``OUT''
    arrows denote the in-mixing and out-mixing rates.  Wiggled arrows
    indicate photo-dissociation. The light green wiggled arrows are
    X-ray induced processes.  Three reactions appear twice with an
    added dashed line, when a reaction is important for two reactants
    or products ($\rm C_2H_2 + C \to C_3 + H_2$, $\rm C_2H_2 + C
    \to C_3H + H$ and ${\rm C_3} + h\nu \to \rm C_2 + C$).  On the left
    side, the reaction paths for \ce{C2O}, \ce{C++}, \ce{CH5+} and
    \ce{H3CS+} are not followed up.}
  \label{fig:C2H2path}
\end{figure*}

Next, we have a look at the hydro-carbon chemistry. We consider a
point in our $\alpha_{\rm mix}\!=\!10^{-3}$ disc model at
$r\!=\!0.15\,$au, at a height where the \ce{C2H2} concentration is
steeply rising towards a plateau in the midplane, that is at a vertical
hydrogen column density of about $N_{\rm\langle
  H\rangle}\!\approx\!1.0(+23)\rm\,cm^{-2}$ on the left plot in
Fig.~\ref{fig:chem_results}.  Points like these are flagged by {\sc
  ProDiMo} as the main emission region for the \ce{C2H2} 13.7\,$\mu$m
emission feature observable with Spitzer/IRS and JWST/MIRI, see
Fig.~\ref{fig:IRspec}.  Here, we have a hydrogen nuclei density of
$\nH\!=\!1.0(+13)\rm\,cm^{-3}$, the gas and dust temperatures are
$T_g\!=\!353\,$K and $T_d\!=\!354\,$K, and the point is optically
thick in the radial direction, $A_V^{\rm rad}\!=\!194$, but semi
optically thin in the vertical direction, $A_V^{\rm ver}\!=\!3.1$,
still resulting in a total UV strength of $\chi\!=\!20000$. The most
abundant species are \ce{H2} $[0.5]$, \ce{H}
$[5.6(-3)]$, \ce{H2O} $[1.7(-4)]$, \ce{CO} $[1.3(-4)]$, and \ce{N2}
$[3.6(-5)]$, no ices.

Figure~\ref{fig:C2H2path} shows how the up-mixing of \ce{CH4} followed
by photo-dissociation leads to much higher concentrations of all
hydro-carbon molecules. The hydro-carbon chemistry forms an almost
self-contained reaction sub-system that is only weakly connected to
other chemicals. Comparing the two depicted cases (with and without
mixing), we see that the same neutral-neutral, ion-neutral and
photo-dissociation reactions connect the different hydro-carbon
species in similar ways, forming sustained reaction cycles. The
various reactions with \ce{O} or \ce{O2} in these cycles form C-O
bonds that, once formed, can hardly be destroyed again. We will denote
these oxidisation reactions as ``burning'' in the following.  On the
left side of Fig.~\ref{fig:C2H2path} (without mixing) these reactions
consume almost all hydro-carbon molecules, until an equilibrium with
neutral carbon atoms is established, on a very low level $n_{\rm
  C}/\nH\!=\!3(-14)$, where further burning of the H--C molecules is
balanced by X-ray destruction of CO and X-ray induced charge exchange
(XA) with \ce{C++}.

On the right side of Fig.~\ref{fig:C2H2path} (with mixing), methane
molecules (\ce{CH4}) that formed further down in the UV-protected
disc layers are mixed up into the semi optically thin region considered
here, where they are quickly photo-dissociated as
\begin{equation}
  {\rm CH_4} ~+~ h\nu ~\longrightarrow
  ~\left\{\begin{array}{lc}
  \rm CH_3 ~+~ H & (15.5\%) \\
  \rm CH_2 ~+~ H_2 & (69.0\%) \\
  \rm CH ~+~ H_2 ~+~ H & (15.5\%)
  \end{array}\right.
\end{equation}
with branching ratios from the UMIST\,2012 database
\citep{McElroy2013}.  This introduces rates of the order of $\rm
10^{-1}\,cm^{-3}\,s^{-1}$ to the hydro-carbon system.  The system
reacts to this input by increasing all concentrations of hydro-carbon molecules
by many orders of magnitude, from about 4 decades for \ce{C} to 15 decades
for \ce{C3H}.  The concentration of acetylene (\ce{C2H2}), which results to
be the most abundant hydro-carbon molecule in this case, increases
from $3.4(-19)$ to $8.3(-7)$, creating a peculiar chemical situation where
carbon can be found in opposite redox states (\ce{C2H2} and \ce{CO2})
in similarly high concentrations.

Again, only the abundant molecules shown with big red arrows in
Fig.~\ref{fig:C2H2path} (\ce{CO}, \ce{CO2} and \ce{CH4}) are
significantly affected by mixing directly. All other molecules just
adjust to those circumstances quickly.  For example, \ce{C2H2} has a
chemical relaxation timescale of about $n_{\rm C2H2}/{\rm
  rate}\approx\rm 8(-7)\times
1(+13)\,cm^{-3}/(3(+4)\,cm^{-3}s^{-1})=270\,$s at this point, which is
much shorter than the mixing timescale $\approx\rm\!2\,yrs$
(Eq.\,\ref{eq:tau_deorb}).

In summary, the \ce{CH4} molecules that are mixed up from deeper
layers by eddy-diffusion are quickly photo-dissociated once they reach
the borderline optically thin layers.  The products of these
photo-processes react further to create a rich hydro-carbon chemistry
in an oxygen-rich situation, where the various hydro-carbon molecules
are ultimately burnt on timescales of a few years at $r\!=\!0.15\,$au,
forming C-O bonds. We actually see in Fig.~\ref{fig:C2H2path} that
there is an over-production of \ce{CO} and \ce{CO2} that is mixed out.

\section{Impact on continuum and line observations}
\label{sec:impact}

In this section, we demonstrate how turbulent mixing can affect
continuum and line observations. The impact is generally quite
substantial, and we have selected to demonstrate three of the mayor
effects in this paper. We do not intend to make any predictions for
particular targets in this paper.

\subsection{Impact on spectral ice features}

\begin{sidewaysfigure*}
  \hspace*{-3mm}
  \resizebox{253mm}{!}{\begin{tabular}{ccccc}
  \hspace*{1mm} {\small $\alpha_{\rm mix}=0$} &
  \hspace*{-1mm} {\small $\alpha_{\rm mix}=10^{-5}$} &
  \hspace*{-1mm} {\small $\alpha_{\rm mix}=10^{-4}$} &
  \hspace*{-1mm} {\small $\alpha_{\rm mix}=10^{-3}$} &
  \hspace*{-1mm} {\small $\alpha_{\rm mix}=10^{-2}$}\\
  \hspace*{-1.5mm} \includegraphics[height=40mm,trim= 0 0 10 0,clip]
          {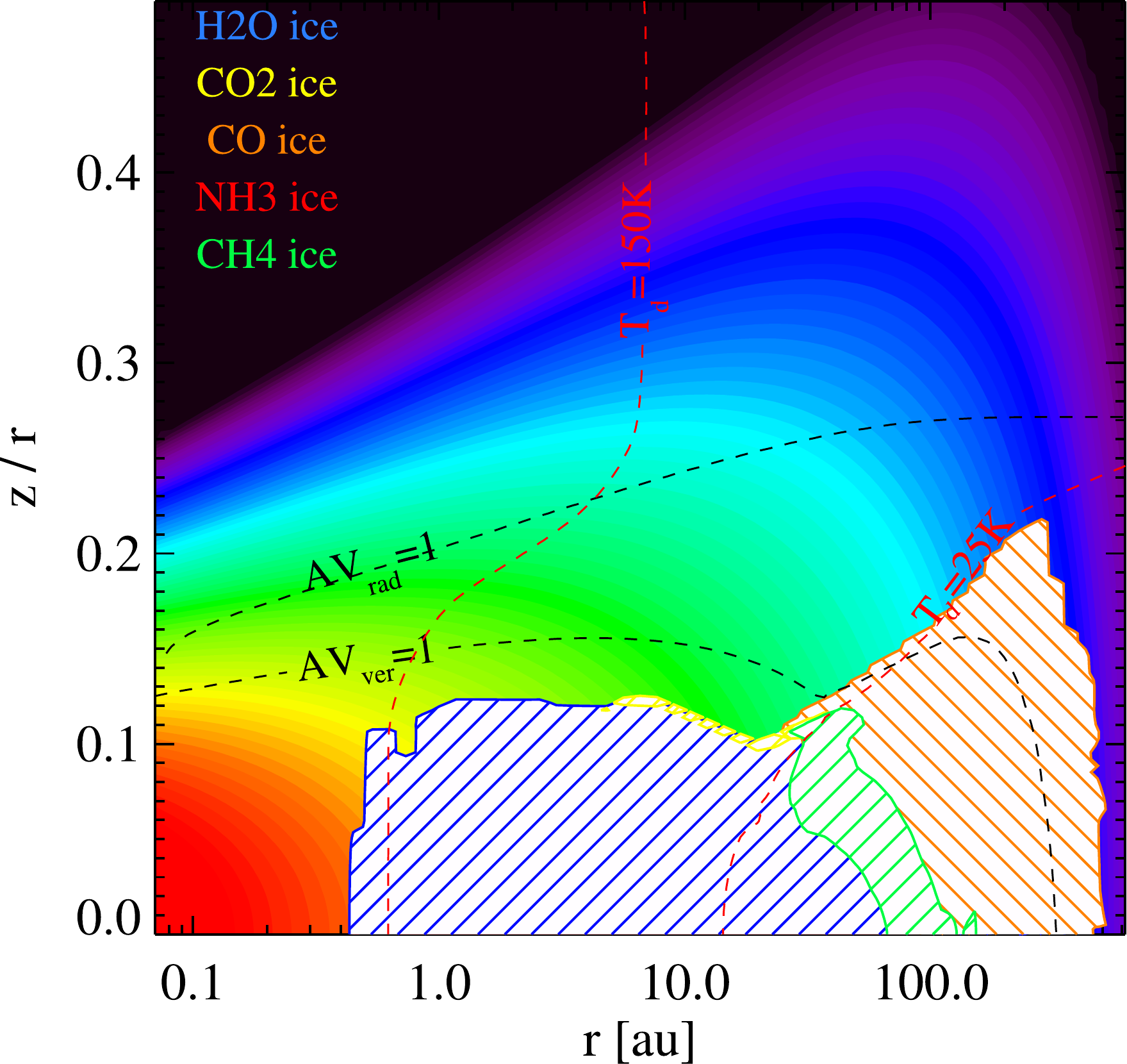} &
  \hspace*{-5.0mm} \includegraphics[height=40mm,trim=68 0 10 0,clip]
          {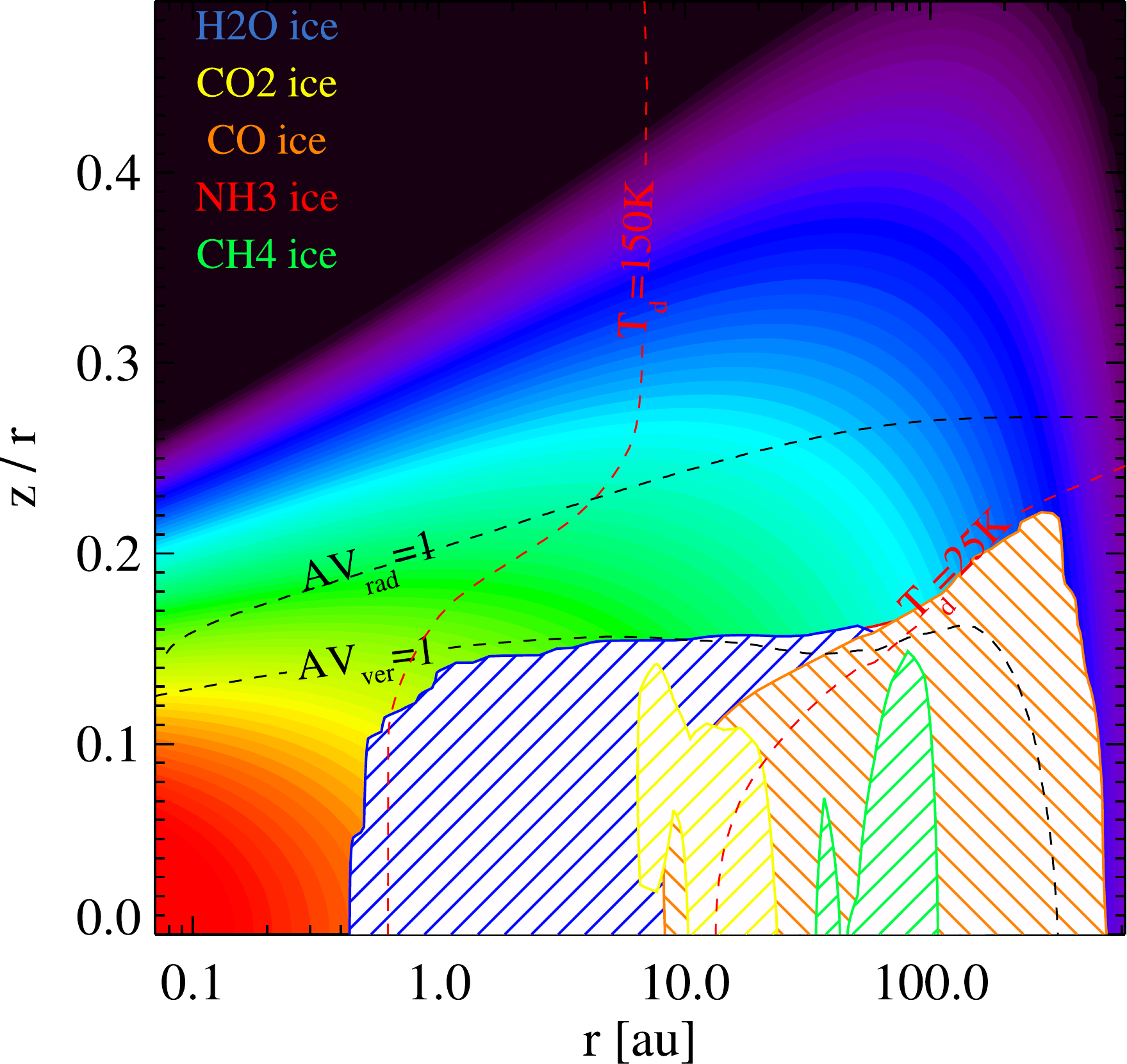} &
  \hspace*{-5.0mm} \includegraphics[height=40mm,trim=68 0 10 0,clip]
          {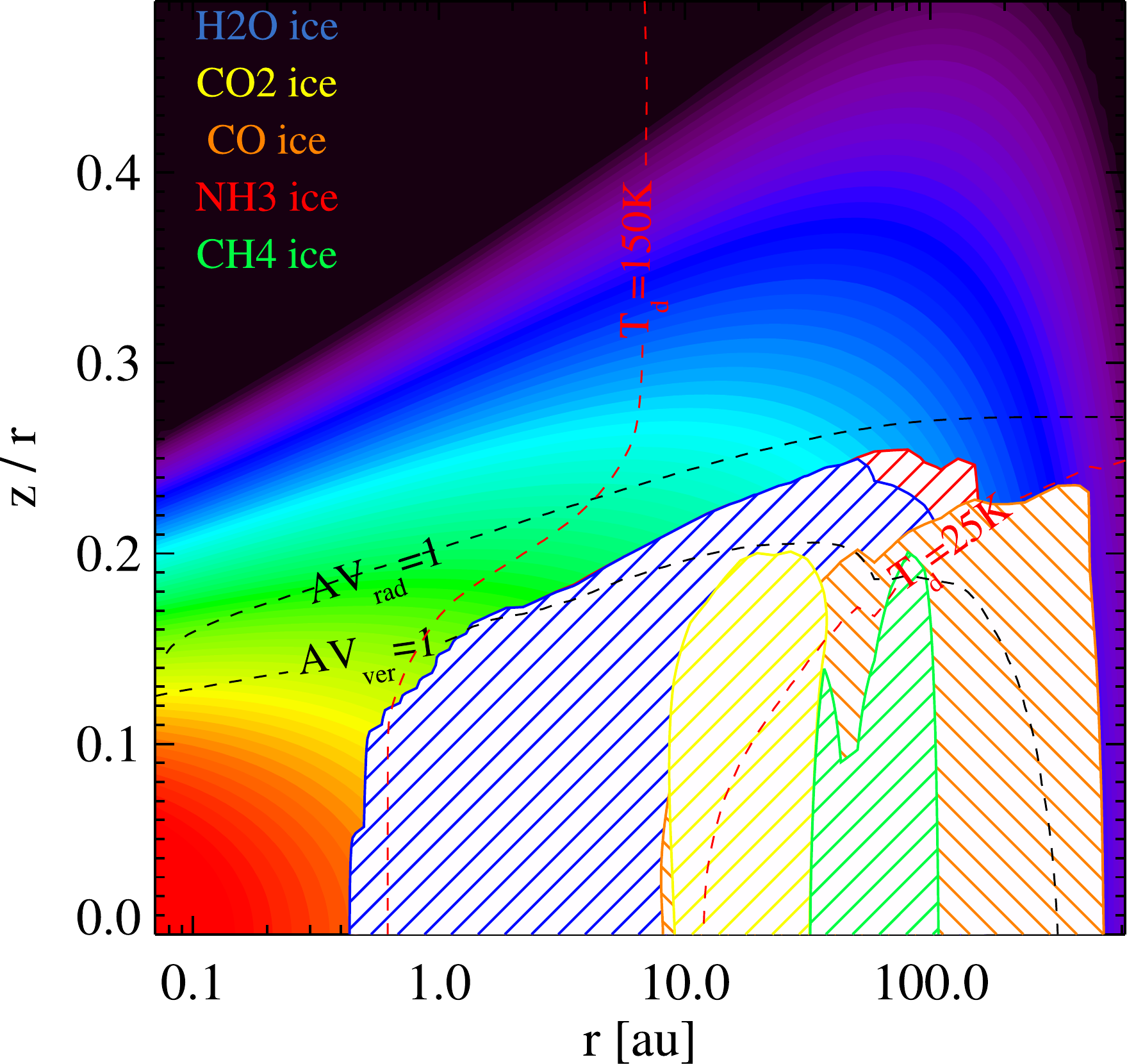} &
  \hspace*{-5.0mm} \includegraphics[height=40mm,trim=68 0 10 0,clip]
          {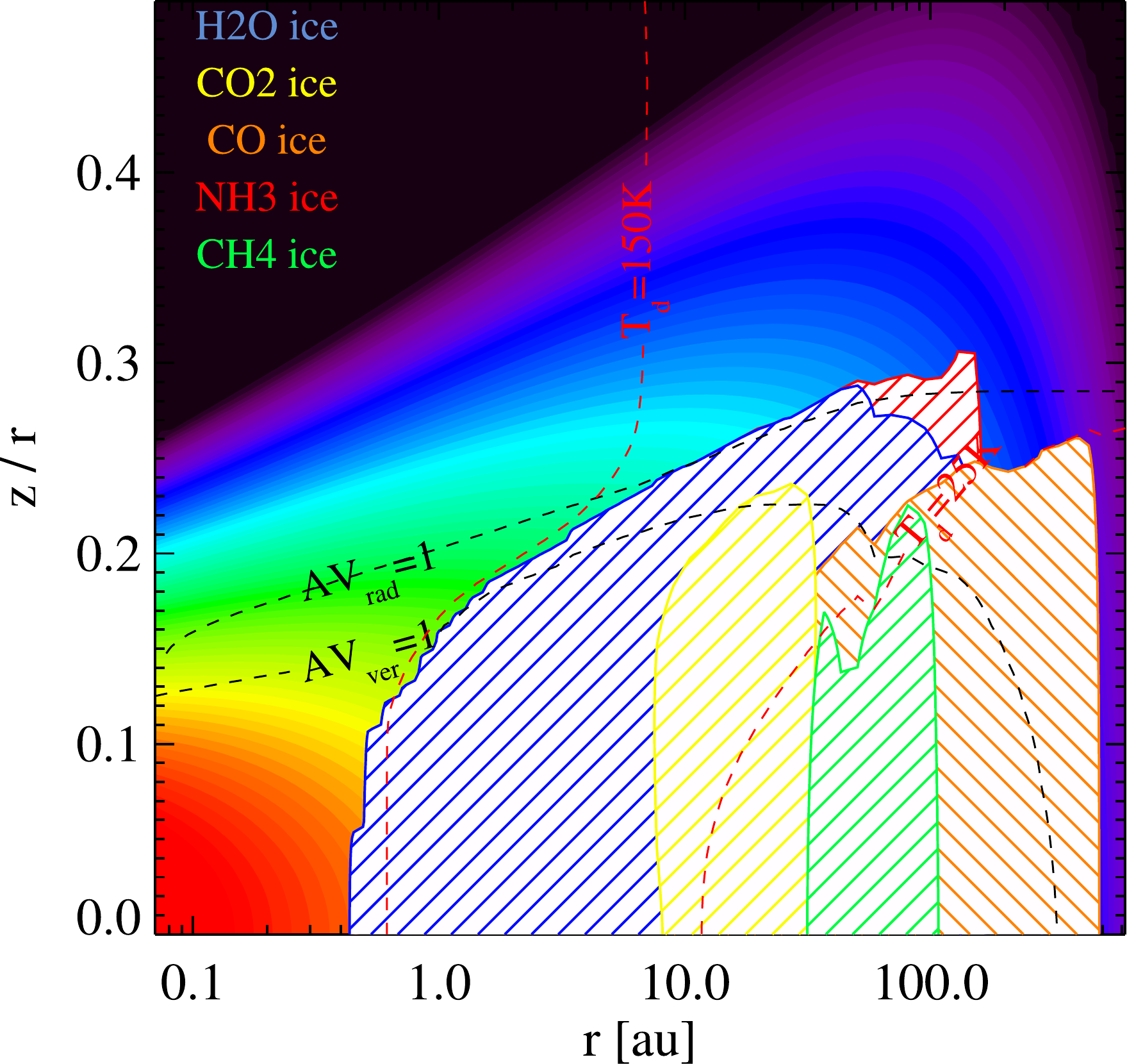} &
  \hspace*{-5.0mm} \includegraphics[height=40mm,trim=68 0 10 0,clip]
          {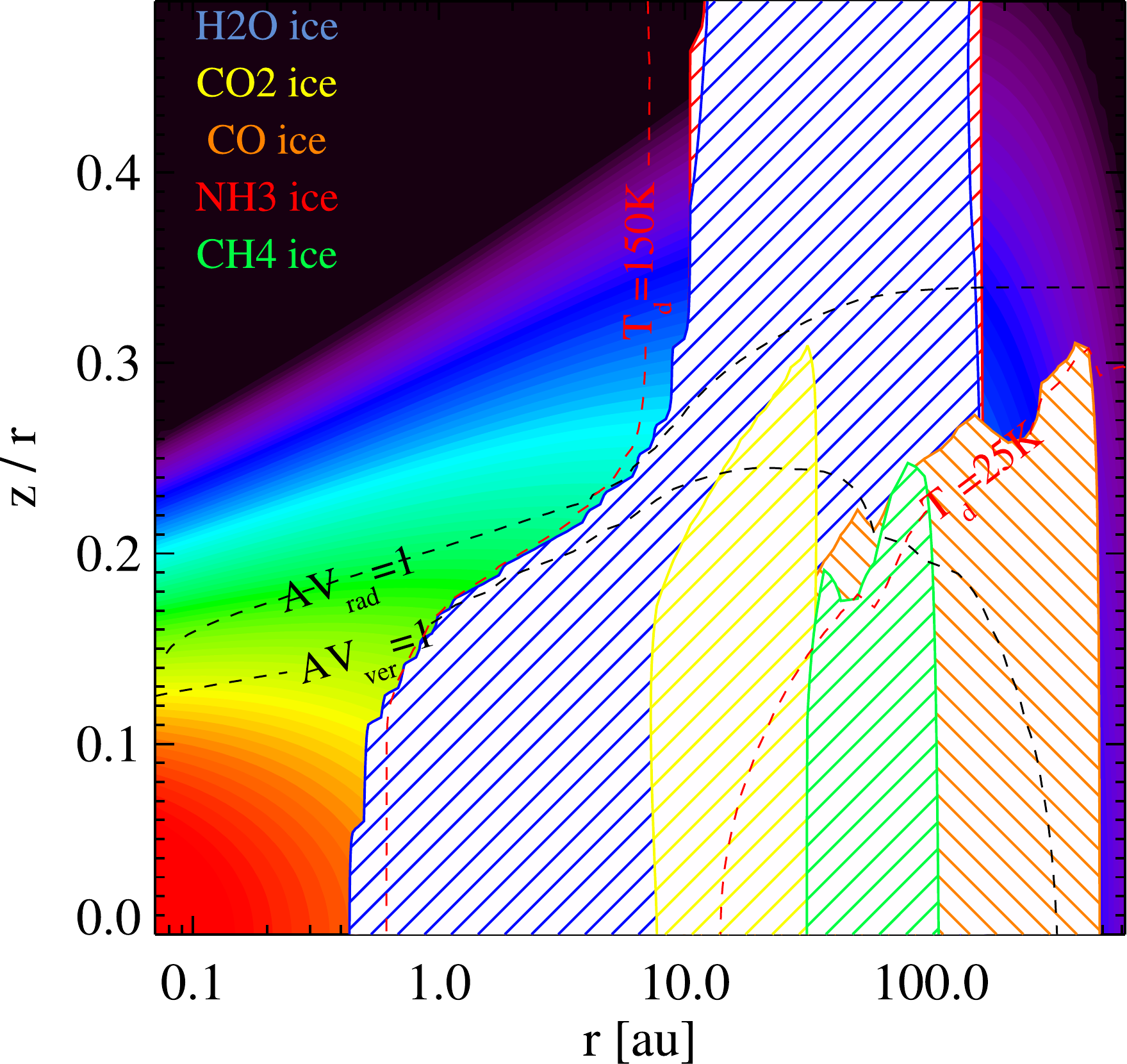} \\*[0mm]
  \hspace*{-2mm} \includegraphics[height=43mm,trim= 0 0 0 0,clip]{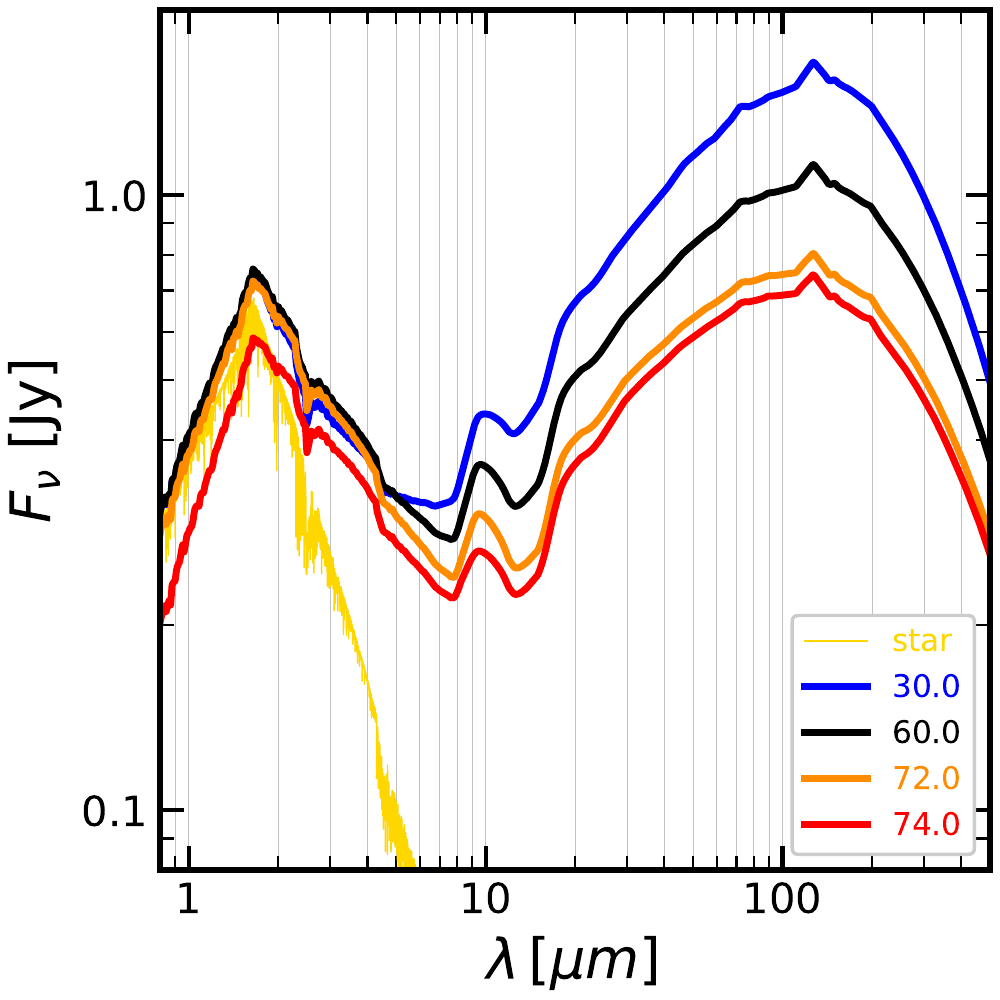} &
  \hspace*{-4.5mm} \includegraphics[height=43mm,trim=45 0 0 0,clip]{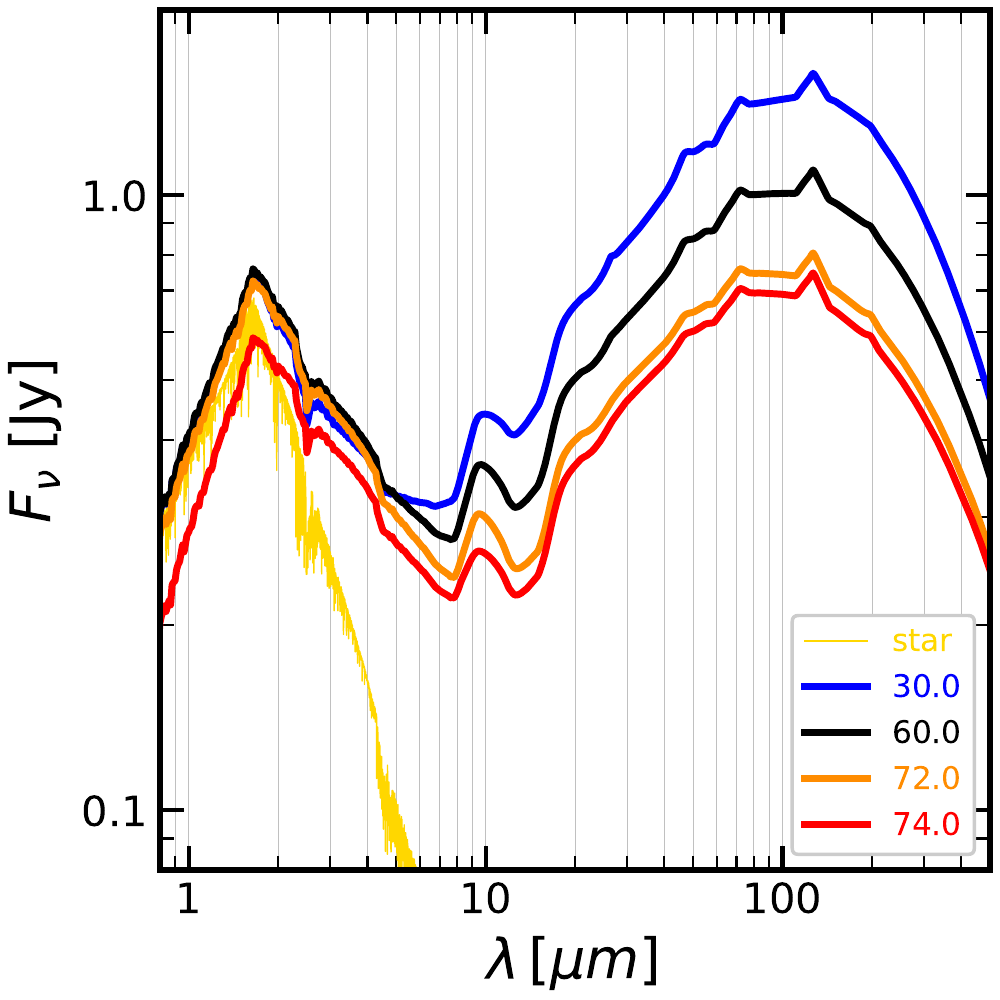} &
  \hspace*{-4.5mm} \includegraphics[height=43mm,trim=45 0 0 0,clip]{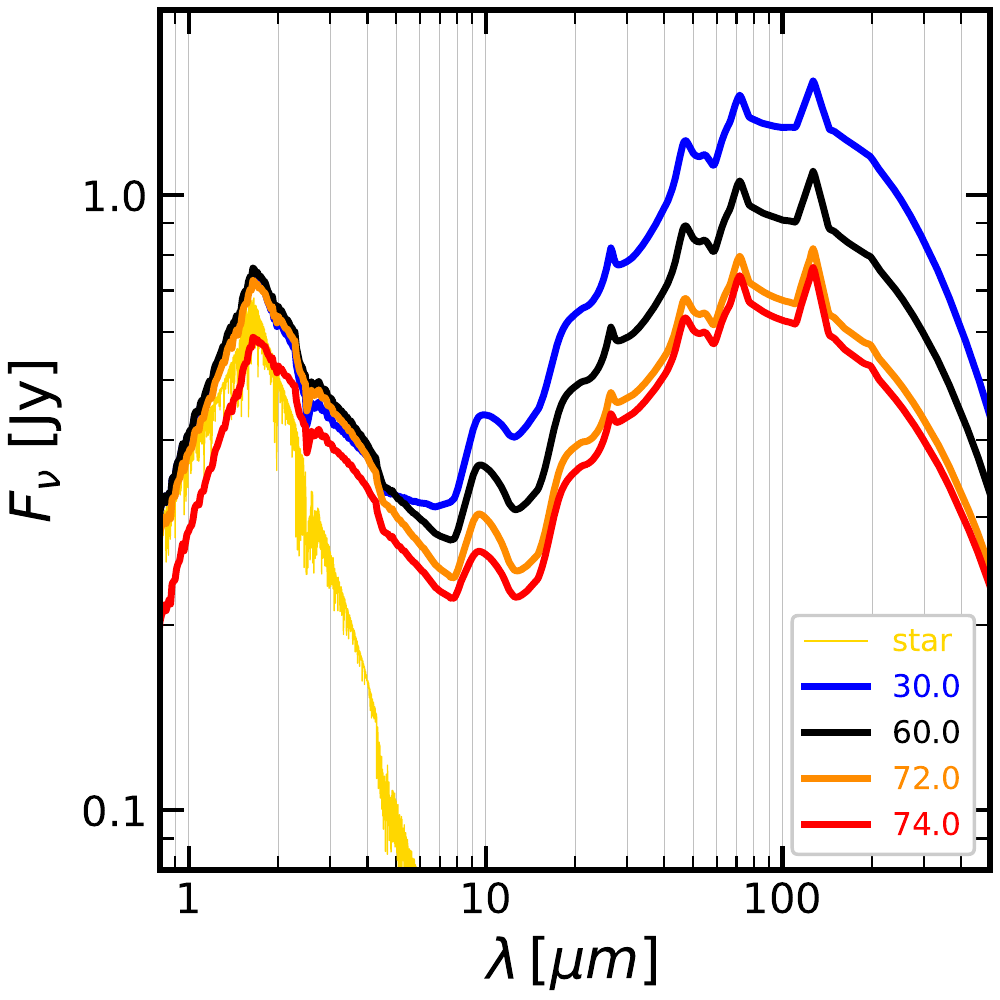} &
  \hspace*{-4.5mm} \includegraphics[height=43mm,trim=45 0 0 0,clip]{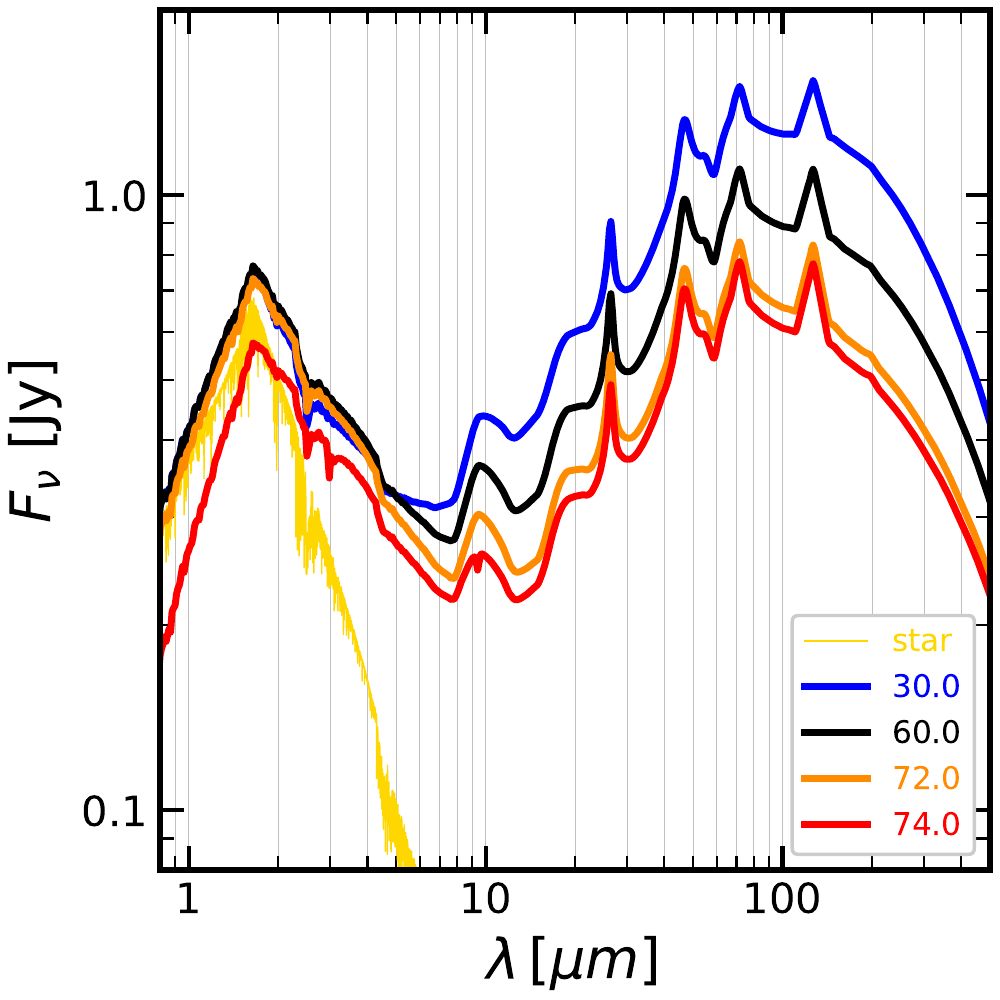} &
  \hspace*{-4.5mm} \includegraphics[height=43mm,trim=45 0 0 0,clip]{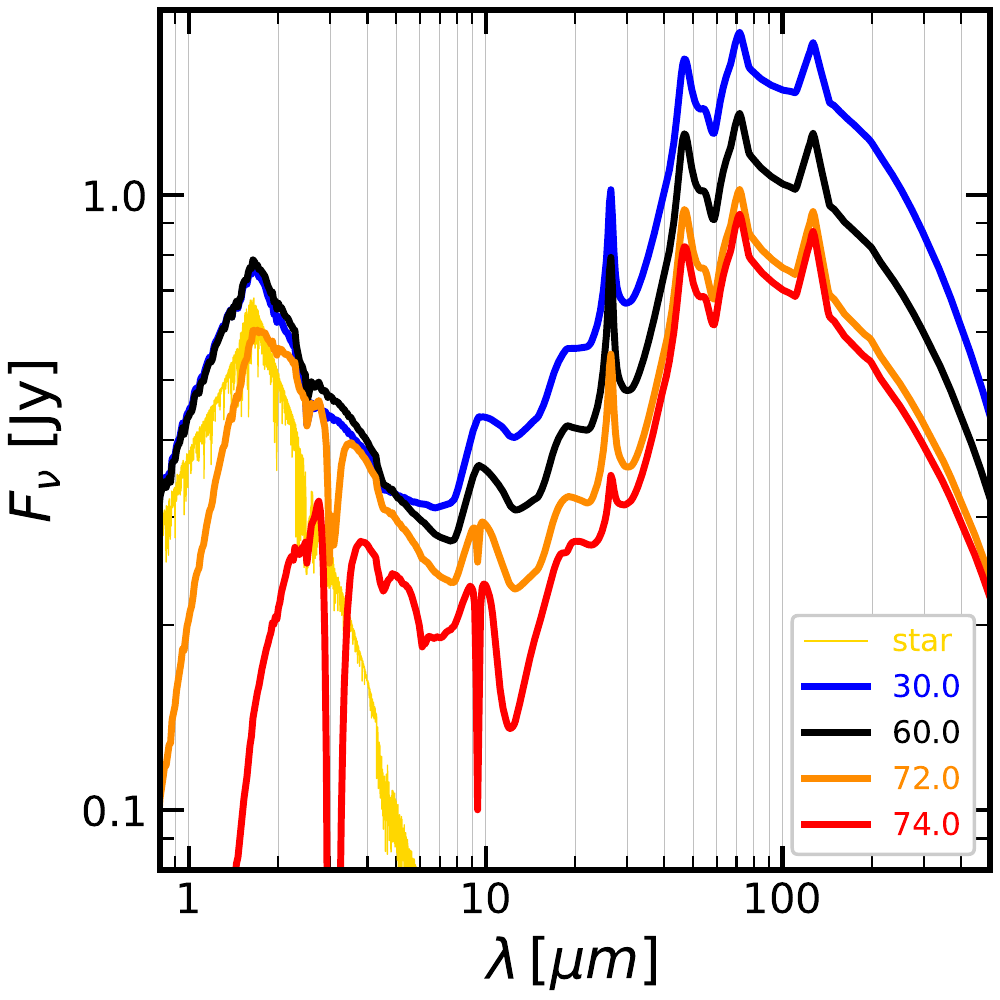} \\*[-1mm]
  \end{tabular}}
  \caption{The {\bf upper row of plots} shows the location of the ices, in
    a series of models with increasing mixing parameter
    $\alpha_{\rm mix}$.  The icy regions are hashed when the
    respective ice concentration, for example $n_{\rm
      H2O\#}/n_{\langle\rm H\rangle}$ for water ice, is larger than
    $10^{-5.5}$. The background colours show the
    hydrogen nuclei density $n_{\rm\langle H\rangle}$ from
    $10^{\,2}\rm\,cm^{-3}$ (black) to $10^{\,16}\rm\,cm^{-3}$
    (red). The black dashed contour lines show the radial visual
    extinction $A_{V,\rm rad}\!=\!1$ and vertical visual extinction
    $A_{V,\rm ver}\!=\!1$, and the red dashed lines show two dust temperature
    contours, 25\,K and 150\,K.  The {\bf lower row of plots} shows the
    respective spectral energy distributions (SEDs) for selected
    inclination angles at a distance of 140\,pc. The yellow lines
    indicates the adopted photospheric spectrum of the naked star. We
    note that the absorption feature around 2.5\,$\mu$m is due to
    molecular absorption in the stellar photosphere, whereas all
    other narrow absorption and emission features are due to ices,
    beside the broad 10$\,\mu$m\,/\,20$\,\mu$m silicate emission features.}
  \label{iceSED}
\end{sidewaysfigure*}

In our models, the up-mixing of icy grains leads to a substantial
increase of the ice concentrations in the observable disc surface, see
details in Sect.~\ref{sec:icemix}.  Figure~\ref{iceSED} shows the
spatial distribution of five selected ice species (\ce{H2O\,\#},
\ce{NH3\,\#}, \ce{CO2\,\#}, \ce{CO\,\#}, and \ce{CH4\,\#}) in the
entire disc.  In the un-mixed case, most ices only populate the
midplane regions below $A_{\rm V,ver}\!\approx\!10$, due to active UV
photo-desorption in the upper disc regions. The only counter-example
is CO\,\# in the outer regions $\ga\!100\,$au, but this region is too
cold to cause any significant far-IR ice emission features.  Since the
spectral energy distribution (SED) is mostly produced by thermal
emission and scattering of the bare (i.e.~ice-less) dust grains above
$A_{\rm V,ver}\!=\!10$, the resulting SED is almost entirely free of
ice features, also at grazing inclination angle ($\approx\!72^{\rm o}$
in this model), which tests the $A_{\rm V,rad}\!\approx\!1$ regions in
absorption.  The ice just sits too deep to cause any notable changes
of the integrated spectral flux.  One would need to observationally
search for the ice features by analysing image data, using the
intensities coming from the dark lane, see discussion in
\citet{Arabhavi2022}.

However, even a weak turbulent mixing with $\alpha_{\rm
  mix}\!=\!10^{-5}$ lifts some of the icy grains upwards, almost
reaching the $A_{\rm V,ver}\!=\!1$ surface, and some weak ice emission
features between 20\,$\mu$m and 200\,$\mu$m start to appear in the
spectrum, which become more clearly observable in the $\alpha_{\rm
  mix}\!=\!10^{-4}$ model. We also see that the $A_{\rm V,ver}\!=\!1$
contour moves upwards with increasing $\alpha_{\rm mix}$ due to the
opacity of the icy grains. So, scattered starlight can now directly
interact with the icy grains.

At $\alpha_{\rm mix}\!=\!10^{-3}$, the ice emission features become
even stronger, and the icy grains reach the $A_{\rm V,rad}\!=\!1$
line, i.e.\ direct starlight can now interact with icy grains.
The first three ice absorption features appear when the disc is
observed under an inclination angle of $\approx\!74^{\rm o}$.

Eventually, the extreme model with $\alpha_{\rm mix}\!=\!10^{-2}$
shows a population of \ce{H2O\,\#} and \ce{NH3\,\#} up to the very top
of the model, accompanied by another considerable lift of the $A_{\rm
  V,ver}\!=\!1$ and $A_{\rm V,rad}\!=\!1$ contours. We note that the
vertical extension of the snowy grains coincides with the dust
temperature $T_{\rm dust}\!=\!150\,$K contour, which means that the
snow extends right up to the water ice sublimation temperature in this
model, whereas the UV photo-desorption becomes an irrelevant process,
see explanations in Sect.~\ref{sec:icemix}. The same is true for the
vertical extension of CO\,\# at larger radial distances, which is
limited by the $T_{\rm dust}\!=\!25\,$K contour.  The SED is full of
strong ice emission features, and clear absorption features for
inclination angles $\ga\!72^{\rm o}$. We note that the strengths of
the ice features depends on how we assume the ices to be distributed
onto the pre-existing bare grains, and our assumption of a layer of
equal thickness on top of each refractory grain creates the strongest
ice opacities, see discussion in \citet{Arabhavi2022}.
 

\subsection{Impact on mid-IR gas emission line spectra}


\begin{figure}
  \centering
  \vspace*{1mm}
  \includegraphics[width=80mm,trim= 4 43 5 30,clip]
                  {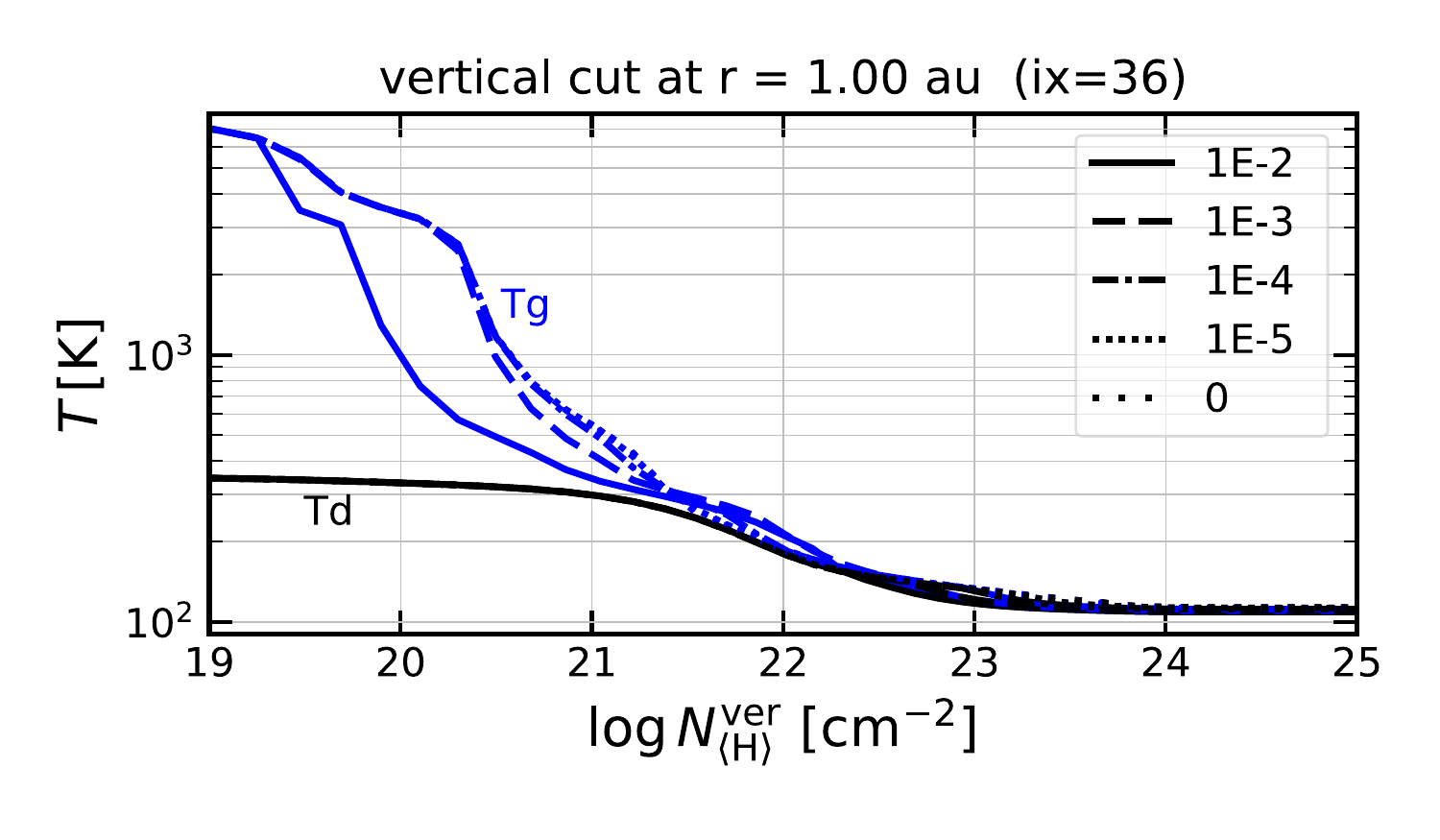}\\*[-1mm]
  \includegraphics[width=80mm,height=54mm,trim=10 43 5 30,clip]
                  {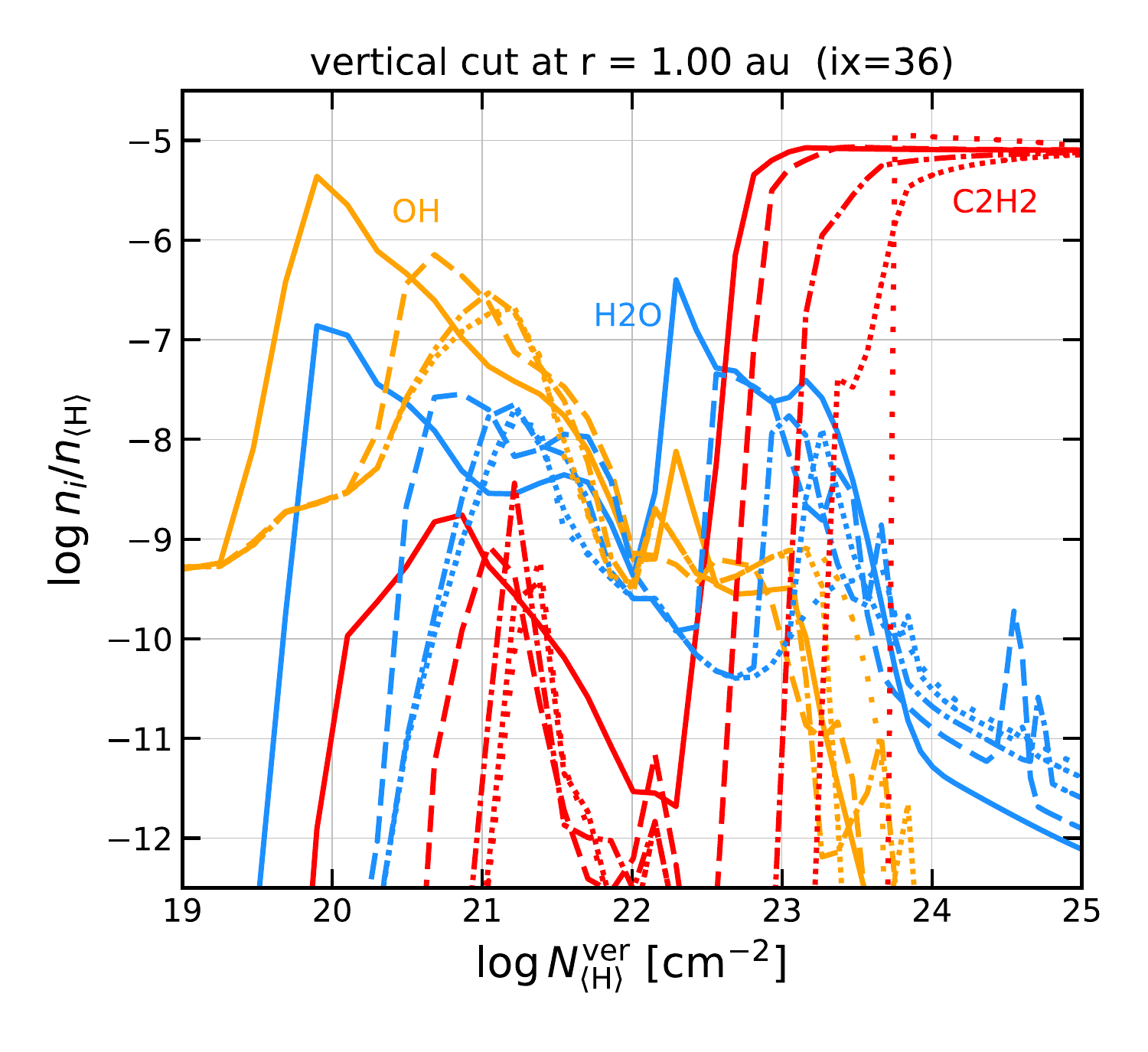}\\*[-1mm]
  \includegraphics[width=80mm,height=57mm,trim=10 10 5 30,clip]
                  {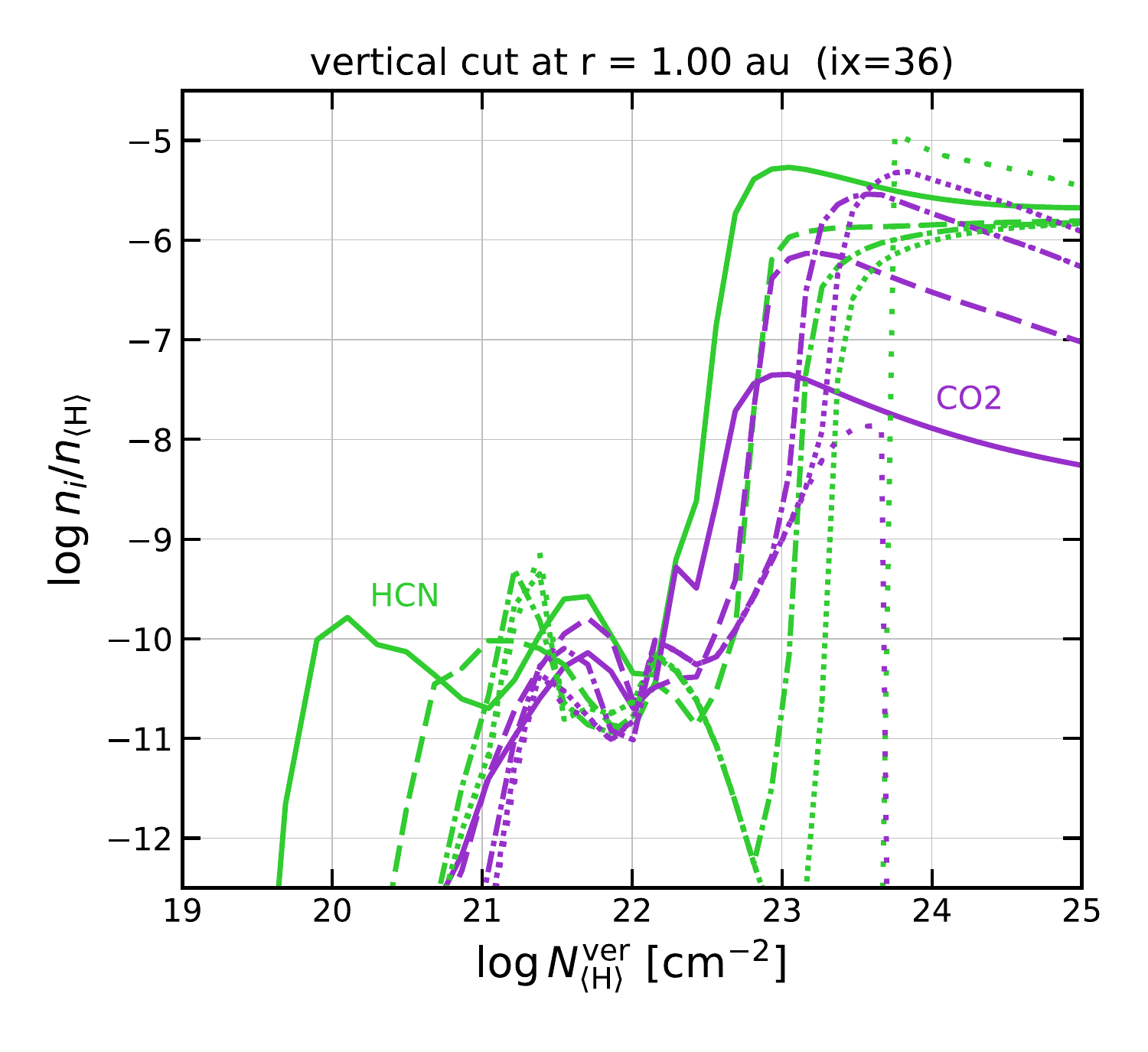}\\*[-3mm]
                  
  \caption{Gas and dust temperatures (upper panel), and concentrations
    of selected IR active molecules (lower panels) along a vertical
    cut at $r\!=\!1$\,au as function of vertical hydrogen nuclei
    column density $N_{\rm\langle H \rangle}$. The results of five
    models are overplotted using different line styles, from
    $\alpha_{\rm mix}\!=\!0$ to $10^{-2}$, see top panel.}
  \label{fig:IRactive}
\end{figure}

Figure~\ref{fig:IRactive} shows the dust and gas temperatures, and the
concentrations of the IR-active molecules OH, \ce{H2O}, \ce{CO2},
\ce{HCN}, \ce{C2H2} along a vertical cut at $r\!=\!1$\,au. We see that
all molecules shift their location of maximum concentration to the
left, i.e.\ upwards towards smaller column densities, but whereas OH,
\ce{H2O}, \ce{HCN} and \ce{C2H2} are enhanced by mixing, \ce{CO2} is
reduced. This is because of the upmixing of \ce{CH4} which leads
  to a rich hydro-carbon chemistry as discussed in
  Sect.~\ref{sec:chempath}, which suppresses the formation of
  \ce{CO2}. The figure also shows that the gas in the upper layers
cools down via increased line emission from OH and \ce{H2O}, the
concentrations of which are enhanced by mixing.

\begin{figure*}
\hspace*{-6mm}    
\begin{tabular}{c}
  \includegraphics[width=187mm]{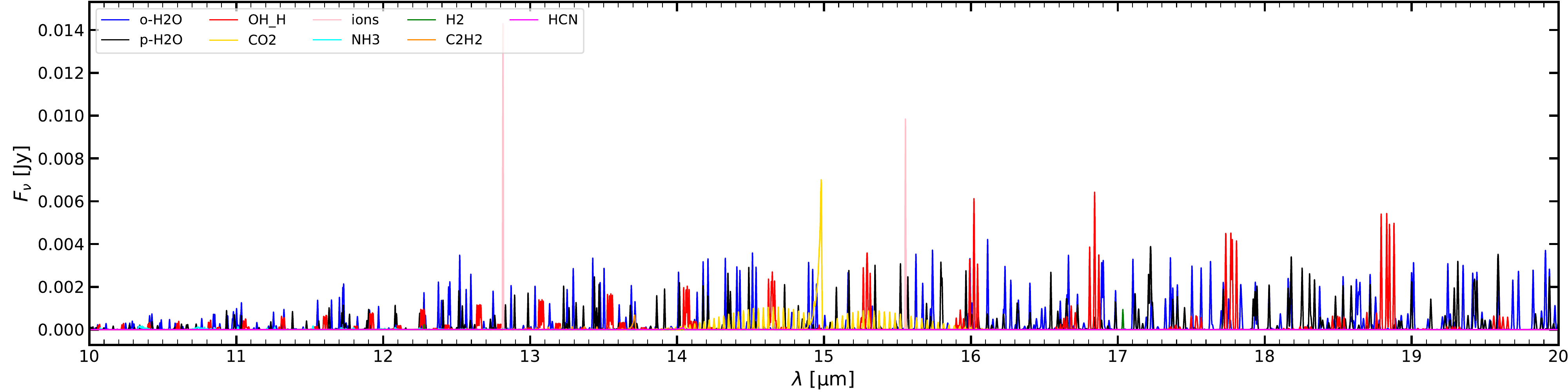} \\[-46mm]
  \hspace{140mm}$\boxed{\alpha_{\rm mix}\!=\!0}$\\[37mm]
  \includegraphics[width=187mm]{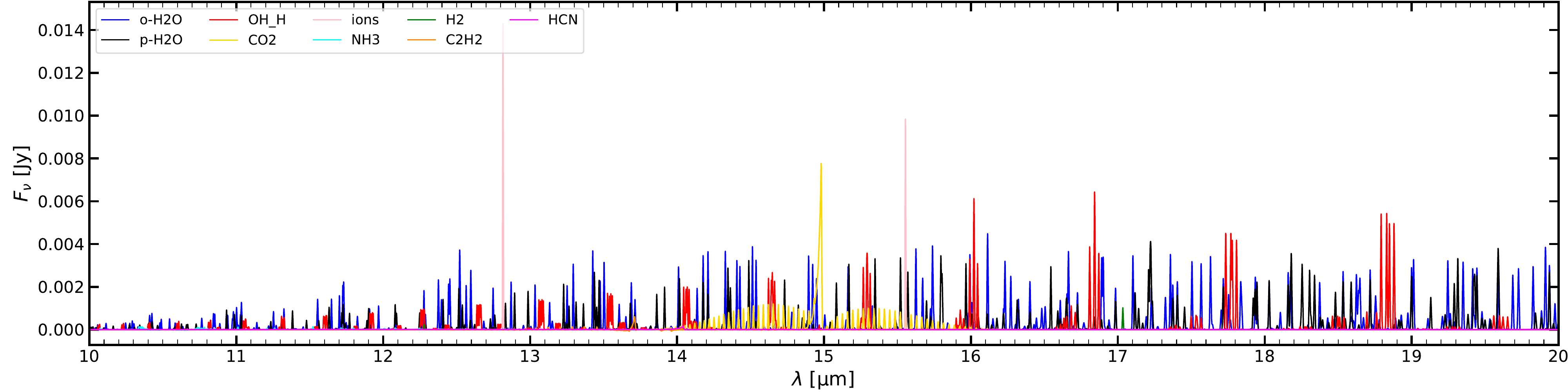} \\[-46mm]
  \hspace{140mm}$\boxed{\alpha_{\rm mix}\!=\!10^{-5}}$\\[36mm]
  \includegraphics[width=187mm]{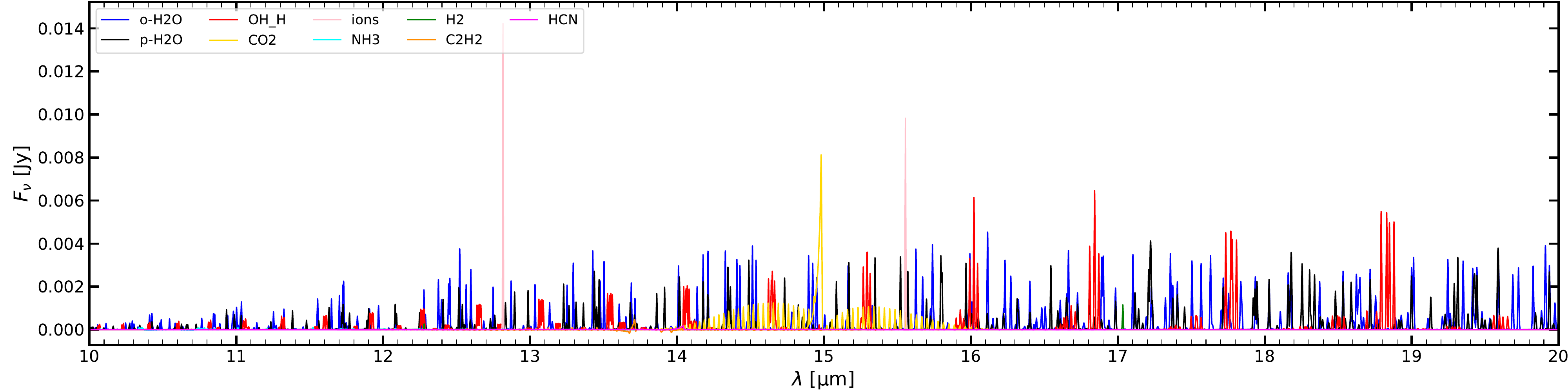} \\[-46mm]
  \hspace{140mm}$\boxed{\alpha_{\rm mix}\!=\!10^{-4}}$\\[36mm]
  \includegraphics[width=187mm]{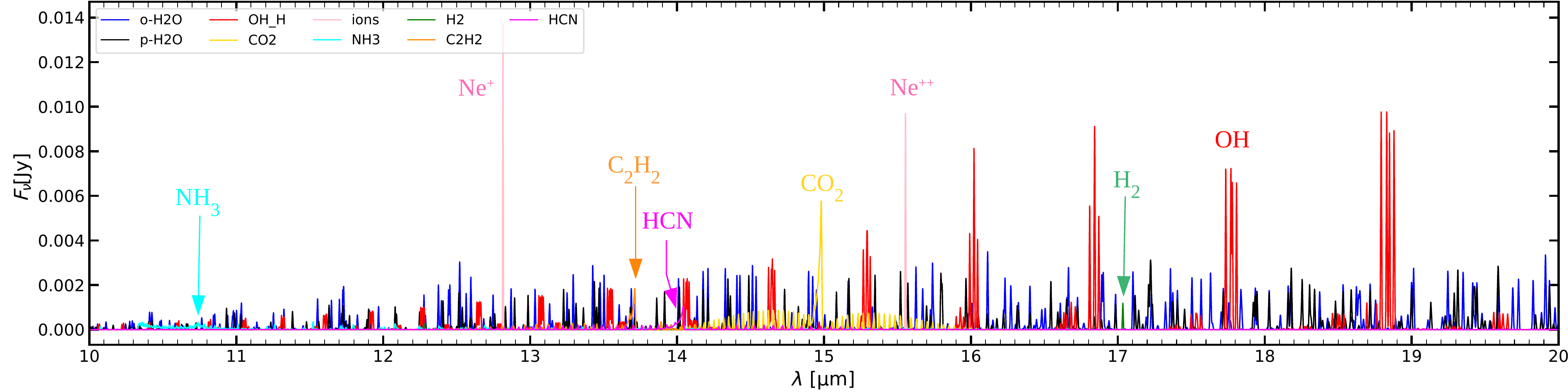} \\[-46mm]
  \hspace{140mm}$\boxed{\alpha_{\rm mix}\!=\!10^{-3}}$\\[36mm]
  \includegraphics[width=187mm]{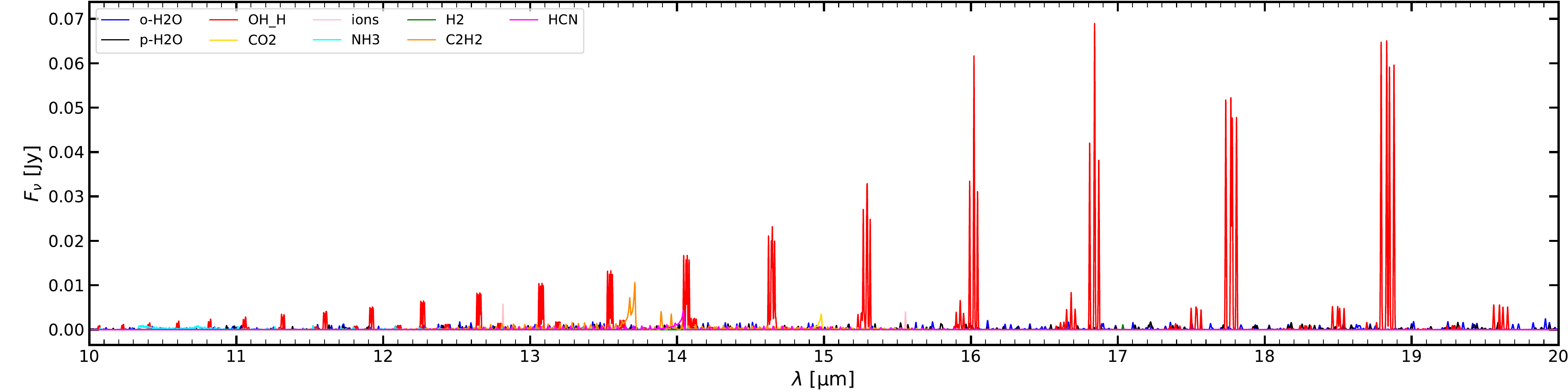} \\[-46mm]
  \hspace{140mm}$\boxed{\alpha_{\rm mix}\!=\!10^{-2}}$\\[39mm]
\end{tabular}
\caption{Continuum-subtracted infrared line emission spectra for 
  disc inclination $45^{\rm o}$ and spectral resolution
  $R\!=\!3000$ with varying mixing parameter $\alpha_{\rm mix}$. Note
  the different scaling of the $y$-axis. Most visible spectral
  features are overlapping \ce{o-H2O} (blue) and \ce{p-H2O} (black)
  emission lines. A few other features emitted from other molecules
  and ions are annotated in one of the plots. Each molecular spectrum
  is first convolved with a Gaussian $R\!=\!3000$ and then overplotted
  with a certain colour, but not co-added.}
\label{fig:IRspec}
\end{figure*}

\begin{figure*}
  \centering
  \resizebox{183mm}{!}{\begin{tabular}{lll}
  \hspace*{29mm} {\small $\alpha_{\rm mix}=0$} &
  \hspace*{21mm} {\small $\alpha_{\rm mix}=10^{-4}$} &
  \hspace*{21mm} {\small $\alpha_{\rm mix}=10^{-2}$}\\
  \hspace*{-0.5mm} \includegraphics[height=58mm,trim=  0 5 5 5,clip]
          {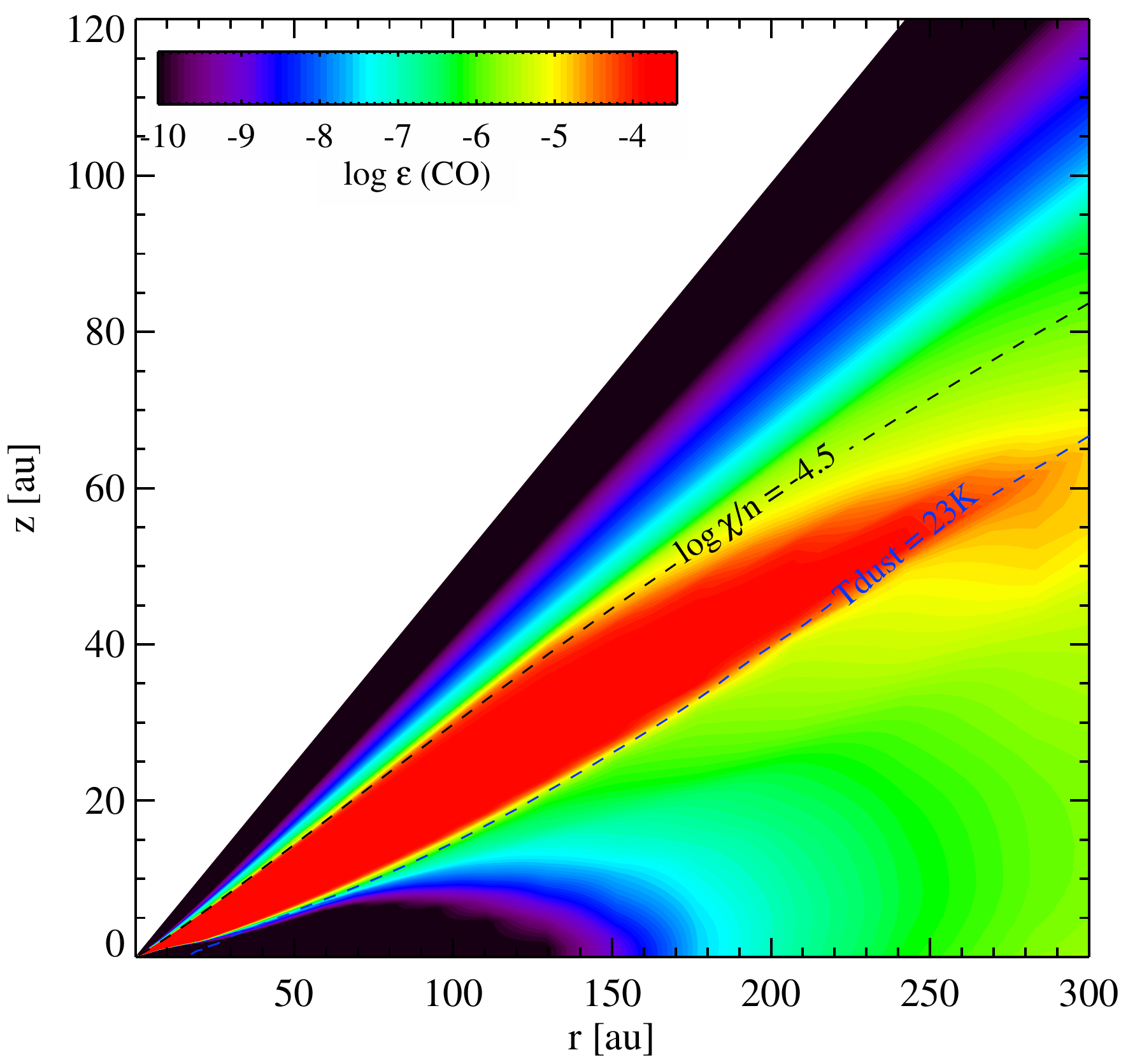} &
  \hspace*{-3mm} \includegraphics[height=58mm,trim= 32 5 5 5,clip]
          {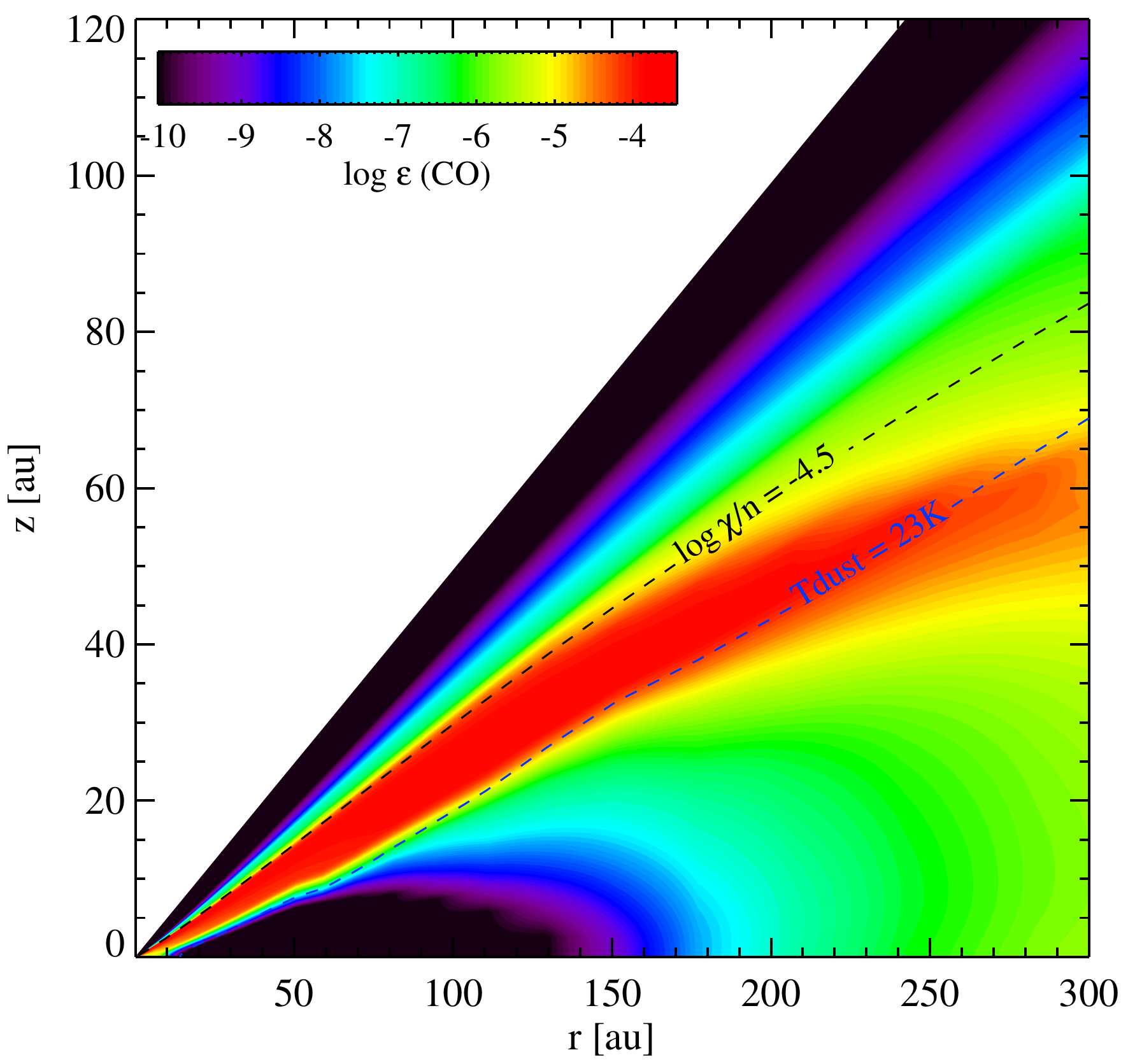} &
  \hspace*{-3mm} \includegraphics[height=58mm,trim= 32 5 5 5,clip]
          {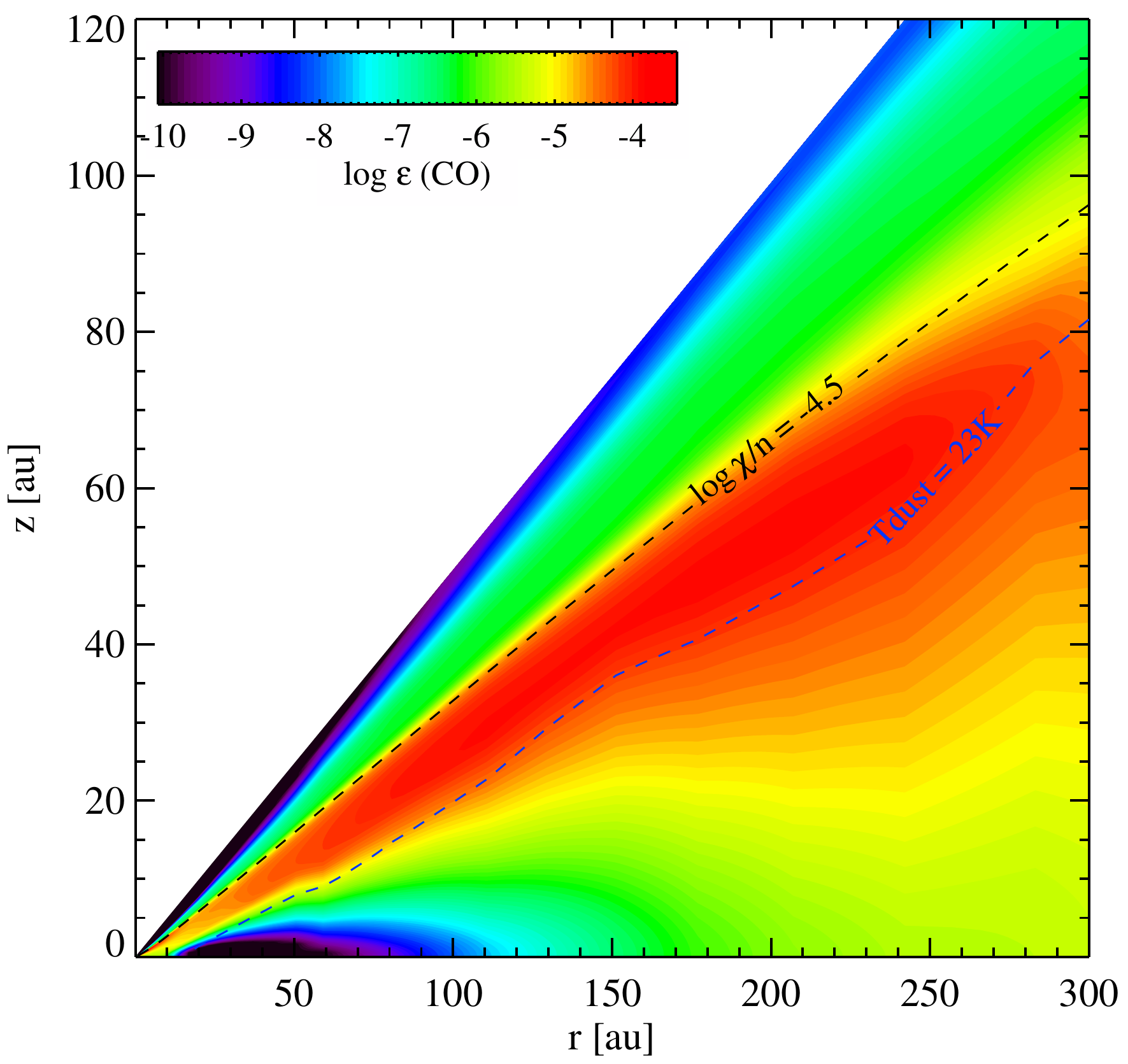} \\*[-0.5mm]
  \hspace*{-2mm} \includegraphics[height=60.5mm,trim= 20 10 12 10,clip]
          {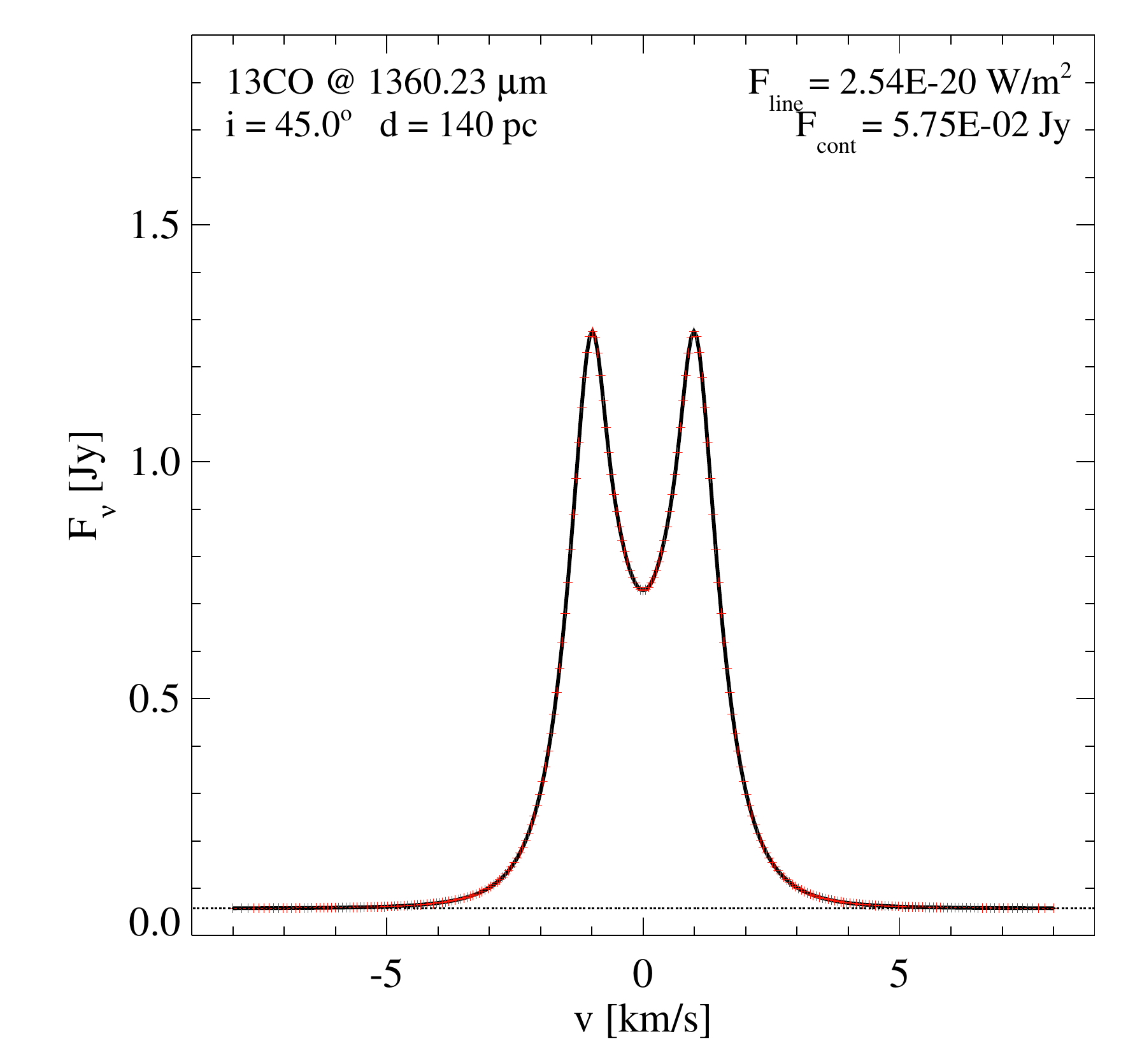} &
  \hspace*{-4mm} \includegraphics[height=60.5mm,trim= 53 10 12 10,clip]
          {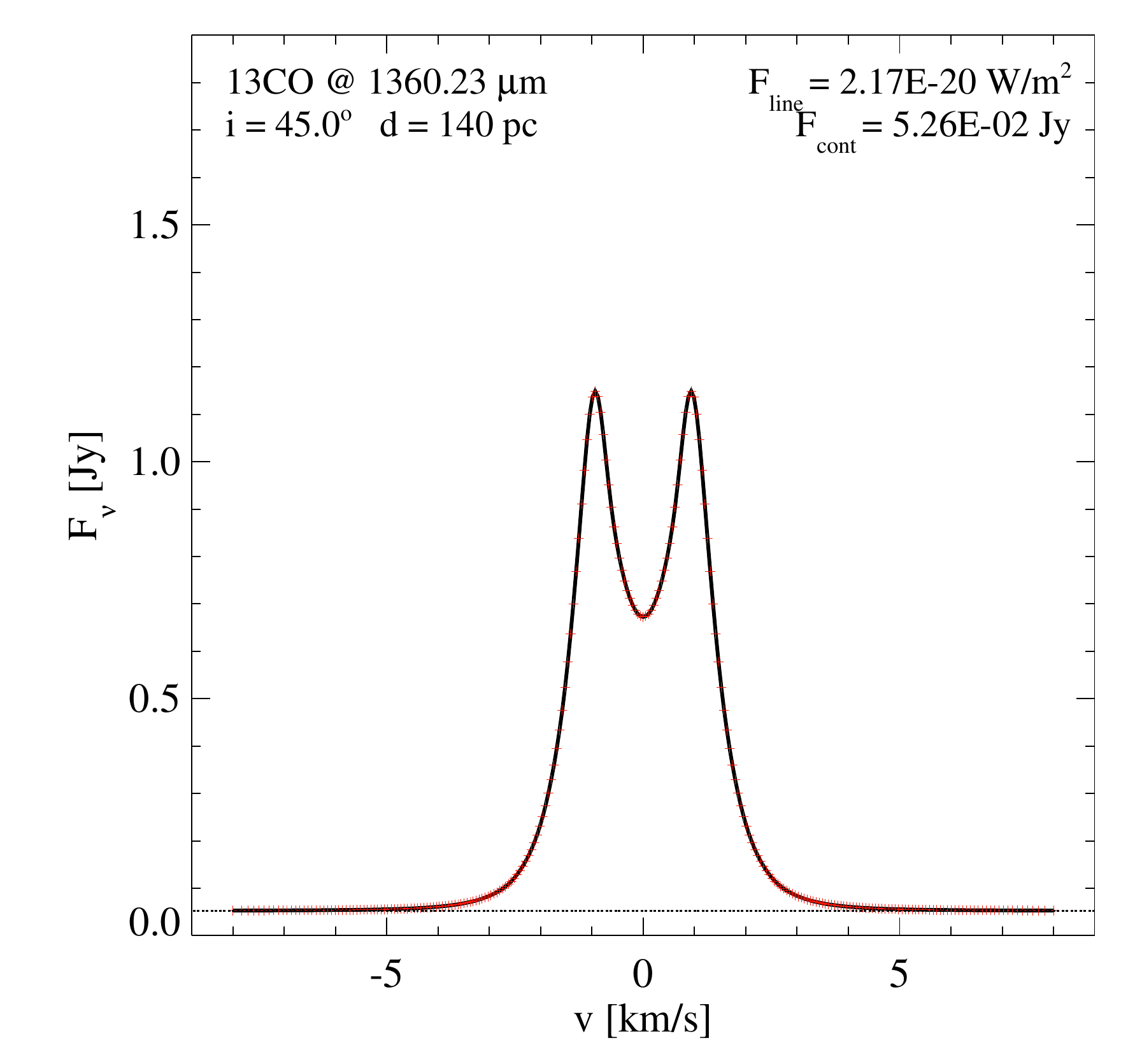} &
  \hspace*{-4mm} \includegraphics[height=60.5mm,trim= 53 10 12 10,clip]
          {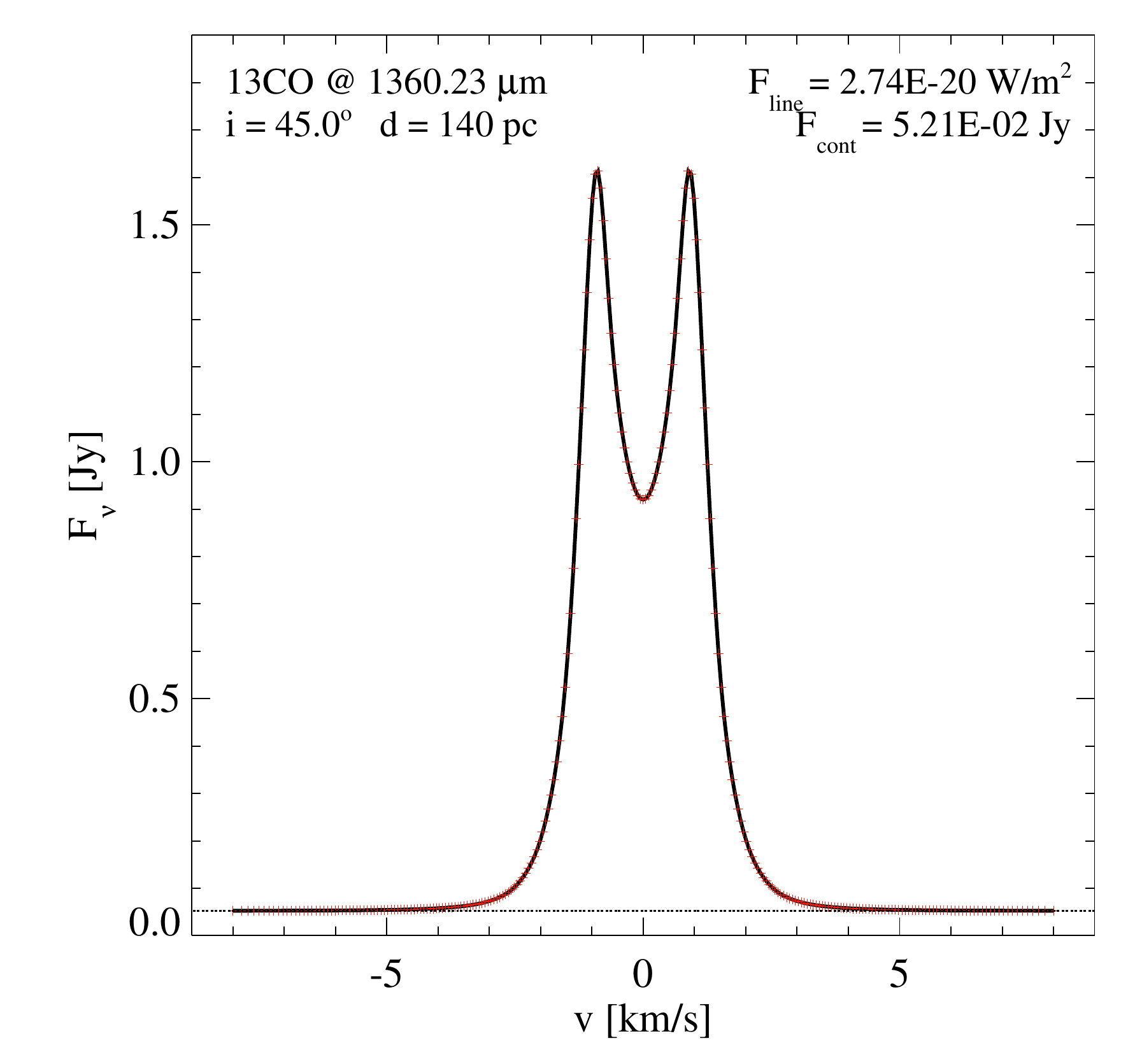} \\*[-0.5mm]      
  \hspace*{-2mm} \includegraphics[height=63mm,trim= 11 10 80 55,clip]
          {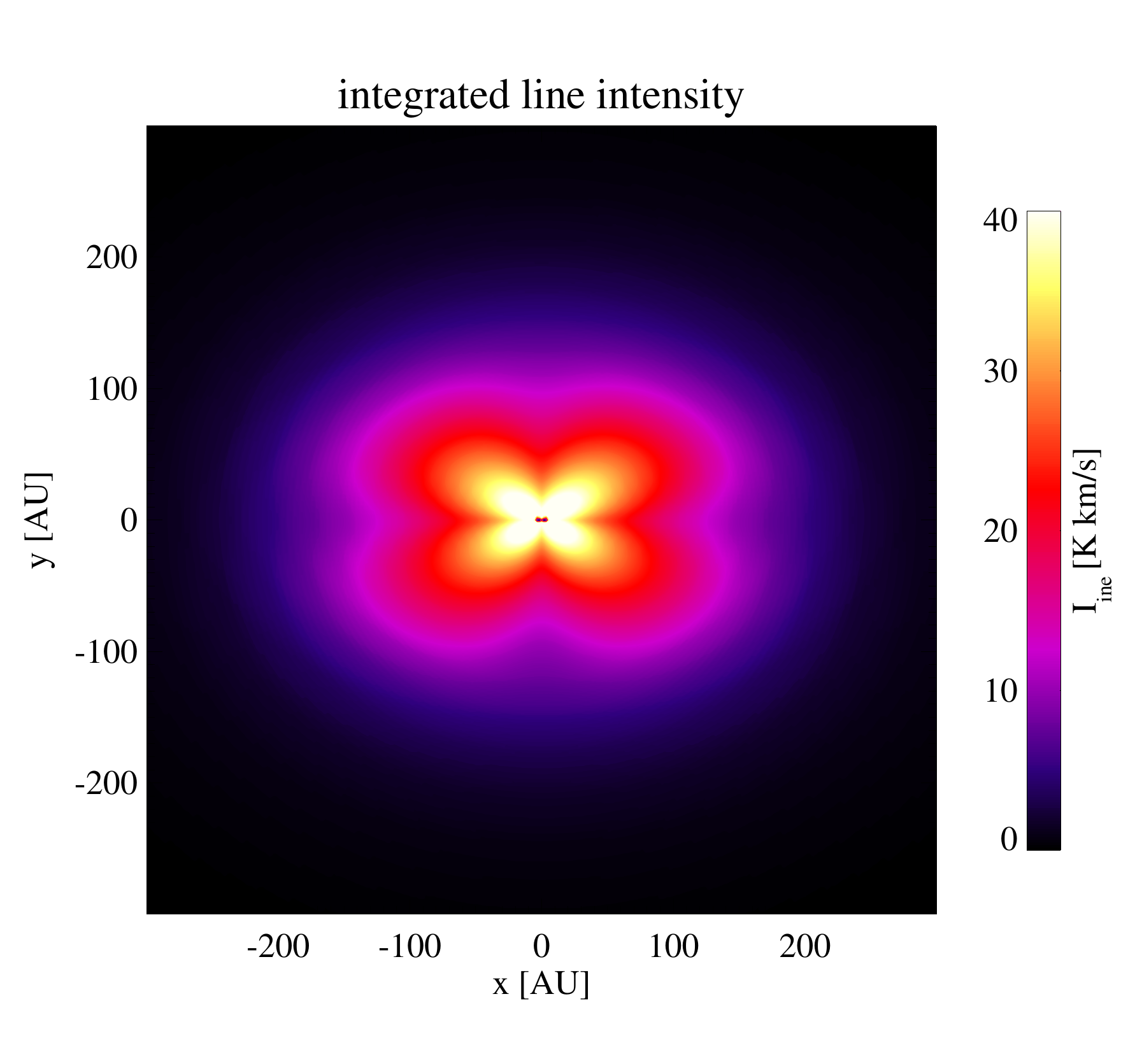} &
  \hspace*{-5mm} \includegraphics[height=63mm,trim= 34 10 80 55,clip]
          {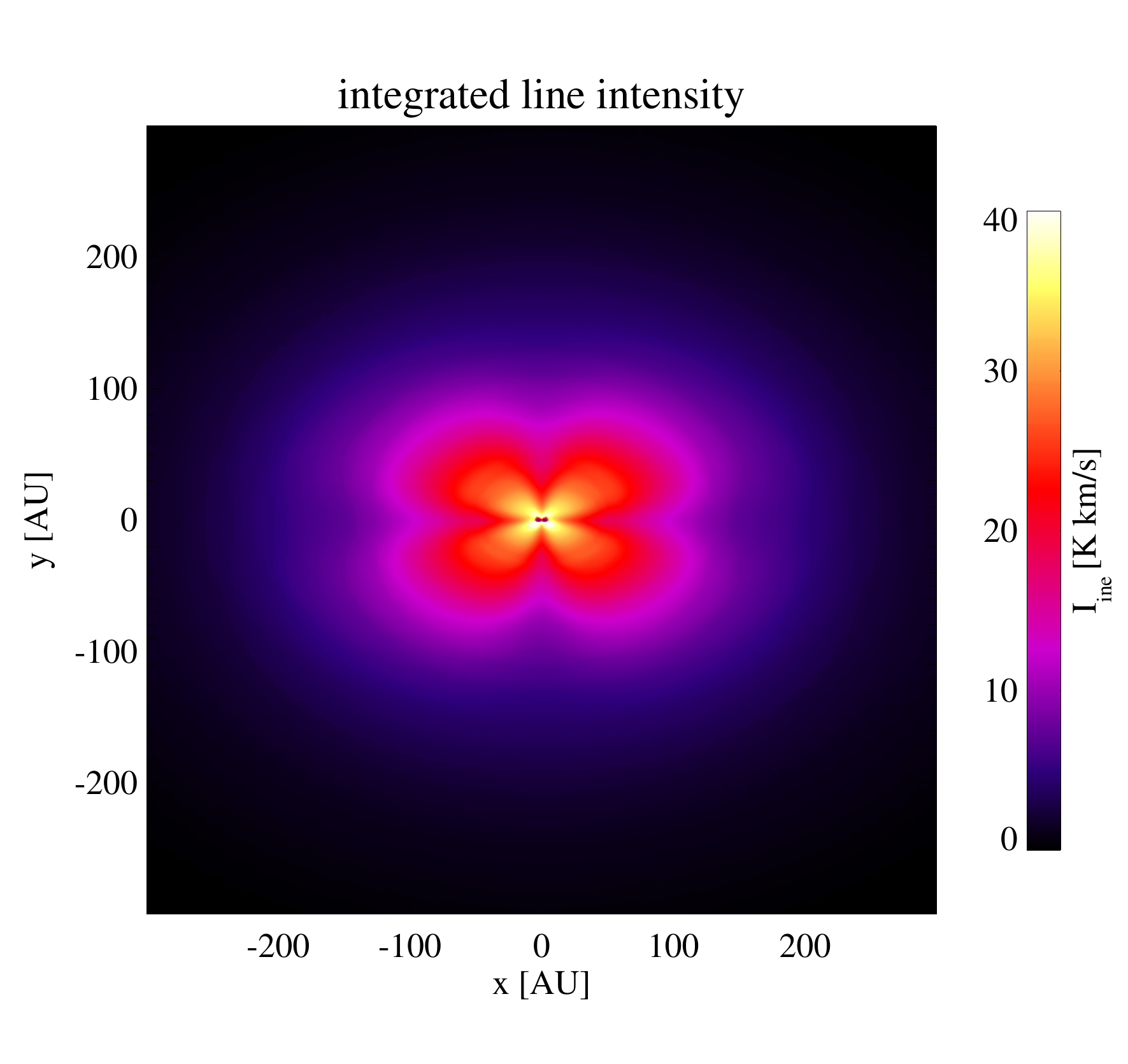} &
  \hspace*{-5mm} \includegraphics[height=63mm,trim= 34 10 05 55,clip]
          {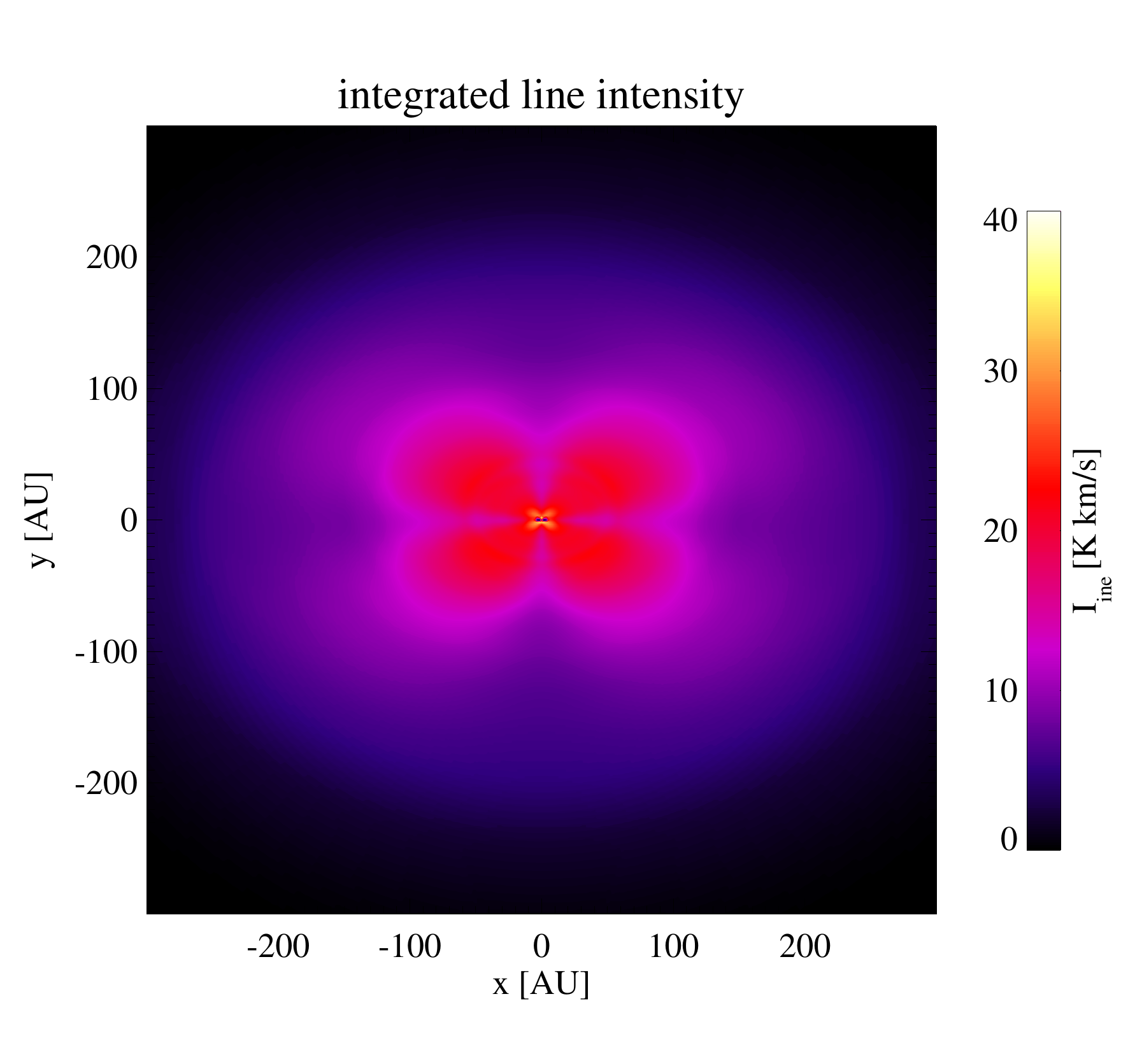} \\*[-3mm]
  \hspace*{-2mm} \includegraphics[height=63mm,trim= 11 10 80 55,clip]
          {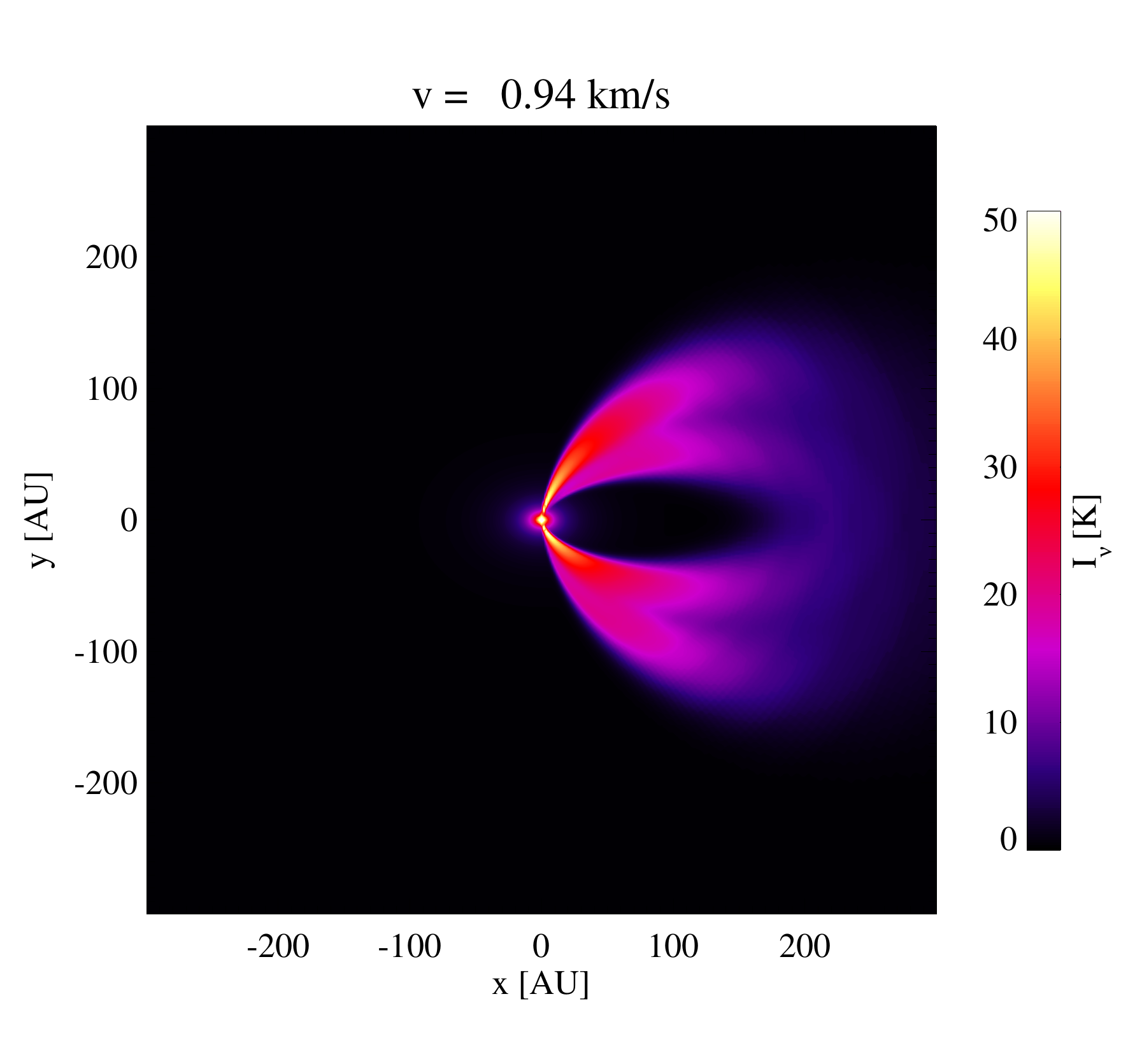} &
  \hspace*{-5mm} \includegraphics[height=63mm,trim= 34 10 80 55,clip]
          {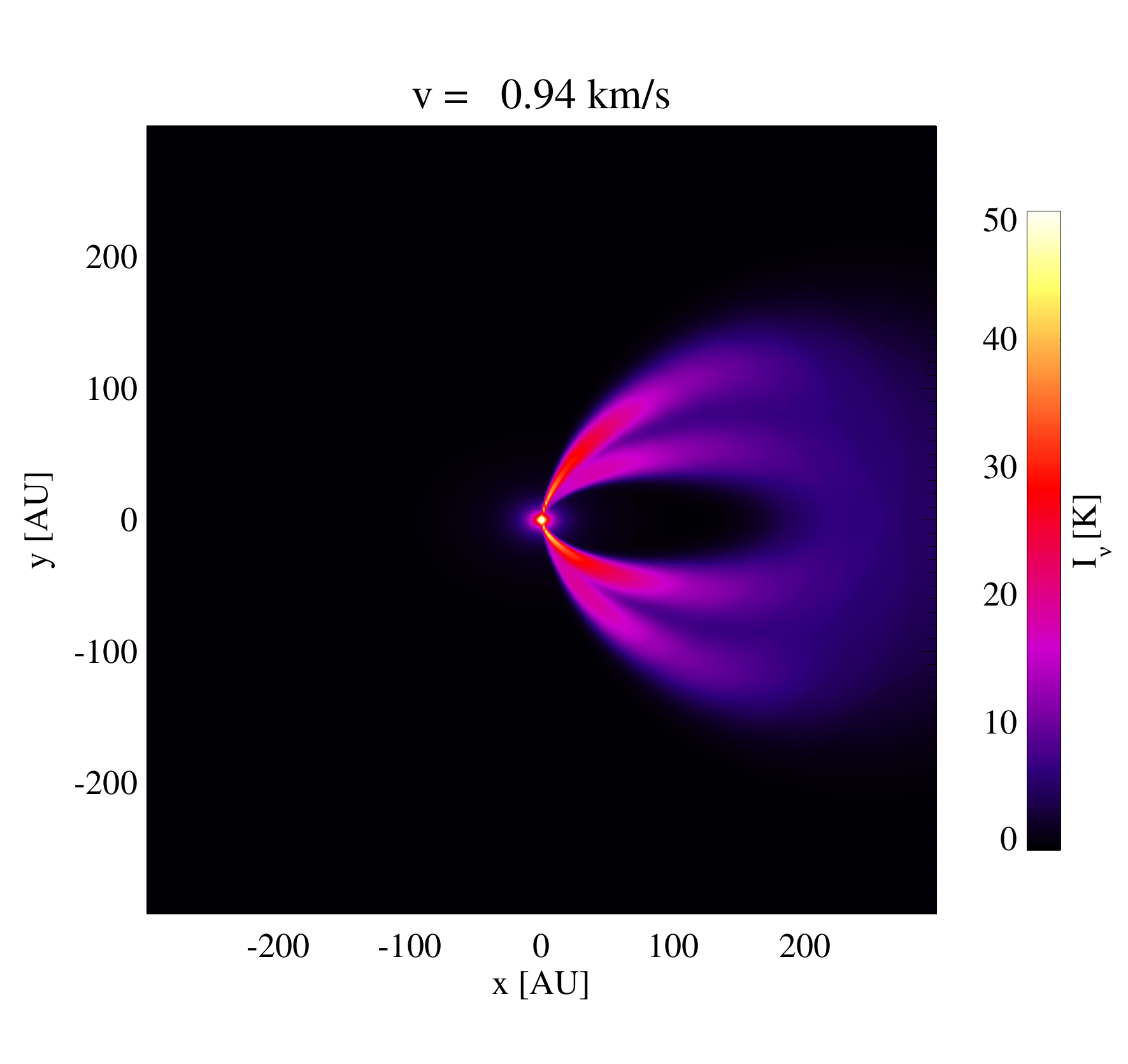} &
  \hspace*{-5mm} \includegraphics[height=63mm,trim= 34 10 05 55,clip]
          {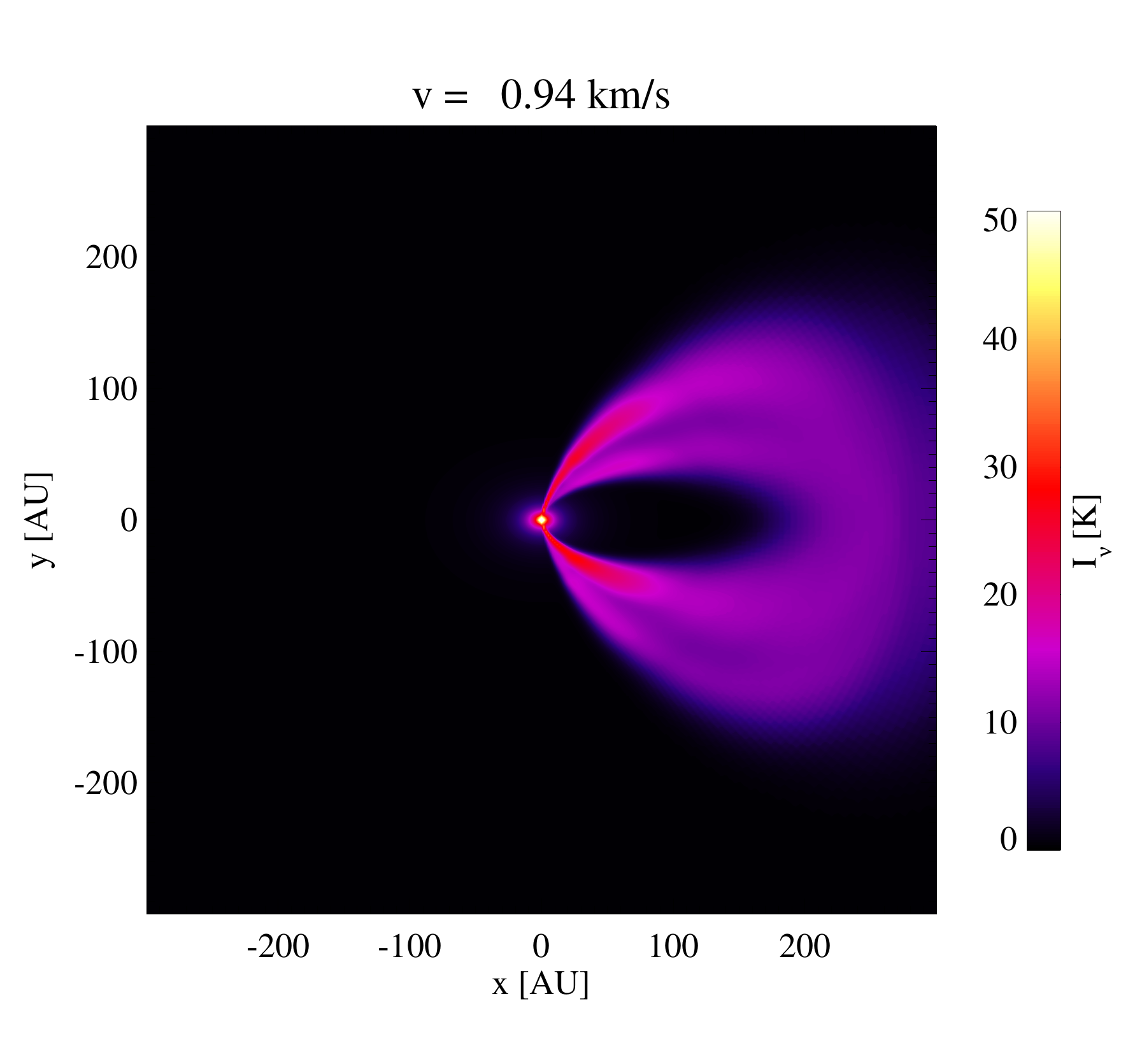} \\*[-4mm]
  \end{tabular}}
  \caption{Influence of vertical mixing on millimetre CO
    observations. The top row shows the CO concentration
    $\epsilon({\rm CO})\!=\!n_{\rm CO}/\nH$ in the disc on linear
    scales for radius $r$ and vertical height $z$. The two dashed
    contour lines enclose the region of maximum CO concentration,
    where $\chi$ is the UV strength and $n=\nH$ is the total hydrogen
    nuclei particle density. Below dust temperature $T_{\rm
      dust}\!\approx\!23\,$K, CO is mostly frozen. The second row
    shows the $^{13}$CO J$=2\!\to\!1$ line velocity profile observed
    from a distance of 140\,pc at disc inclination $45^{\rm o}$. The
    third row shows the integrated line map after continuum
    subtraction and the bottom row shows a channel map at radial
    velocity $v+0.94\,$km/s. To plot the intensities in Kelvin, we
    consider $I_\nu\,c^2/(2\nu^2)/k$, where
    $I_\nu\rm\,[erg/cm^2/s/Hz/sr]$ is the spectral intensity, $c$ the
    speed of light, $\nu$ the frequency and $k$ the Boltzmann
    constant.}
  \label{fig:CO}
\end{figure*}

Figure \ref{fig:IRspec} shows a series of mid-IR gas emission line
spectra between 10\,$\mu$m and 20\,$\mu$m computed from our series of
T\,Tauri disc models with $\alpha_{\rm mix}\!=\!0$ to $10^{-2}$, see
Sect.~\ref{sec:model} These spectra have been calculated with the {\sl
  Fast Line Tracer} {\sc FLiTs} developed by Michiel Min
\citep[see][]{Woitke2018} on top of the gas \& dust temperature
structure, the continuum dust opacities, the molecular abundances and
the non-LTE excitation populations calculated by {\sc ProDiMo} as
described in the previous sections.  Due to our lack of sufficient
collisional data, the populational numbers of the molecules \ce{OH},
\ce{CO2}, \ce{HCN}, \ce{C2H2} and \ce{NH3} are assumed to be in LTE.
{\sc FLiTs} calculates the entire spectrum on a wavelength grid that
corresponds to a velocity resolution of 1\,km/s.  We have calculated
the spectra separately for each molecule listed in the legend of
Fig.~\ref{fig:IRspec}. In contrast, the IR lines of the atoms and ions
\ce{Ne+}, \ce{Ne++}, \ce{Ar+}, \ce{Ar++}, \ce{Mg+}, \ce{Fe+},
\ce{Si+}, \ce{S}, \ce{S+} and \ce{S++} are calculated in one go. For
more details and explanations, see \citet{Woitke2018}.  The resulting
continuum-subtracted spectra are then convolved with a $R\!=\!3000$
Gaussian before plotting, to simulate the spectra that we can expect
from JWST/MIRI.

As discussed in \citet{Woitke2018}, these predicted spectra are
broadly consistent with Spitzer/IRS observational data of T\,Tauri
stars concerning line colours and ratios.  However, since we assumed a
dust/gas ratio of 100 here, instead of 1000 in this paper, and since we only
applied a very mild dust settling with $\alpha_{\rm settle}\!=\!10^{-2}$
according to \citet{Dubrulle1995}, the dust is still covering most of
the line emitting regions, and the predicted IR lines are generally
too weak, by a factor of about 10-100.  It also means that the lines that
are emitted from higher disc layers (e.g.\ by \ce{Ne+} and \ce{OH}) are
stronger than those emitted from lower layers (e.g. \ce{H2O}, \ce{HCN}
and \ce{C2H2}).  Therefore, we can only discuss the trends here that
are caused by the introduction of mixing.

We see little changes between $\alpha_{\rm settle}\!=\!0$ and
$\alpha_{\rm settle}\!=\!10^{-4}$. Only at $\alpha_{\rm
  settle}\!=\!10^{-3}$, the OH lines start to become significantly stronger, and
the \ce{C2H2} and \ce{HCN} emission features at 13.7\,$\mu$m and
14\,$\mu$m, respectively, start to be enhanced and become about
comparable to the \ce{CO2} feature at 15\,$\mu$m, as is observed for
many T\,Tauri stars \citep{Pontoppidan2010}.  For $\alpha_{\rm
  settle}\!=\!10^{-2}$, the OH lines become too strong with respect to
the \ce{H2O} lines, similar to the findings of \citet{Heinzeller2011}.
\ce{C2H2} 13.7\,$\mu$m is now the strongest spectral feature between
10\,$\mu$m and 20\,$\mu$m after the various OH ro-vibrational
features, which is also in disagreement with Spitzer observations.

\subsection{Impact on CO lines and channel maps}


The upper row of Fig.\,\ref{fig:CO} shows the resulting CO
concentration on linear radial and vertical scales in the same model
series.  In our disc models, carbon results to be mostly present in
form of CO, with a maximum concentration of $n_{\rm
  CO}/\nH\!\approx\!\epsilon_{\rm C}\!=\!10^{\,-3.85}$, in a vertical
layer that is limited by photo-dissociation on top (ionisation
parameter $\chi/\nH\!\approx\!10^{\,-4.5}$) and by ice formation at
the bottom ($T_{\rm dust}\!\approx\!23\,$K).  The $^{13}$CO
$J\!\!=\!2\!\to\!1$ line at 1360.2\,$\mu$m is mostly emitted from this
vertical layer, and 70\% of the line flux is emitted radially from
about 50\,au to 200\,au in the disc model with $\alpha_{\rm
  mix}\!=\!0$. Beyond about 250\,au the line becomes optically thin.
Figure~\ref{fig:CO} shows the resulting line fluxes and double-peaked
velocity profiles, the velocity-integrated line maps, and selected
channel maps in the lowest panel.

We see two major effects of the mixing on the spectral appearance of
the CO in millimetre lines. First, the CO line emission becomes weaker
with increasing $\alpha_{\rm mix}$ at small radii.
Figure~\ref{fig:chem_results} shows that at 100\,au (right column) the
disc cools down slightly with increasing $\alpha_{\rm mix}$.  This is
because of the additional presence of icy grains in higher disc layers
due to mixing, see Sect.~\ref{sec:icemix}.  Because of their enhanced
opacity, the icy grains increase the vertical optical depths and also
scatter away the starlight more efficiently.  As a consequence, the
height below which CO freezes out increases, and hence the vertical CO
column density decreases.  Figure~\ref{fig:CO} shows a clear trend of
the CO emissions from the inner disc regions becoming weaker with
increasing $\alpha_{\rm mix}$, see line maps, while the vertical layer
of maximum CO concentration becomes thinner, see concentrations.

Second, the CO line emissions intensify with increasing $\alpha_{\rm
  mix}$ at large radii. This is a direct consequence of downward CO
mixing.  For large $\alpha_{\rm mix}$ the CO mixing timescale becomes
shorter than the ice adsorption timescale in the outer regions, which
leads to a more gradual decline of the CO concentration towards the
midplane, and hence larger CO column densities in the outermost disc
regions.  The lower boundary of the vertical layer of maximum CO
concentration becomes less well defined for large $\alpha_{\rm
  mix}$. This leads to a more ``diffuse'' spectral appearance in
millimetre CO channel maps, see lower row of plots in
Fig.\,\ref{fig:CO}.  Without mixing, we obtain clearly separated line
emissions from two lobes, one from the upper disc half, and one from
the lower disc half.  With strong mixing $\alpha_{\rm
  mix}\!=\!10^{-2}$, the two lobes visually merge beyond about
150\,au. \citet{Cleeves2016} and \citet{Pinte2018} noticed that for
the case of the giant disc of IM\,Lupi, where the CO lines are
detectable out to $\sim\!1000$\,au, it becomes impossible to identify
two separate CO layers beyond $\sim\!450$\,au. \citet{Pinte2018}
achieved a distinctively better fit when adding diffuse CO emission
from the outer midplane. We note, however, that even without mixing,
the CO concentration in the midplane increases with radius as the
freeze-out timescale of CO increases with falling density whereas the
photo-dissociation timescale is density-independent.

In the model with the strongest mixing $\alpha_{\rm mix}\!=\!10^{-2}$,
we furthermore observe that the CO is also vertically more extended
toward to top.  This is partly caused by an upward shift of the
$\chi/n\!=\!10^{-4.5}$ contour line due to the increased dust
opacities of the icy grains, but at the very top, the CO is actually
mixed up quicker than it can be photo-dissociated. In summary, we
confirm the finding of \citet{Semenov2006} that the CO density can be
enhanced by mixing above and below the layer of maximum concentration,
but we disagree with \citet{Semenov2011} that the CO is not responsive
to mixing.

\section{Summary and Conclusions}
\label{sec:summary}

\subsection{Summary of chemical mechanisms}
  
\begin{figure}[!t]
  \centering
  \vspace*{-2mm}
  \includegraphics[width=80mm,trim=0 0 0 0,clip]{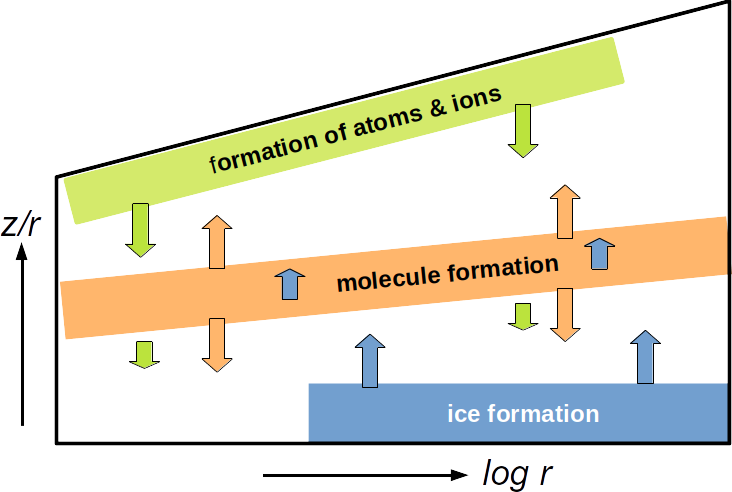} 
  \caption{Schematic sketch of the disc regions responsible for the
    formation of atoms, molecules and ices, and their vertical
    diffusive transport.}
  \label{fig:sketch}
\end{figure} 

Figure~\ref{fig:sketch} summarises the principle mechanisms that we
have identified in our disc models by means of a physical sketch. The
mixing is immediately relevant for the most abundant species, such as
\ce{H2}, \ce{CH4}, \ce{CO}, the neutral atoms in higher layers, and
the ices in the midplane.  With ``immediately relevant'', we mean that
the transport rates can locally dominate over the chemical formation
and destruction rates.  The abundant species, whose concentrations are
limited by element abundance constraints, are transported away from
their regions of maximum concentration, both up and down, and in the
locations where these abundant chemicals are finally destroyed, for
example by photo-reactions, their reaction products fuel the local
chemistry.  The release of these fragments has important consequences
for all other, chemically interlinked but less abundant molecules,
which quickly adjust their concentrations to these changed
circumstances.  This generally creates a more active chemistry, with a
richer mixture of ionised, atomic, molecular and ice species and new
chemical pathways that are not relevant in the un-mixed case.

This situation differs somewhat from exoplanet atmospheres as
e.g.\ discussed by \citet{Moses2011}, \citet{Zahnle2016} and
\citet{Barth2021}. In
exoplanet atmospheres, the chemical factories are located in the
deepest layers, where temperatures and densities are high.  From
there, molecules are mixed upward, resulting in constant
concentrations at intermediate heights, until photo-dissociation
destroys them at high altitudes. In discs, the main chemical factories
are located in the warm intermediate layers. From there, molecules are
mixed upwards and downwards.  The dense midplane is too cold to have
an active chemistry.  Instead, molecules freeze out there, and the icy
grains are mixed up.

\subsection{Summary of impact on spectral observations}

Concerning the mid-IR molecular spectra, we see increased molecular
abundances in the upper layers observable with Spitzer and JWST, in
particular for OH, \ce{H2O}, \ce{HCN} and \ce{C2H2} in our models when
we include mixing.  The chemical response to mixing in these layers
happens on short timescales, of the order of hours or days, and
liberates more heat via exothermic reactions. However, at the same
time the increased abundances of the IR active molecules lead to
increased line cooling and a decrease of gas temperatures in those
high layers. Despite the strong effects on both molecular abundances
and gas temperature structure, the calculated IR spectra are
surprisingly robust. We need to assume strong mixing $\alpha_{\rm
  mix}\!\ga\!10^{-3}$ to produce notable spectral changes. The
spectral features of OH, \ce{HCN} and \ce{C2H2} are enhanced by
mixing, whereas the \ce{CO2} feature at 15\,$\mu$m is reduced.

Concerning the CO millimetre emission lines, mixing reduces the line
intensities at small radii but enhances them at larger radii, leading
to an overall more flat line intensity profile in ALMA line observations.
For intense mixing, CO starts to populate the outer midplane
regions, which leads to a distinct, more diffuse appearance of the CO
in velocity channel maps.  However, it is likely that such
differences in CO channel maps can also be produced in
alternative ways in our models, for example by changing the column
density profile $N_{\rm\langle H\rangle}(r)$ of the disc.

\subsection{Conclusions about the $\alpha$ viscosity parameter}

The parameter $\alpha_{\rm mix}$ describes the strength of the
vertical turbulence according to the \cite{Shakura1973}
$\alpha$-model.  Our mixed models with $\alpha_{\rm
  mix}\!\ga\!10^{-3}$ show strong ice emission features between
20\,$\mu$m and 200\,$\mu$m, which neither Spitzer nor Herschel have
seen for T\,Tauri stars.  However, the $\alpha_{\rm
  mix}\!\approx\!10^{-5}$ and $10^{-4}$ models predict weak ice features
that might require the signal/noise ratio of JWST to be detectable.
Without mixing, it would be difficult to explain such detections
with our models.

Concerning the mid-IR molecular emission lines, our models with
$\alpha_{\rm mix}\!\approx\!10^{-3}$ predict more equal strengths of
the \ce{CO2}, HCN and \ce{C2H2} features between 13.7\,$\mu$m and
15\,$\mu$m, which is difficult to achieve in our models otherwise.
However, for $\alpha_{\rm mix}\!\approx\!10^{-2}$, the OH and
\ce{C2H2} emission features become unrealistically strong.

The detection of layered CO emission in ALMA channel maps, in form
of two distinct lobes emitted by the lower and the upper disc halves,
with no direct emission from the in-between midplane, allows us to put
some constraint on the $\alpha$ viscosity parameter, because for
large $\alpha_{\rm mix}$, CO would be expected to populate the midplane.
However, that constraint is weak, about $\alpha_{\rm mix}\!<\!10^{-2}$.

We conclude that models with a modest choice of $\alpha_{\rm
  mix}\!\approx\!10^{-5}$ to $10^{-3}$ might work best to produce
realistic continuum and line observations. However, a constant value
of $\alpha_{\rm mix}$ is certainly unrealistic in the first place,
as the turbulence is expected to vary with distance and height, see
e.g.\ figure 9 of \citet{Thi2019}.

\begin{acknowledgements}
   P.\,W.\ acknowledges funding from the European Union
   H2020-MSCA-ITN-2019 under Grant Agreement no.\,860470 (CHAMELEON).
\end{acknowledgements}

\bibliographystyle{aa}
\bibliography{references}

\begin{thebibliography}{46}
\expandafter\ifx\csname natexlab\endcsname\relax\def\natexlab#1{#1}\fi

\bibitem[{{Aikawa} {et~al.}(1996){Aikawa}, {Miyama}, {Nakano}, \&
  {Umebayashi}}]{Aikawa1996}
{Aikawa}, Y., {Miyama}, S.~M., {Nakano}, T., \& {Umebayashi}, T. 1996, \apj,
  467, 684

\bibitem[{{Arabhavi} {et~al.}(2022){Arabhavi}, {Woitke}, {Cazaux}, {Kamp},
  {Rab}, \& {Thi}}]{Arabhavi2022}
{Arabhavi}, A.~M., {Woitke}, P., {Cazaux}, S.~M., {et~al.} 2022, arXiv
  e-prints, arXiv:2208.12739

\bibitem[{{Aresu} {et~al.}(2011){Aresu}, {Kamp}, {Meijerink}, {Woitke}, {Thi},
  \& {Spaans}}]{Aresu2011}
{Aresu}, G., {Kamp}, I., {Meijerink}, R., {et~al.} 2011, \aap, 526, A163

\bibitem[{{Barth} {et~al.}(2021){Barth}, {Helling}, {St{\"u}eken}, {Bourrier},
  {Mayne}, {Rimmer}, {Jardine}, {Vidotto}, {Wheatley}, \& {Fares}}]{Barth2021}
{Barth}, P., {Helling}, C., {St{\"u}eken}, E.~E., {et~al.} 2021, \mnras, 502,
  6201

\bibitem[{{Charnay} {et~al.}(2015){Charnay}, {Meadows}, \&
  {Leconte}}]{Charnay2015}
{Charnay}, B., {Meadows}, V., \& {Leconte}, J. 2015, \apj, 813, 15

\bibitem[{{Cleeves} {et~al.}(2016){Cleeves}, {{\"O}berg}, {Wilner}, {Huang},
  {Loomis}, {Andrews}, \& {Czekala}}]{Cleeves2016}
{Cleeves}, L.~I., {{\"O}berg}, K.~I., {Wilner}, D.~J., {et~al.} 2016, \apj,
  832, 110

\bibitem[{{Draine} \& {Bertoldi}(1996)}]{Draine1996}
{Draine}, B.~T. \& {Bertoldi}, F. 1996, \apj, 468, 269

\bibitem[{{Dubrulle} {et~al.}(1995){Dubrulle}, {Morfill}, \&
  {Sterzik}}]{Dubrulle1995}
{Dubrulle}, B., {Morfill}, G., \& {Sterzik}, M. 1995, \icarus, 114, 237

\bibitem[{{Flaherty} {et~al.}(2015){Flaherty}, {Hughes}, {Rosenfeld},
  {Andrews}, {Chiang}, {Simon}, {Kerzner}, \& {Wilner}}]{Flaherty2015}
{Flaherty}, K.~M., {Hughes}, A.~M., {Rosenfeld}, K.~A., {et~al.} 2015, \apj,
  813, 99

\bibitem[{{Heinzeller} {et~al.}(2011){Heinzeller}, {Nomura}, {Walsh}, \&
  {Millar}}]{Heinzeller2011}
{Heinzeller}, D., {Nomura}, H., {Walsh}, C., \& {Millar}, T.~J. 2011, \apj,
  731, 115

\bibitem[{{Hu} {et~al.}(2012){Hu}, {Seager}, \& {Bains}}]{Hu2012}
{Hu}, R., {Seager}, S., \& {Bains}, W. 2012, \apj, 761, 166

\bibitem[{{Hunten}(1973)}]{Hunten1973}
{Hunten}, D.~M. 1973, Journal of Atmospheric Sciences, 30, 1481

\bibitem[{{Ilgner} {et~al.}(2004){Ilgner}, {Henning}, {Markwick}, \&
  {Millar}}]{Ilgner2004}
{Ilgner}, M., {Henning}, T., {Markwick}, A.~J., \& {Millar}, T.~J. 2004, \aap,
  415, 643

\bibitem[{{Jeans}(1967)}]{Jeans1967}
{Jeans}, J. 1967, An introduction to the kinetic theory of gases (Cambridge
  University Press)

\bibitem[{{Kamp} {et~al.}(2017){Kamp}, {Thi}, {Woitke}, {Rab}, {Bouma}, \&
  {M{\'e}nard}}]{Kamp2017}
{Kamp}, I., {Thi}, W.~F., {Woitke}, P., {et~al.} 2017, \aap, 607, A41

\bibitem[{{Kreidberg} {et~al.}(2014){Kreidberg}, {Bean}, {D{\'e}sert}, {Line},
  {Fortney}, {Madhusudhan}, {Stevenson}, {Showman}, {Charbonneau},
  {McCullough}, {Seager}, {Burrows}, {Henry}, {Williamson}, {Kataria}, \&
  {Homeier}}]{Kreidberg2014}
{Kreidberg}, L., {Bean}, J.~L., {D{\'e}sert}, J.-M., {et~al.} 2014, \apjl, 793,
  L27

\bibitem[{{Krijt} {et~al.}(2016){Krijt}, {Ciesla}, \& {Bergin}}]{Krijt2016}
{Krijt}, S., {Ciesla}, F.~J., \& {Bergin}, E.~A. 2016, \apj, 833, 285

\bibitem[{{Krijt} {et~al.}(2018){Krijt}, {Schwarz}, {Bergin}, \&
  {Ciesla}}]{Krijt2018}
{Krijt}, S., {Schwarz}, K.~R., {Bergin}, E.~A., \& {Ciesla}, F.~J. 2018, \apj,
  864, 78

\bibitem[{{Kuramoto} {et~al.}(2013){Kuramoto}, {Umemoto}, \&
  {Ishiwatari}}]{Kuramoto2013}
{Kuramoto}, K., {Umemoto}, T., \& {Ishiwatari}, M. 2013, Earth and Planetary
  Science Letters, 375, 312

\bibitem[{{Lesur} {et~al.}(2022){Lesur}, {Ercolano}, {Flock}, {Lin}, {Yang},
  {Barranco}, {Benitez-Llambay}, {Goodman}, {Johansen}, {Klahr}, {Laibe},
  {Lyra}, {Marcus}, {Nelson}, {Squire}, {Simon}, {Turner}, {Umurhan}, \&
  {Youdin}}]{Lesur2022}
{Lesur}, G., {Ercolano}, B., {Flock}, M., {et~al.} 2022, arXiv e-prints,
  arXiv:2203.09821

\bibitem[{{Manara} {et~al.}(2016){Manara}, {Rosotti}, {Testi}, {Natta},
  {Alcal{\'a}}, {Williams}, {Ansdell}, {Miotello}, {van der Marel}, {Tazzari},
  {Carpenter}, {Guidi}, {Mathews}, {Oliveira}, {Prusti}, \& {van
  Dishoeck}}]{Manara2016}
{Manara}, C.~F., {Rosotti}, G., {Testi}, L., {et~al.} 2016, \aap, 591, L3

\bibitem[{{McElroy} {et~al.}(2013){McElroy}, {Walsh}, {Markwick}, {Cordiner},
  {Smith}, \& {Millar}}]{McElroy2013}
{McElroy}, D., {Walsh}, C., {Markwick}, A.~J., {et~al.} 2013, \aap, 550, A36

\bibitem[{{Moses} {et~al.}(2011){Moses}, {Visscher}, {Fortney}, {Showman},
  {Lewis}, {Griffith}, {Klippenstein}, {Shabram}, {Friedson}, {Marley}, \&
  {Freedman}}]{Moses2011}
{Moses}, J.~I., {Visscher}, C., {Fortney}, J.~J., {et~al.} 2011, \apj, 737, 15

\bibitem[{{Mulders} {et~al.}(2017){Mulders}, {Pascucci}, {Manara}, {Testi},
  {Herczeg}, {Henning}, {Mohanty}, \& {Lodato}}]{Mulders2017}
{Mulders}, G.~D., {Pascucci}, I., {Manara}, C.~F., {et~al.} 2017, \apj, 847, 31

\bibitem[{{Nomura} \& {Millar}(2005)}]{Nomura2005}
{Nomura}, H. \& {Millar}, T.~J. 2005, \aap, 438, 923

\bibitem[{{{\"O}berg} \& {Bergin}(2021)}]{Oberg2021}
{{\"O}berg}, K.~I. \& {Bergin}, E.~A. 2021, \physrep, 893, 1

\bibitem[{{Parmentier} {et~al.}(2013){Parmentier}, {Showman}, \&
  {Lian}}]{Parmentier2013}
{Parmentier}, V., {Showman}, A.~P., \& {Lian}, Y. 2013, \aap, 558, A91

\bibitem[{{Pinte} {et~al.}(2018){Pinte}, {M{\'e}nard}, {Duch{\^e}ne}, {Hill},
  {Dent}, {Woitke}, {Maret}, {van der Plas}, {Hales}, {Kamp}, {Thi}, {de
  Gregorio-Monsalvo}, {Rab}, {Quanz}, {Avenhaus}, {Carmona}, \&
  {Casassus}}]{Pinte2018}
{Pinte}, C., {M{\'e}nard}, F., {Duch{\^e}ne}, G., {et~al.} 2018, \aap, 609, A47

\bibitem[{{Pinte} {et~al.}(2022){Pinte}, {Teague}, {Flaherty}, {Hall},
  {Facchini}, \& {Casassus}}]{Pinte2022}
{Pinte}, C., {Teague}, R., {Flaherty}, K., {et~al.} 2022, arXiv e-prints,
  arXiv:2203.09528

\bibitem[{{Pontoppidan} {et~al.}(2010){Pontoppidan}, {Salyk}, {Blake},
  {Meijerink}, {Carr}, \& {Najita}}]{Pontoppidan2010}
{Pontoppidan}, K.~M., {Salyk}, C., {Blake}, G.~A., {et~al.} 2010, \apj, 720,
  887

\bibitem[{{Rab} {et~al.}(2020){Rab}, {Kamp}, {Dominik}, {Ginski}, {Muro-Arena},
  {Thi}, {Waters}, \& {Woitke}}]{Rab2020}
{Rab}, C., {Kamp}, I., {Dominik}, C., {et~al.} 2020, \aap, 642, A165

\bibitem[{{Rimmer} \& {Helling}(2016)}]{Rimmer2016}
{Rimmer}, P.~B. \& {Helling}, C. 2016, \apjs, 224, 9

\bibitem[{{Schr{\"a}pler} \& {Henning}(2004)}]{Schraepler2004}
{Schr{\"a}pler}, R. \& {Henning}, T. 2004, \apj, 614, 960

\bibitem[{{Semenov} \& {Wiebe}(2011)}]{Semenov2011}
{Semenov}, D. \& {Wiebe}, D. 2011, \apjs, 196, 25

\bibitem[{{Semenov} {et~al.}(2006){Semenov}, {Wiebe}, \&
  {Henning}}]{Semenov2006}
{Semenov}, D., {Wiebe}, D., \& {Henning}, T. 2006, \apjl, 647, L57

\bibitem[{{Shakura} \& {Syunyaev}(1973)}]{Shakura1973}
{Shakura}, N.~I. \& {Syunyaev}, R.~A. 1973, \aap, 24, 337

\bibitem[{{Thi} {et~al.}(2020{\natexlab{a}}){Thi}, {Hocuk}, {Kamp}, {Woitke},
  {Rab}, {Cazaux}, \& {Caselli}}]{Thi2020a}
{Thi}, W.~F., {Hocuk}, S., {Kamp}, I., {et~al.} 2020{\natexlab{a}}, \aap, 634,
  A42

\bibitem[{{Thi} {et~al.}(2020{\natexlab{b}}){Thi}, {Hocuk}, {Kamp}, {Woitke},
  {Rab}, {Cazaux}, {Caselli}, \& {D'Angelo}}]{Thi2020b}
{Thi}, W.~F., {Hocuk}, S., {Kamp}, I., {et~al.} 2020{\natexlab{b}}, \aap, 635,
  A16

\bibitem[{{Thi} {et~al.}(2019){Thi}, {Lesur}, {Woitke}, {Kamp}, {Rab}, \&
  {Carmona}}]{Thi2019}
{Thi}, W.~F., {Lesur}, G., {Woitke}, P., {et~al.} 2019, \aap, 632, A44

\bibitem[{{Wiesemann}(2014)}]{Wiesemann2014}
{Wiesemann}, K. 2014, arXiv e-prints, arXiv:1404.0509

\bibitem[{{Woitke} {et~al.}(2009){Woitke}, {Kamp}, \& {Thi}}]{Woitke2009a}
{Woitke}, P., {Kamp}, I., \& {Thi}, W. 2009, \aap, 501, 383

\bibitem[{{Woitke} {et~al.}(2016){Woitke}, {Min}, {Pinte}, {Thi}, {Kamp},
  {Rab}, {Anthonioz}, {Antonellini}, {Baldovin-Saavedra}, {Carmona}, {Dominik},
  {Dionatos}, {Greaves}, {G{\"u}del}, {Ilee}, {Liebhart}, {M{\'e}nard},
  {Rigon}, {Waters}, {Aresu}, {Meijerink}, \& {Spaans}}]{Woitke2016}
{Woitke}, P., {Min}, M., {Pinte}, C., {et~al.} 2016, \aap, 586, A103

\bibitem[{{Woitke} {et~al.}(2018){Woitke}, {Min}, {Thi}, {Roberts}, {Carmona},
  {Kamp}, {M{\'e}nard}, \& {Pinte}}]{Woitke2018}
{Woitke}, P., {Min}, M., {Thi}, W.~F., {et~al.} 2018, \aap, 618, A57

\bibitem[{{Youdin} \& {Lithwick}(2007)}]{Youdin2007}
{Youdin}, A.~N. \& {Lithwick}, Y. 2007, \icarus, 192, 588

\bibitem[{{Zahnle} {et~al.}(2016){Zahnle}, {Marley}, {Morley}, \&
  {Moses}}]{Zahnle2016}
{Zahnle}, K., {Marley}, M.~S., {Morley}, C.~V., \& {Moses}, J.~I. 2016, \apj,
  824, 137

\bibitem[{{Zhang} \& {Showman}(2018)}]{Zhang2018}
{Zhang}, X. \& {Showman}, A.~P. 2018, \apj, 866, 1

\end{thebibliography}

\appendix
\section{Numerical implementation of mixing}
\label{app:numerical}

\subsection{Numerical scheme}

Let us consider height grid points $\{z_k\,|\,k=1,...,K\}$.  The basic
reaction-diffusion equation (Eq.\,\ref{eq:chemsys}) is numerically
discretised as
\begin{eqnarray}
  \pabl{n_{i,k}}{t} = P_{i,k} - L_{i,k} -
  \frac{\Phi_{i,k+\half}-\Phi_{i,k-\half}}{z_{k+\half}-z_{k-\half}}\ ,
  \label{chemsys}
\end{eqnarray}  
where the notation means $x_{k\pm \half} = \frac{1}{2}(x_k+x_{k\pm
  1})$, and
\begin{align}
  \Phi_{i,k+\half} =& -(K_{k+\half}+D_{i,k+\half})\,n_{k+\half}\,
  \frac{\frac{n_{i,k+1}}{n_{k+1}}-\frac{n_{i,k}}{n_k}}{z_{k+1}-z_k}\nonumber\\
                 & +\,D_{i,k+\half}\,n_{i,k+\half}\,
                    \frac{(\mu_{k+\half}-m_i)\,g_{k+\half}}{kT_{k+\half}} \\
  \Phi_{i,k-\half} =& -(K_{k-\half}+D_{i,k-\half})\,n_{k-\half}\,
  \frac{\frac{n_{i,k}}{n_{k}}-\frac{n_{i,k-1}}{n_{k-1}}}{z_k-z_{k-1}}\nonumber\\
                  & +\,D_{i,k-\half}\,n_{i,k-\half}\,
                    \frac{(\mu_{k-\half}-m_i)\,g_{k-\half}}{kT_{k-\half}}
\end{align}
Assuming that all quantities are given and constant except for $n_{i,k-1}$,
$n_{i,k}$ and $n_{i,k+1}$ we write
\begin{equation}
  -\,\frac{\Phi_{i,k+\half}-\Phi_{i,k-\half}}{z_{k+\half}-z_{k-\half}}
  = {\cal A}_{i,k}\,n_{i,k+1} ~-~ {\cal B}_{i,k}\,n_{i,k}
                           ~+~ {\cal C}_{i,k}\,n_{i,k-1} 
\end{equation}
with
\begin{align}
  {\cal A}_{i,k} =&~ \textstyle\frac{1}{z_{k+\half}-z_{k-\half}}\bigg(
  +(K_{k+\half}+D_{i,k+\half})\,\frac{n_{k+\half}}{n_{k+1}(z_{k+1}-z_k)}\nonumber\\
  &\hspace*{14mm}\textstyle
           -\,D_{i,k+\half}\,\frac{1}{2}\,\frac{(\mu_{k+\half}-m_i)\,g_{k+\half}}
                                       {kT_{k+\half}}\bigg) \label{rateA}\\
  {\cal B}_{i,k} =&~ \textstyle\frac{1}{z_{k+\half}-z_{k-\half}}\bigg(
  +(K_{k+\half}+D_{i,k+\half})\,\frac{n_{k+\half}}{n_k(z_{k+1}-z_k)}\nonumber\\
  &\hspace*{14mm}\textstyle
           +\,D_{i,k+\half}\,\frac{1}{2}\,\frac{(\mu_{k+\half}-m_i)\,g_{k+\half}}
        {kT_{k+\half}}\nonumber\\
  &\hspace*{14mm}\textstyle
  +\;(K_{k-\half}+D_{i,k-\half})\,\frac{n_{k-\half}}{n_k(z_k-z_{k-1})}\nonumber\\
  &\hspace*{14mm}\textstyle
  -\,D_{i,k-\half}\,\frac{1}{2}\,\frac{(\mu_{k-\half}-m_i)\,g_{k-\half}}
        {kT_{k-\half}}\bigg) \label{rateB}\\ 
  {\cal C}_{i,k} =&~ \textstyle\frac{1}{z_{k+\half}-z_{k-\half}}\bigg(
  +(K_{k-\half}+D_{i,k-\half})\,\frac{n_{k-\half}}{n_{k-1}(z_k-z_{k-1})}\nonumber\\
  &\hspace*{14mm}\textstyle
  +\,D_{i,k-\half}\,\frac{1}{2}\,\frac{(\mu_{k-\half}-m_i)\,g_{k-\half}}
        {kT_{k-\half}}\bigg) \label{rateC}
\end{align}
The apparent $\pm$ asymmetries in the $(K+D)$ and $D$ terms are
correct; they reflect the different nature of mixing, which is
modelled as divergence of a gradient in concentration ($\rm 2^{nd}$
order), and gravitational de-mixing, which is modelled as divergence
of a mass-dependent flux ($\rm 1^{st}$ order).

\subsection{Boundary condition}
We apply zero-flux boundary conditions at $k\!=\!1$ and $k\!=\!K$
\begin{eqnarray}
  \Phi_{i,\half} ~=~ \Phi_{i,K+\half} ~=~ 0
\end{eqnarray}
In practise, this means that all terms in Eqs.~(\ref{rateA}) to
(\ref{rateC}) with $K_{k-\half}$ or
$D_{i,k-\half}$ are dropped at $k\!=\!1$, and all terms with $K_{k+\half}$ or
$D_{i,k+\half}$ are dropped at $k\!=\!K$. For the pre-factor, we
approximate $z_{k+\half}-z_{k-\half}\approx z_2-z_1$ at $k=1$ and
$z_{k+\half}-z_{k-\half}\approx z_K-z_{K-1}$ at $k=K$.

\begin{figure}
  \centering
  \includegraphics[width=87mm,trim=12 10 2 15,clip]{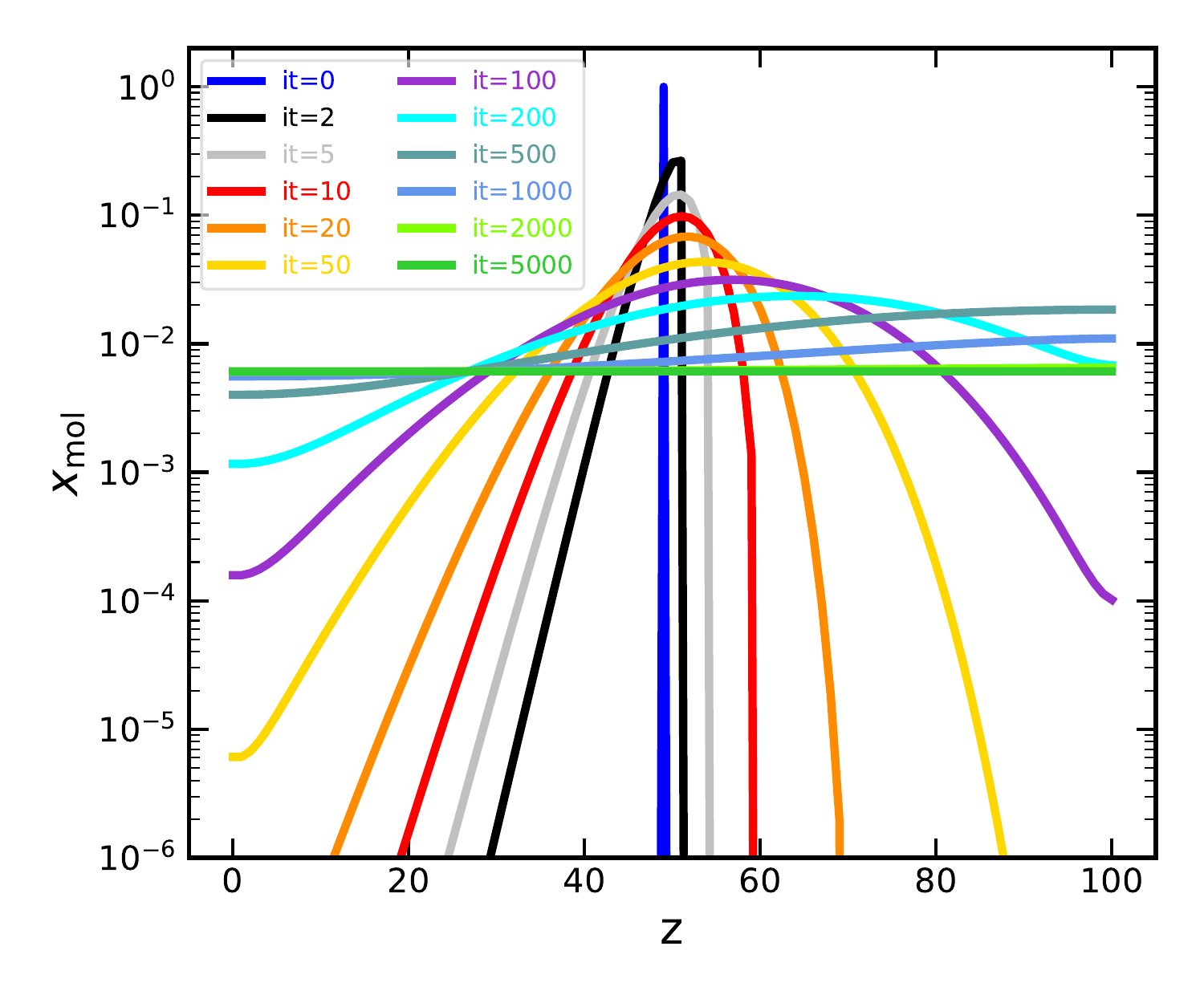}\\[-2mm]
  \includegraphics[width=87mm,trim=0 20 3 10,clip]{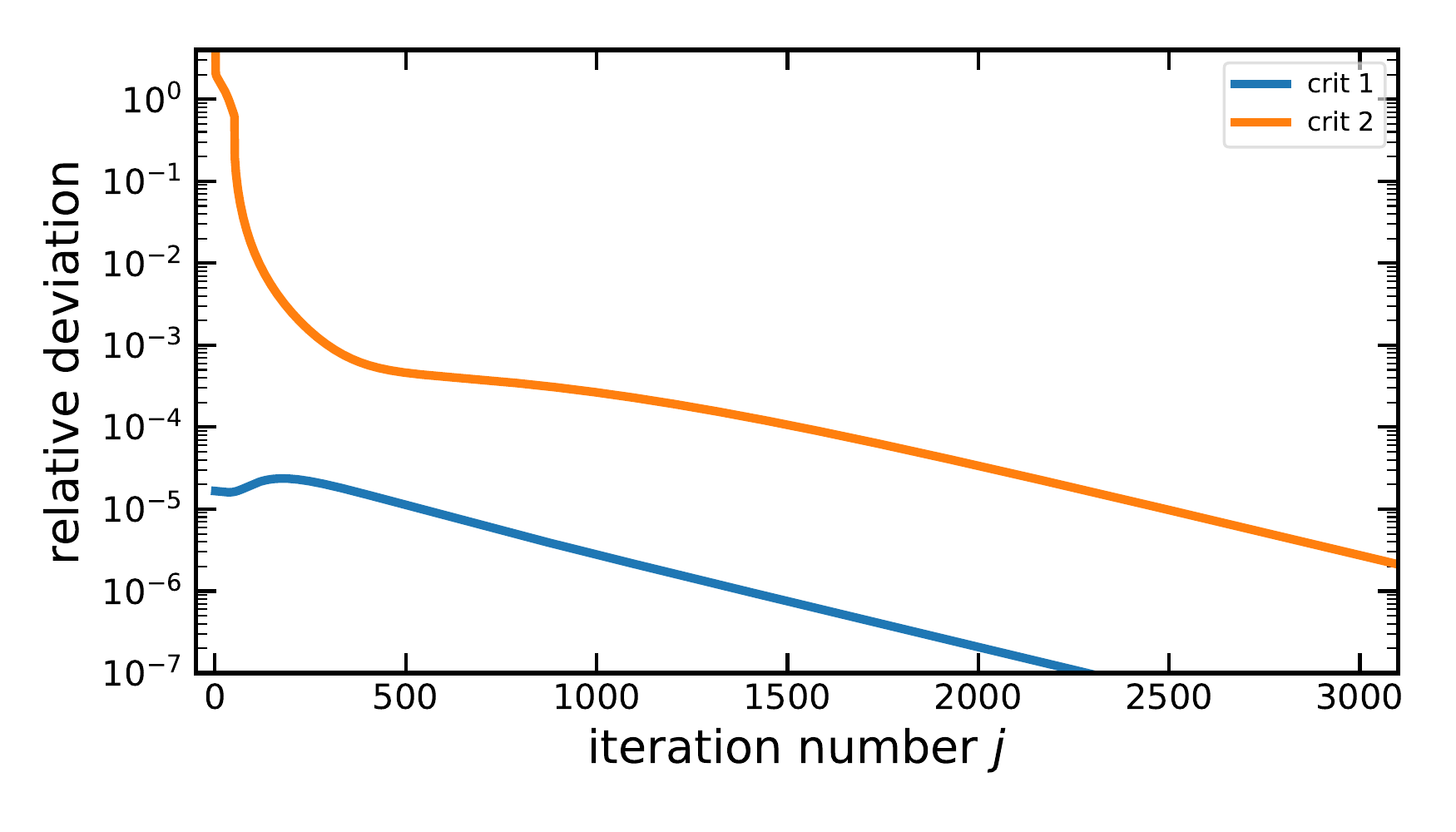}
  \caption{Test of the iterative numerical method to solve the
    diffusion equation without chemical sources and sinks. We consider
    a $\delta$-peak for the initial concentration $x_{\rm mol}=n_{i,k}/n_k$ at the
    start of the iterations in a Gaussian density structure, see
    text. The upper figure shows the concentrations as function of the
    number of iterations carried out. The lower figure shows the relative
    deviations between iterations, see Eqs.\,(\ref{eq:crit1}) and
    (\ref{eq:crit2}).}
  \label{fig:verify}
\end{figure}

\subsection{Initial condition}
The very first disc iteration is computed with zero mixing rates.
In all forthcoming iterations, the mixing rates are activated.

\subsection{Test of the numerical solution method}
\label{app:convergence}
To verify our numerical implementation, we made a quick test to check
the capabilities of our simple iterative scheme.  We considered a
Gaussian density profile on an equidistant spatial grid with 101
points $k\!=\!1 ... 101$
\begin{align}
  z_k    =&~ k-1\\
  n_k    =&~ \exp\Big(-\Big(\frac{z_k}{20}\Big)^2\Big) \\
  K_k    =&~ 1\\
  D_{i,k} =&~ 0
\end{align}
and choose, for a single test species $i\!=\!1$, the following
initial condition
\begin{equation}
  n_{i,k} = \left\{\begin{array}{ll}
  1 & ,\ k=50\\
  0 & ,\ \mbox{otherwise}
  \end{array}\right. 
\end{equation}
at the start of the iteration. Then we copied the exact
implementation of the vertical mixing in {\sc ProDiMo} to an external
tool where we omit the chemical source and sink terms $P_{i,k}$
and $L_{i,k}$.  In a loop from $k\!=\!K$ down to 1, we then
\begin{itemize}
  \item[1)] calculate ${\cal A}_{i,k}$, ${\cal B}_{i,k}$ and ${\cal
    C}_{i,k}$, and\\[-7pt]
  \item[2)] solve the rate equation without chemical sources and sinks
    $n_{i,k} = \frac{1}{{\cal B}_{i,k}}
    \left({\cal A}_{i,k}\,n_{i,k+1} + {\cal C}_{i,k}\,n_{i,k-1}\right)$.
\end{itemize}
This numerical scheme mimics the downward sweep in {\sc ProDiMo} while
solving the chemistry in the time-independent case from top to
bottom. Figure~\ref{fig:verify} shows the resulting concentrations
$n_{i,k}/n_k$ as function of $z$, where $z\!=\!0$ represents the
midplane. The results after selected numbers of iterations are plotted
with different colours as indicated. This plot can be compared to our
results with chemical sources and sinks plotted in Fig.~\ref{fig:iteration1}
in the main part of the paper.

The figure indicates that up to 5000 iterations are required to reach
a constant concentration as physically expected in this setup.  The
final value of that constant concentration results to be 0.61\%, which
is close to the predicted value of 0.63\% that we get from a numerical
integral of $n_{ik}(z)$ over $z$ in the initial state, divided by the
numerical integral of $n(z)$. We note that in real applications that
include the chemical source terms, the chemicals will rarely travel
many scale heights. Instead, there are normally some dominant, but
spatially restricted formation and destruction regions, and the model
must properly describe the diffusive transport between those
regions, hence we expect a couple of hundred iterations to be
sufficient.

Figure~\ref{fig:verify} demonstrates the numerical behaviour during
the course of the iterations:
\begin{itemize}
  \item The first few tens of iterations are featured by an asymmetry,
    because the resultant $n_{i,k+1}$ is used already for the determination of
    $n_{i,k}$ whereas $n_{i,k-1}$ is not.  Therefore, the
    information of a certain molecule being mixed downward is
    travelling faster downwards than it travels upwards.
  \item After some hundreds of iterations, however, the upper regions
    fill in quicker than the lower regions.  This is because of the
    density gradient in the system, i.e.\ a large reservoir in the denser
    lower disc regions is able to more quickly fill in the low-density
    regions above than vice versa. 
\end{itemize}


\end{document}